\documentclass[twocolumn,twoside]{IEEEtran}

\usepackage[table,xcdraw,dvipsnames]{xcolor} 
\definecolor{Gray}{HTML}{F0F0F0}
\definecolor{myblue}{RGB}{173,216,230}  
\definecolor{myred}{RGB}{244,177,131} 
\definecolor{myyellow}{RGB}{255,226,166}  
\definecolor{mygreen}{RGB}{198,230,162} 
\definecolor{mygray}{RGB}{242,242,242} 
\colorlet{shadecolor}{yellow}

\usepackage{amsmath, amsthm, amsfonts, amssymb, amsbsy} 
\usepackage[cmex10]{mathtools} 
\usepackage{bm} 

\usepackage{array}
\usepackage{booktabs} 
\usepackage{makecell} 
\usepackage{multirow} 
\usepackage{tabularx} 
\usepackage{threeparttable} 
\usepackage{stfloats} 

\usepackage[pdftex]{graphicx}
\graphicspath{{../pdf/}{../jpeg/}}
\DeclareGraphicsExtensions{.pdf,.jpeg,.png}
\usepackage{epstopdf} 
\usepackage{subfig} 
\usepackage{caption} 
\usepackage[figuresright]{rotating} 

\usepackage{ragged2e} 
\usepackage{setspace} 
\renewcommand{\baselinestretch}{1.1}
\usepackage{soul} 
\usepackage{framed} 
\usepackage{microtype} 
\usepackage{url} 

\usepackage[noend]{algpseudocode}
\usepackage{algorithmicx,algorithm}

\usepackage{cases} 
\usepackage{dsfont} 
\usepackage{balance} 
\usepackage{textcomp} 
\usepackage{amscd} 

\usepackage{eqparbox} 
\usepackage{cite} 
\usepackage{multicol} 

\usepackage[breaklinks]{hyperref} 
\usepackage{flushend,cuted} 

\newcommand{\RNum}[1]{\uppercase\expandafter{\romannumeral #1\relax}}

\begin{document}
\title{Multibeam High Throughput Satellite: Hardware Foundation, Resource Allocation, and Precoding}

\author{Rui Chen,~\IEEEmembership{Member,~IEEE,} Wen-Xuan Long,~\IEEEmembership{Member,~IEEE,} Bing-Qian Wang,~\IEEEmembership{Graduate Student Member,~IEEE,} Yuan He,~\IEEEmembership{Graduate Student Member,~IEEE,} Rui-Jin Sun,~\IEEEmembership{Member,~IEEE,}  Nan Cheng,~\IEEEmembership{Senior Member,~IEEE,}  Gan Zheng,~\IEEEmembership{Fellow,~IEEE,} and Dusit Niyato,~\IEEEmembership{Fellow,~IEEE}

\thanks{Rui Chen, Bing-Qian Wang, Yuan He, Rui-Jin Sun and Nan Cheng are with the State Key Laboratory of ISN, Xidian University, Xi'an 710071, China, Xi'an 710068, China (e-mail: rchen@xidian.edu.cn; wangbingqian@stu.xidian.edu.cn; xdyuanhe@gmail.com; sunruijin@xidian.edu.cn; nancheng@xidian.edu.cn).}

\thanks{Wen-Xuan Long is with the Dipartimento di Ingegneria dell’Informazione,
University of Pisa, 56126 Pisa, Italy (e-mail: wenxuan.long@ing.unipi.it).}

\thanks{Gan Zheng is with the School of Engineering, University of Warwick, CV4 7AL Coventry, U.K. (e-mail: gan.zheng@warwick.ac.uk).} %

\thanks{Dusit Niyato is with the College of Computing and Data Science, Nanyang
Technological University, Singapore. (email: dniyato@ntu.edu.sg).}
}

\maketitle
\begin{abstract}

With its wide coverage and uninterrupted service, satellite communication is a critical technology for next-generation 6G communications. High throughput satellite (HTS) systems, utilizing multipoint beam and frequency multiplexing techniques, enable satellite communication capacity of up to Tbps to meet the growing traffic demand. Therefore, it is imperative to review the-state-of-the-art of multibeam HTS systems and identify their associated challenges and perspectives. Firstly, we summarize the multibeam HTS hardware foundations, including ground station systems, on-board payloads, and user terminals. Subsequently, we review the flexible on-board radio resource allocation approaches of bandwidth, power, time slot, and  joint allocation schemes of HTS systems to optimize resource utilization and cater to non-uniform service demand. Additionally, we survey multibeam precoding methods for the HTS system to achieve full-frequency reuse and interference cancellation, which are classified according to different deployments such as single gateway precoding, multiple gateway precoding, on-board precoding, and hybrid  on-board/on-ground  precoding. Finally, we disscuss the challenges related to Q/V band link outage, time and frequency synchronization of gateways, the accuracy of channel state information (CSI), payload light-weight development, and the application of deep learning (DL). Research on these topics will contribute to enhancing the performance of HTS systems and finally delivering high-speed data to areas underserved by terrestrial networks.


\end{abstract}


\section{Introduction}

\begin{figure}[t]
\begin{center}
\includegraphics[width=8.5cm]{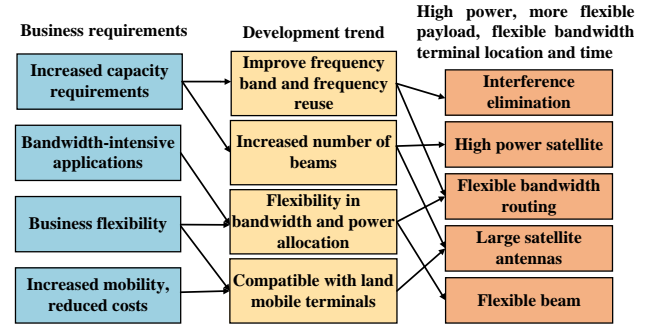}
\end{center}
\caption{The development direction of satellite communication technology.}%
\label{Develop}
\end{figure}

\IEEEPARstart{W}ith the evolution of mobile communications, terrestrial wireless networks have advanced rapidly each decade since the 1980s, providing high-speed access to much of the world’s population \cite{andrews2014will,10648786}. However, their coverage remains limited in remote areas and during natural disasters due to infrastructure vulnerability \cite{perez2019signal,lin2021supporting,an2016secure}. Satellite communication (SatCom), offering global coverage and high reliability, is thus seen as a key complement to terrestrial networks in 6G systems \cite{giordani2020satellite,giordani2020non,chen2020system,khan2022ris,dang2020should}. To satisfy growing data demands \cite{long2021promising,chen2022reconfigurable,10555049,10614084}, SatCom is evolving from traditional services to high-capacity applications, with high throughput satellite (HTS) systems enabling terabit-level transmission to support future military, civilian, and commercial needs \cite{RN02}, as shown in Fig. \ref{Develop}.

In 2008, Northern Sky Research (NSR), an American aerospace consulting firm, introduced the concept of HTS. It defined HTS as a satellite employing multiple spot beams and frequency reuse technologies, capable of achieving a total system throughput several times greater than that of traditional fixed satellite service (FSS) satellites using the same spectrum resources \cite{WXYG201611012}. Since then, the industry has gradually reached a consensus on the definition of HTS: characterized by spot beams and frequency reuse. HTS can operate in any frequency band, with throughput varying depending on the allocated spectrum and reuse factor. It can support a wide range of commercial satellite communication services, including fixed, broadcast, and mobile services.

\begin{figure}[t]
\centering
\subfloat[Traditional FSS satellites]{\includegraphics[height=4cm,width=4cm]{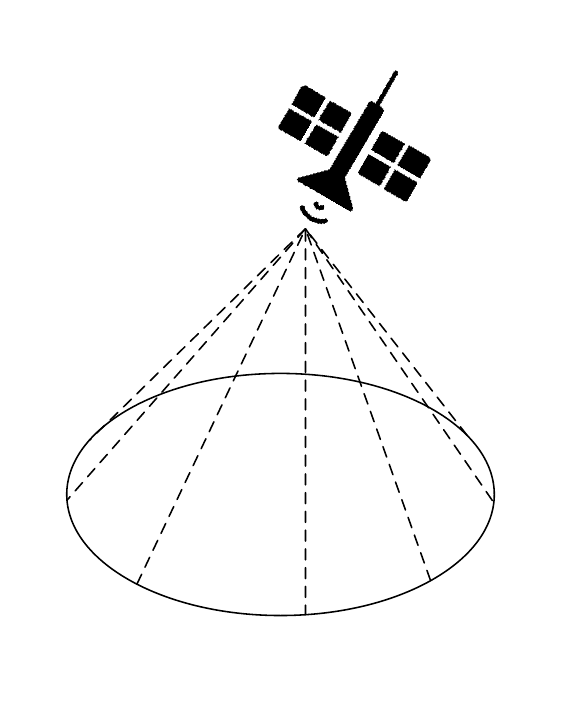}}
\subfloat[Multibeam HTS]{\includegraphics[height=4cm,width=4cm]{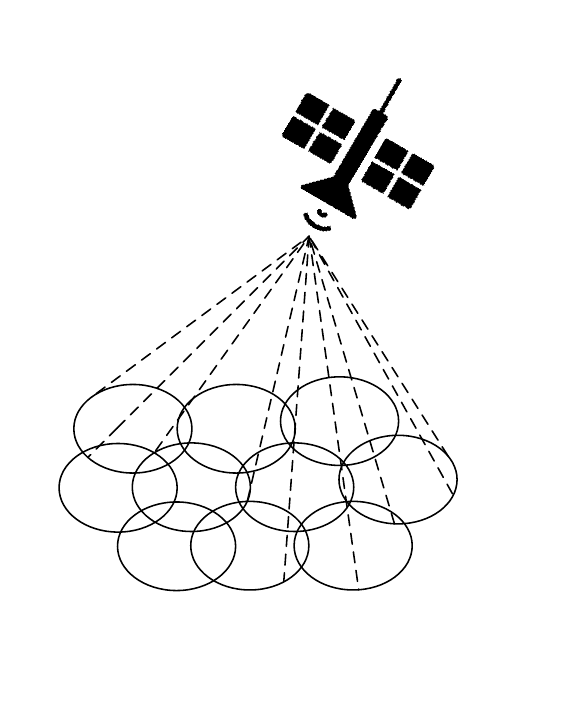}}
\caption{Comparison of beam coverage between traditional FSS satellites and multibeam satellites
 \cite{nguyen2020overview}.}
\label{Typical commercial satellites and multibeam HTS configurations}
\end{figure}

One of the most fundamental features of HTS systems is the use of multiple spot beams combined with aggressive frequency reuse. Unlike traditional FSS systems, which typically employ a single wide beam to cover a large area, HTS systems utilize multiple narrowly focused spot beams. These spot beams, generated by advanced multibeam antennas, enable targeted and efficient transmission between the satellite and ground terminals. As illustrated in Fig. \ref{Typical commercial satellites and multibeam HTS configurations} and summarized in Table \ref{Comparison}, multibeam HTS systems typically deploy between 100 and over 3000 beams, representing a 10 to 150 times increase compared to the 1 to 20 beams used in traditional FSS systems. The coverage area of a typical spot beam ranges from 300 to 700 km, which is significantly smaller than the approximately 2000 km coverage of a conventional FSS beam \cite{10555049,10614084}. This dramatic difference stems from the core architectural design of HTS, which enhances system capacity through aggressive frequency reuse enabled by spot beams. The implementation of such a design relies heavily on key technologies such as digital beamforming, software-defined payloads, and dense gateway deployment, marking a fundamental shift in satellite communications toward higher-density coverage and lower-cost transmission.

Multibeam technology  serves as a cornerstone of modern HTS architecture, enabling a shift from traditional wide beam systems to highly efficient, dense beam configurations.  While traditional FSS satellites operate predominantly in geostationary earth orbit (GEO) with a single beam and limited throughput ($\sim$1–10 Gbps), HTS systems leverage multiple narrow beams across GEO, MEO, and LEO orbits to achieve system-level performance improvements \cite{WXYG201611012}. This transition enables several performance enhancements, including reduced spillover loss, increased spectral efficiency, improved power utilization, and more dynamic resource allocation \cite{10563886}. Over the past two decades, multibeam satellites have evolved toward HTS systems with greater capacity, efficiency, and density.

Most currently deployed multibeam HTS systems are based in GEO, benefiting from simpler deployment and wide-area coverage. Since 2015, numerous emerging HTS constellation projects have been proposed, particularly in non-GEO orbits. For consumer-oriented internet access services, GEO HTS satellites are typically preferred due to their sufficient coverage and lower deployment complexity \cite{WXYG201706017}. In contrast, non-GEO HTS constellations are better suited for commercial applications such as VSAT networks, trunking, and cellular backhaul, where lower latency and wider coverage are essential \cite{nguyen2020overview}. Mobile applications may employ either GEO or non-GEO systems, depending on specific service requirements. Detailed information about GEO HTS systems and non-GEO HTS  constellations is provided in Section \RNum{2}.B.

\subsection{Key Techniques of Multibeam HTS}
 \begin{figure*}[]
\begin{center}
\includegraphics[width=14cm]{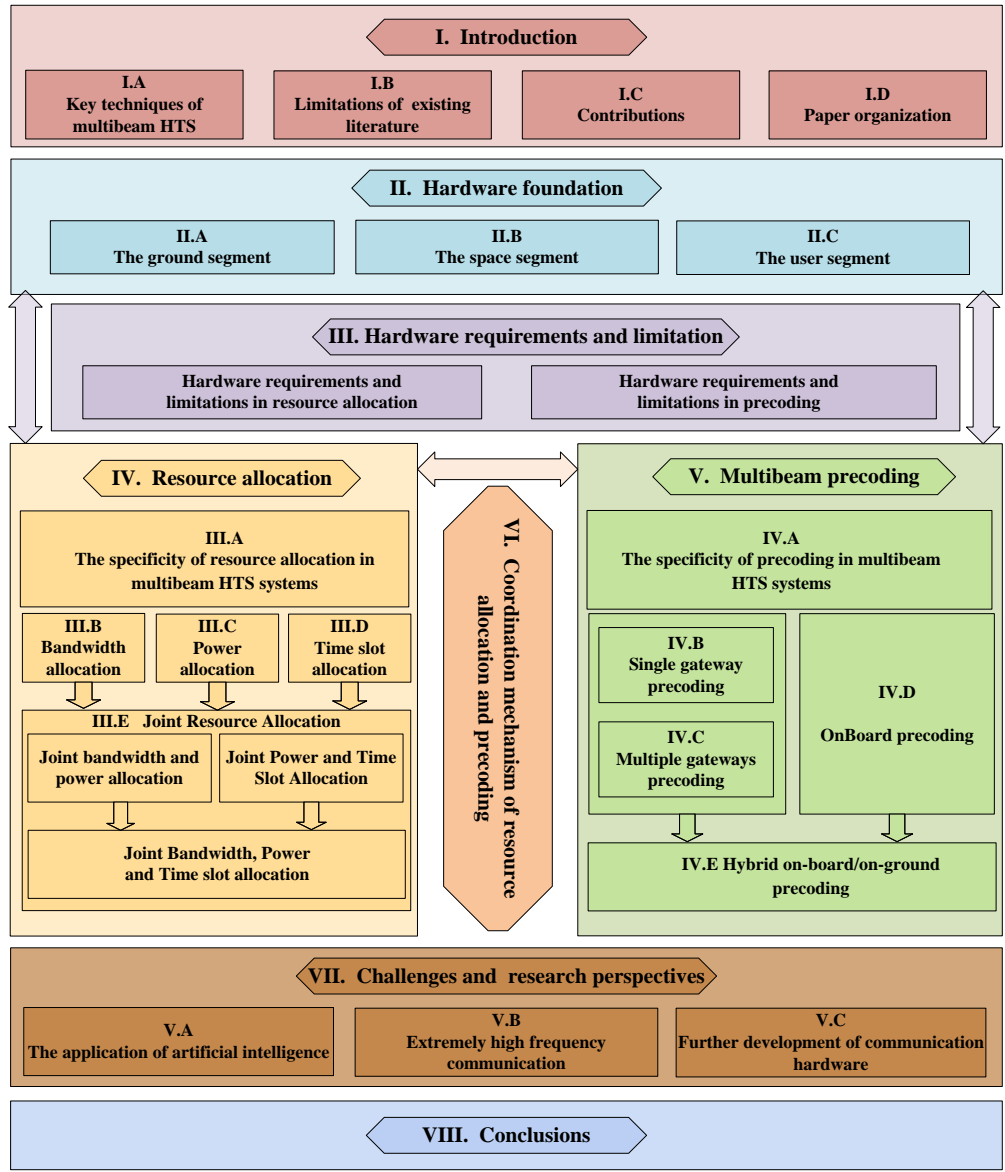}
\end{center}
\caption{Organization of the review article.}%
\label{SectionContact}
\end{figure*}

\begin{table}[t]
\centering
\renewcommand\arraystretch{1.1} 
\caption{Comparison of Traditional FSS Satellites and Multibeam HTS  Satellites \cite{nguyen2020overview}.}
\resizebox{1\linewidth}{!}{
\begin{tabular}{c|l|l}
\hline \hline
\textbf{Comparison factor}                                                        & \multicolumn{1}{c|}{\textbf{\begin{tabular}[c]{@{}c@{}}Traditional FSS \\ satellite\end{tabular}}} & \multicolumn{1}{c}{\textbf{\begin{tabular}[c]{@{}c@{}}Multibeam HTS \\ satellite\end{tabular}}}                  \\ \hline
\begin{tabular}[c]{@{}c@{}}Operational \\ frequency band\end{tabular}             & \begin{tabular}[c]{@{}l@{}}C-band, Ku-band, \\ Ka-band\end{tabular}                                & Ku-band, Ka-band                                                                                                 \\ \hline
Number of beams & \begin{tabular}[c]{@{}l@{}}Large single beam \\($\sim$1–20)\end{tabular} & \begin{tabular}[c]{@{}l@{}}Multiple narrow beams\\ ($\sim$100–3000+)\end{tabular}  \\ \hline 

Typical orbit type                                                                & GEO                                                                                                & \begin{tabular}[c]{@{}l@{}}GEO, MEO, LEO  (including \\hybrid  architectures)\end{tabular}                     \\ \hline
\begin{tabular}[c]{@{}c@{}}Throughput\\ capability (Gbps)\end{tabular}            & $\sim$1–10                                                                                    & \begin{tabular}[c]{@{}l@{}}$\sim$5–300+ (with frequency\\ reuse in multiple spot beam)\end{tabular}              \\ \hline
\begin{tabular}[c]{@{}c@{}}Typical cost \\ including\\  launch (USD)\end{tabular} & $\sim$200–300                                                                                      & \begin{tabular}[c]{@{}l@{}}$\sim$300–500 (cost can be\\ twice of regular satellite)\end{tabular}                 \\ \hline
\begin{tabular}[c]{@{}c@{}}Multiple access\\  technology\end{tabular}             & FDMA, TDMA                                                                                         & MF-TDMA                                                                                                          \\ \hline
Services                                                                     & \begin{tabular}[c]{@{}l@{}}Early space communi-\\ cations, TV broadcasts.\end{tabular}              & \begin{tabular}[c]{@{}l@{}}Broadband services,\\ remote area connectivity,\\ dedicated formobility.\end{tabular} \\ \hline \hline
\end{tabular}
}
\label{Comparison}
\end{table}


Multibeam HTS systems significantly enhance communication capacity while effectively reducing per-unit bandwidth costs. This performance improvement is enabled by advanced hardware foundation such as satellite antennas, and on-board payloads, which provide the physical infrastructure for high-throughput operation.  However, the dense deployment of beams inevitably results in significant inter-beam interference. Moreover, dynamic and uneven user demand poses additional challenges to system-level resource scheduling, thereby increasing hardware complexity and operational costs.
Resource allocation in multibeam HTS systems is inherently challenging due to three main factors:  dynamically changing user distributions,  non-uniform traffic demands, and limited onboard processing capabilities. These issues often lead to inefficient utilization of spectrum and power resources, while further intensifying inter-beam interference in densely packed beam configurations.

Traditional resource management approaches struggle to adapt to such dynamic environments in real time. To overcome these limitations, precoding techniques have been introduced as an advanced signal processing solution. By considering both interference patterns and resource allocation outcomes, precoding enables more adaptive and efficient use of system resources, significantly improving overall performance in interference-limited scenarios.
To provide a structured overview, the key technologies of multibeam HTS systems can be introduced from three main perspectives: hardware foundation, resource allocation, and precoding. Fig. 
 \ref{SectionContact} illustrates the interrelationship among these components.

\begin{figure}[t]
\begin{center}
\includegraphics[width=9cm]{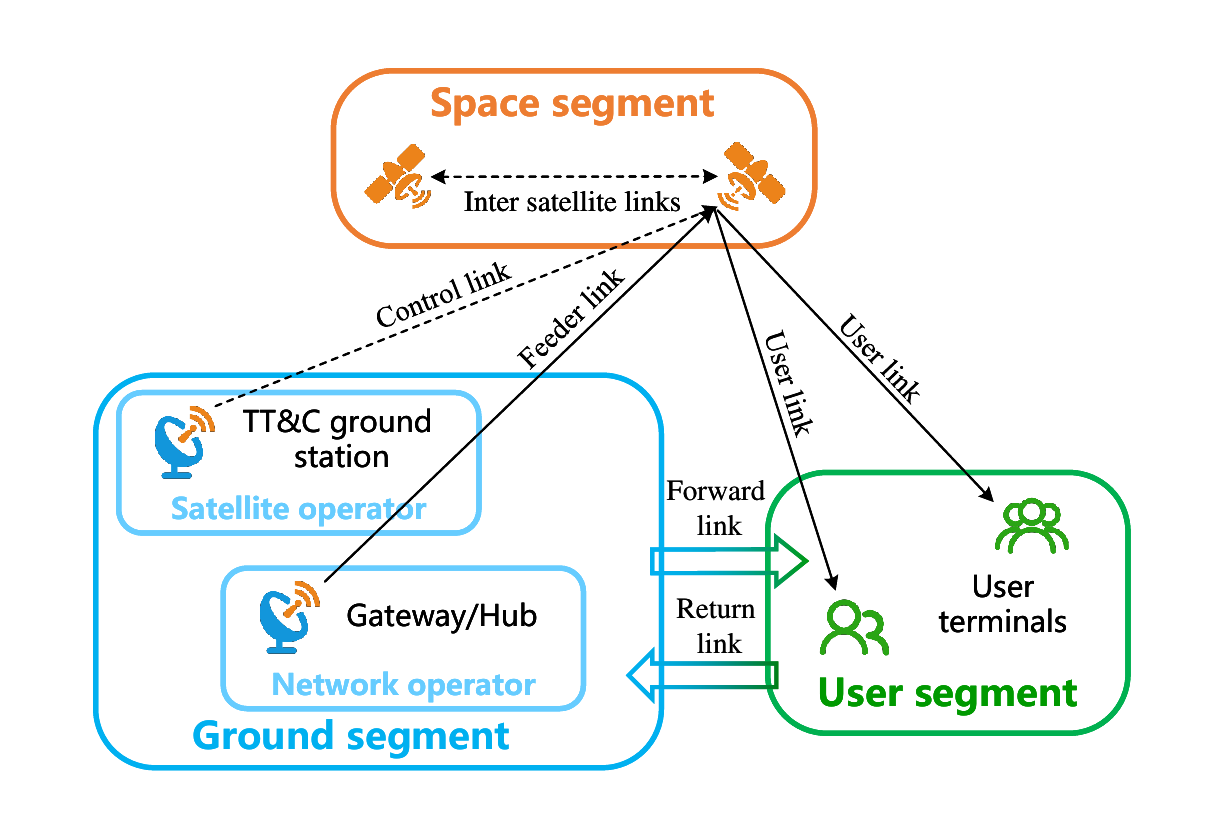}
\end{center}
\caption{A process on satellite communication system architecture, where the basic composition is proposed. In the basic composition, space segment is used for indicating satellite networks that relay signals between the earth and user terminals; Ground segment denotes a variety of ground facilities encompassing gateway  stations and large installations; User segment denotes the terminals positioned on diverse platforms, including stationary and mobile ones.}
\label{Satcom system}
\end{figure}
\emph{1) Hardware Foundation:}
The hardware foundation of multibeam HTS systems serves as the bedrock that enables their high-capacity and wide-coverage capabilities. In terms of hardware foundation, the multibeam HTS system is composed of the three key segments: the ground segment, the space segment, and the user segment, as depicted in Fig. \ref{Satcom system}.  
The ground segment consists of earth stations operated by satellite and network operators. Earth stations are composed of four main hardware components: the RF part, the IF part, the baseband part and the support part \cite{kodheli2020satellite}. In the space segment, a satellite consists of the satellite platform and the payload. There are presently sophisticated HTS platforms in operation, such as the DFH-5 \cite{DFH5} and BSS-702 series \cite{BSS702}. The payloads consist of antennas and transponders. Satellite antennas have developed from omnidirectional low gain antennas to high gain and multiple-point beams antennas \cite{elbert2008introduction}. Especially, a phased array antenna have attracted a lot of attention due to their advantages of wide scanning range, fast scanning speed, and multibeam generation ability. A phased array antenna can steer its beam electronically by changing the phase of the current at each element. Therefore, the beam of a large fixed phased array antenna can be quickly steered from one direction to another without the need to mechanically position a large and heavy antenna. A typical microwave radar phased array radar may have several thousand independent radiating elements, allowing the beam to be switched from one direction to another in a few microseconds or less\cite{6529010}. The transponders have two main types including transparent transponders and regenerative transponders \cite{kolawole2017satellite}. The transparent transponders include analog and digital bent-pipe transponders. The user segment consists of fixed user terminals (UTs) and mobile UTs \cite{ilvcev2017global}. The fixed UTs mainly include butterfly antennas, low noise amplifiers (LNAs), frequency converters, orthogonal mode transducers (OMTs), and indoor units. The mobile UTs include smaller antennas for portability, screen display systems for user interface and sufficient memory for data storage.

\emph{2) Resource Allocation:} 
Based on the aforementioned hardware foundation, efficient resource allocation is crucial for maximizing the performance of multibeam HTS systems. HTS utilizes multiple point beams and frequency multiplexing technology to achieve high communication capacity. By combining bandwidth and frequency multiplexing, HTS can achieve capacities of hundreds of Gbps or even Tbps, which is tens or even hundreds of times higher than traditional satellites. Most current HTS systems, such as ViaSat-3, operate in geosynchronous orbit, providing large coverage and low constellation complexity advantages \cite{hu2012propagation}. 
However, equating bandwidth per beam can result in a lack of flexibility in the satellite system, leading to wasted bandwidth resources in areas with low traffic demands and a shortage of bandwidth resources in areas with high traffic demands. Therefore, it is essential to allocate flexible bandwidth per beam based on the different service requirements and channel conditions of the served users, to prevent unnecessary resource wastage and maximize resource utilization efficiency, ensuring a specific quality of service (QoS) \cite{ha2010performance}. Flexible bandwidth allocation can also enable local resilient networks to quickly access the public network via satellite during disasters when there is a huge demand for high-capacity beams \cite{minoli2015innovations}. HTS achieves flexibility in bandwidth allocation per beam and increased load flexibility by installing digital channelizers as payloads \cite{kyrgiazos2014gateway,kaneko2017construction}. 

In addition to bandwidth resources, on-board power resources are flexibly allocated to each beam to counteract different environmental propagation losses with digital beamforming \cite{xiao2018joint}. Digital beamforming allows the shape of each beam and the power acquired to be adjusted according to user requests, resulting in optimal beam gain \cite{freedman2014advantages}. For time resource allocation, a hopping beam technique with time slice principle \cite{villanti2007differential} is selected in each time slot to allocate resources and fulfill the scenario of uneven service demand while reducing the number of amplifiers without changing the coverage area of HTS \cite{hu2019deep,zhang2018resource}. In summary, the allocation of limited resources is an effective way to achieve better performance under the current frequency multiplexing scheme \cite{han2008resource}.


\emph{3) Precoding:}
To further enhance the performance of multibeam HTS systems, precoding techniques play a vital role in mitigating interference and maximizing system capacity. The forward link in a multibeam HTS system is similar to a multi-user multiple-input multiple-output (MU-MIMO) broadcast channel. In MU-MIMO systems, precoding can effectively mitigate the interference problem of the broadcast channel. The precoding matrix aims to maximize the system sum rate by keeping the channel vectors of the individual users in the transmitted signal as orthogonal to each other as possible, thereby minimizing interference \cite{zhang2009networked}. 

In existing research, there are two primary approaches to serving multiple users per beam: unicast and multicast. Most precoding literature for multibeam satellite communications assumes an ideal scenario where a single gateway can handle all users. However, this assumption is unrealistic as it implies large feeder link bandwidth requirements, which are unavailable under current frequency allocations (Ka-band)\cite{ZGTH202303001012}. Satellite communications generally prioritize minimizing payload complexity. Consequently, most computations must be performed at the gateway, and the pre-coded signals should be fed to the satellite. In this case, the feeder link must support the satellite's entire traffic load, leading to significant bandwidth demands, which become even larger when full frequency reuse is deployed.
Deploying multiple gateways can effectively mitigate this issue, with each gateway using the entire bandwidth to serve a subset of beams within the satellite's coverage. Multiple gateway precoding requires a CSI feedback mechanism between the gateways and the satellite \cite{joroughi2018board}.

On-board precoding (OBP) addresses the challenge of calculating precoding matrices in real-time based on CSI variations as a solution to reduce feeder link bandwidth and mitigate interference \cite{bandi2018sparsity}. However, implementing OBP is challenging due to payload complexity constraints. Therefore, hybrid  on-board/on-ground  precoding techniques, combining ground-based precoding with on-board beamforming \cite{2010Hybrid}, have become an attractive option. These techniques offer the advantages of high spectral efficiency, manageable satellite payload complexity, and mitigation of feeder link shortages.

\subsection{Prior Related Surveys}
 {In recent years, the benefits of multibeam HTS systems have stimulated active research, and there have been several surveys/reviews \cite{guan2019review, WXDT202004002, TXWL202204012, 10648786, ZGTH202303001012, 10256078} in multibeam HTS systems. In the following, we briefly review the relevant surveys. Afterwards, a comparison between these reviews and our work in this paper is summarized at a glance in Table \ref{survey compare} in order to point out the distinctive contribution of our survey.}

 {Guan et al. \cite{guan2019review} provided an overview of HTS, integrating both market and technical considerations, and examined key metrics and trade-offs associated with these satellites. At that time, the application of HTS in China began to emerge. To seize the transformative opportunities brought by new technologies, Pang  et al. \cite{WXDT202004002} analyzed data and conducted studies on key technologies to discuss the development trends of HTS in China from both market and technical perspectives. In addition, Zhang et al. \cite{TXWL202204012} further introduced the key technologies and development directions of HTS. By employing multibeam technology, frequency reuse, and adaptive modulation and coding, HTS systems significantly enhance communication capacity and quality, providing stable and reliable satellite communication services for terminal users. While many surveys focus on physical-layer techniques, networking aspects such as programmable payloads, dynamic scheduling, and inter-gateway coordination are equally important. Yahia et al. \cite{10648786} notably provided a comprehensive overview of these networking aspects in HTS systems.}

 {Regarding interference management, multibeam precoding has been recognized for its high spectral and energy efficiency enabling full frequency reuse. Yi et al. \cite{ZGTH202303001012} introduced precoding techniques based on satellite system design, including both single gateway and multi-gateway systems; however, hybrid on-board/on-ground precoding was not addressed. To further detail the precoding techniques in HTS systems, Khammassi et al. \cite{10256078} presented an overview and classification of recent precoding methods from two main perspectives: problem formulation and  system design.}

 {Although existing surveys often discuss satellite constellations, hardware, resource allocation, and precoding, they typically address these aspects in isolation, without analyzing their interactions or the trade-offs involved in deploying satellites across different orbital regimes \cite{guan2019review, WXDT202004002, TXWL202204012, 10648786}.  The lack of integration obscures how hardware limitations such as phase noise, raveling wave tube amplifier (TWTA) nonlinearity, and restricted on-board memory bandwidth impact resource allocation and precoding \cite{guan2019review, WXDT202004002}. While bandwidth, power, and time slot allocation are commonly considered, joint optimization across these dimensions is often neglected\cite{ZGTH202303001012}. In addition, few studies  provide a comprehensive classification of precoding methods across practical deployment architectures, including single gateway, multiple gateway, on-board, and hybrid on-board/on-ground systems.}

\begin{table*}[t]
\centering
\renewcommand{\arraystretch}{1.2} 
\caption{Survey and Tutorial Papers Related to the Multibeam High Throughput Satellite Communication Systems.}
\resizebox{0.9\linewidth}{!}{
\begin{tabular}{cclcccccc}
\hline \hline  
\multicolumn{1}{c|}{\textbf{survey}} & \multicolumn{1}{c|}{\textbf{year}} & \multicolumn{1}{c|}{\textbf{Main topics}}                                                                                                   & \multicolumn{1}{c|}{\textbf{\begin{tabular}[c]{@{}c@{}}Hardware \\ restriction\end{tabular}}} & \multicolumn{1}{c|}{\textbf{\begin{tabular}[c]{@{}c@{}}Hardware \\ foundation\end{tabular}}} & \multicolumn{1}{c|}{\textbf{\begin{tabular}[c]{@{}c@{}}Hardware \\ requirement\end{tabular}}} & \multicolumn{1}{c|}{\textbf{\begin{tabular}[c]{@{}c@{}}Resource \\ allocation\end{tabular}}} & \multicolumn{1}{c|}{\textbf{\begin{tabular}[c]{@{}c@{}}Impact between \\ resource allocation \\ and precoding\end{tabular}}} & \textbf{precoding} \\ \hline
\multicolumn{1}{c|}{{[}43{]}}        & \multicolumn{1}{c|}{2019}          & \multicolumn{1}{l|}{\begin{tabular}[c]{@{}l@{}}Market and cost performance \\ analysis of HTS\end{tabular}}                                 & \multicolumn{1}{c|}{$\checkmark$}                                                                        & \multicolumn{1}{c|}{$\Delta$}                                                                     & \multicolumn{1}{c|}{$\Delta$}                                                                      & \multicolumn{1}{c|}{$\Delta$}                                                                     & \multicolumn{1}{c|}{}                                                                                                        &                    \\ \hline
\multicolumn{1}{c|}{{[}44{]}}        & \multicolumn{1}{c|}{2020}          & \multicolumn{1}{l|}{\multirow{2}{*}{\begin{tabular}[c]{@{}l@{}}Key technologies and future\\ trends prospects of HTS.\end{tabular}}}        & \multicolumn{1}{c|}{$\Delta$}                                                                      & \multicolumn{1}{c|}{$\Delta$}                                                                     & \multicolumn{1}{c|}{$\checkmark$}                                                                        & \multicolumn{1}{c|}{$\Delta$}                                                                     & \multicolumn{1}{c|}{}                                                                                                        &                    \\ \cline{1-2} \cline{4-9} 
\multicolumn{1}{c|}{{[}45{]}}        & \multicolumn{1}{c|}{2022}          & \multicolumn{1}{l|}{}                                                                                                                       & \multicolumn{1}{c|}{}                                                                         & \multicolumn{1}{c|}{$\checkmark$}                                                                       & \multicolumn{1}{c|}{}                                                                         & \multicolumn{1}{c|}{$\Delta$}                                                                     & \multicolumn{1}{c|}{}                                                                                                        &                    \\ \hline
\multicolumn{1}{c|}{{[}39{]}}        & \multicolumn{1}{c|}{2023}          & \multicolumn{1}{l|}{\multirow{2}{*}{\begin{tabular}[c]{@{}l@{}}Precoding technology in \\ multibeam HTS systems.\end{tabular}}}             & \multicolumn{1}{c|}{}                                                                         & \multicolumn{1}{c|}{}                                                                        & \multicolumn{1}{c|}{$\checkmark$}                                                                        & \multicolumn{1}{c|}{}                                                                        & \multicolumn{1}{c|}{}                                                                                                        & $\checkmark$                  \\ \cline{1-2} \cline{4-9} 
\multicolumn{1}{c|}{{[}46{]}}        & \multicolumn{1}{c|}{2024}          & \multicolumn{1}{l|}{}                                                                                                                       & \multicolumn{1}{c|}{}                                                                         & \multicolumn{1}{c|}{$\checkmark$}                                                                       & \multicolumn{1}{c|}{$\checkmark$}                                                                        & \multicolumn{1}{c|}{$\Delta$}                                                                     & \multicolumn{1}{c|}{$\Delta$}                                                                                                     & $\checkmark$                  \\ \hline
\multicolumn{1}{c|}{{[}2{]}}         & \multicolumn{1}{c|}{2025}          & \multicolumn{1}{l|}{\begin{tabular}[c]{@{}l@{}}Evolution of  HTS systems \\ enabled by programmable\\  regenerative payloads.\end{tabular}} & \multicolumn{1}{c|}{$\Delta$}                                                                      & \multicolumn{1}{c|}{$\checkmark$}                                                                       & \multicolumn{1}{c|}{$\checkmark$}                                                                        & \multicolumn{1}{c|}{$\Delta$}                                                                     & \multicolumn{1}{c|}{}                                                                                                        &     \multicolumn{1}{c}{$\Delta$}                 \\ \hline
\multicolumn{1}{c|}{This paper}      & \multicolumn{1}{c|}{2025}          & \multicolumn{1}{l|}{\begin{tabular}[c]{@{}l@{}}Hardware foundation, resource \\ allocation and precoding \\ in multibeam HTS.\end{tabular}} & \multicolumn{1}{c|}{$\checkmark$}                                                                        & \multicolumn{1}{c|}{$\checkmark$}                                                                       & \multicolumn{1}{c|}{$\checkmark$}                                                                        & \multicolumn{1}{c|}{$\checkmark$}                                                                       & \multicolumn{1}{c|}{$\checkmark$}                                                                                                       & $\checkmark$                  \\ \hline
\multicolumn{9}{l}{\begin{tabular}[c]{@{}l@{}}Note:\\ The $\checkmark$ symbol indicates that this aspect is covered in detail in the reference, the $\Delta$ symbol means that this aspect is onlymentioned brietiy or with other\\ conents but not discussed comprehensively in a single secion in the reference, and the biank meas that this aspect is not covered at all in the reference.\end{tabular}}                                                                                          \\ \hline     \hline    
\end{tabular}
}
\label{survey compare}
\end{table*}

\subsection{Contributions}

This paper presents  a comprehensive review of multibeam HTS systems, focusing on hardware architecture and constraints, dynamic resource allocation mechanisms \cite{1019649081,palacin2017multibeam, lei2020deep}, and multibeam precoding from the perspective of single gateway precoding, multiple gateways precoding, on-board precoding and hybrid on-board/on-ground precoding. Key challenges and potential research directions are identified to establish a foundational framework for a unified theoretical and technical paradigm. The main contributions are summarized as follows:

\begin{itemize}
\item We propose a structured review framework that categorizes multibeam HTS systems into three core components: hardware foundation, resource allocation, and precoding. As shown in Fig. \ref{SectionContact}, this framework highlights the interactions and coordinated evolution among these components, providing an integrated perspective that is often missing in existing literature.

\item We analyze hardware restrictions, such as limited data caching, beamforming network (BFN) complexity, TWTA nonlinearity, frequency offset, and phase mismatch, that affect resource allocation and precoding but are rarely discussed in existing literature.

\item We summarize the hardware compatibility and architectural differences between GEO HTS systems and non-GEO constellations, filling gaps left by fragmented discussions in previous surveys.

\item We provide a detailed comparison of resource allocation strategies, covering bandwidth, power, and time slot allocation. Particular attention is given to joint optimization approaches that consider multiple resource dimensions simultaneously, such as bandwidth and power, power and time slot, and three-dimensional scheduling.

\item We present a systematic classification and comparative analysis of precoding techniques in practical deployment scenarios, including single gateway, multiple gateway, on-board, and hybrid on-board/on-ground architectures. In contrast to studies that focus primarily on mathematical modeling, this review emphasizes the importance of balancing algorithmic performance with deployment feasibility under hardware constraints.
\end{itemize}

\subsection{Paper Organization}
The structure of the remaining sections in this paper is organized as follows, as illustrated in Fig. \ref{SectionContact}. The list of acronyms, which facilitates comprehension, is provided in Table \ref{acro}.
Section \RNum{2} presents an introduction to the hardware foundation of multibeam HTS, including a discussion on the different types of satellite transponders and antennas that are relevant to HTS. 
Section \RNum{3} presents the hardware limitations that impact resource allocation and precoding in multibeam HTS systems, as well as the specific hardware capabilities required to support these functions effectively.
In Section \RNum{4}, we focus on introducing flexible resource allocation algorithms, including bandwidth allocation, power allocation, time slot allocation, and joint resource allocation.
Section \RNum{5} begins by introducing precoding algorithms in the single gateway system model, which serves as preliminary evidence for the rationality of precoding algorithms in HTS. We then review precoding algorithms in the multiple gateways model, on-board model, and hybrid on-board/on-ground model.
Section \RNum{6} explores the coordination mechanism of resource allocation and precoding in multibeam HTS systems. Resource allocation influences the interference environment that precoding must mitigate, while precoding technique helps optimize resource usage by reducing inter-beam interference.
In Section \RNum{7}, we present some challenges and research perspectives that could shape the direction of further advancements in the field of HTS.
Conclusions follow in Section \RNum{8}.

\begin{table*}[]
  \centering
  \caption{List of  Abbreviations.}
\begin{tabular}{c|c|c|c}
\hline\hline

\textbf{Acronyms} &\textbf{Description}  & \textbf {Acronyms} &\textbf{Description}           \\ \hline
AI       & Artificial intelligence  & MF       & Matched filter \\
BER      & Bit error rate & MFB      & Multiple feed per beam  \\
BFN      & Beamforming network  & MF-TDMA   & Multi-frequency time division multiple access  \\
CCI      & Co-channel interference& MIMO     & Multiple-input multiple-output     \\
CDMA      & Code division multiple access & MMSE     & Minimum mean square error \\
CSI       & Channel state information  & MODCOD & Modulation coding    \\
DFT      & Discrete fourier transform & MU-MIMO  & Multi-user multiple-input multiple-output   \\
DL       & Deep learning  & NSR  &  Northern sky research  \\
DPC      & Dirty paper coding   & OBP   & On-board precoding   \\
DPSS     & Discrete prolate spheroidal sequences& OMT     & Orthogonal mode transducer  \\
DRL      & Deep reinforcement learning &  PCA      & Principal component analysis    \\
DSP      &  Digital signal processing  & PSO& Particle swarm optimization  \\
DVB      & Digital video broadcasting  &   QoS      & Quality of service  \\
EHF      & Extremely high-frequency     &  RF       & Radio frequency\\
EIRP     & Effective isotropic radiated power  & RL   & Reinforcement learning   \\ 
ESA      & European space agency & R-ZF     & Regularized zero forcing  \\ 
FAPI     & Fast approximated power iteration  &SA       & Simulated annealing  \\ 
FCC      &  Federal communications commission &   SatCom  & Satellite communication  \\ 
FDMA     & Frequency division multiple access & SCA      & Sequential convex approximation  \\ 
FEC      & Forward error correction code & SDR      & Software-defined radio \\
FFT      & Fast fourier transform  & SFB      & Single feed per beam\\
FSS      & Fixed satellite service & SINR & Signal to interference plus noise ratio  \\ 
GA       & Genetic algorithms     & SISO     & Single-input single-output\\ 
GEO      & Geostationary earth orbit  &  SLNR     & Signal-to-leakage-and-noise ratio  \\ 
GSR      & Greedy sparse recovery  & SMSE     & Sum mean square error \\ 
HEO      & High earth orbit   &   SNR      & Signal to noise ratio   \\
HTS      & High throughput satellite  &  SVD      & Singular value decomposition  \\ 
IF       & Intermediate frequency   & TDMA      & Time division multiple access   \\
IMOB     &\begin{tabular}[c]{@{}c@{}} Improved multibeam opportunity beamforming\end{tabular}  &  THP & Tomlinson-harashima precoding   \\
IoT      & Internet of Things  & TT\& C    & Telemetry, tracking, and Control\\ 
ITU      & International telecommunication union   &TWTA     & Traveling wave tube amplifier  \\ 
JPCB     & Joint power control and beamforming     & UTs       & User terminals \\
LEO & Low earth orbit &  VP       & Vector perturbation \\ 
LNA      & Low noise amplifier &  VSAT     & Very small aperture terminal station \\ 
MEO      & Medium earth orbit   &  ZF       & Zero forcing \\ 
\hline\hline

\end{tabular}
\label{acro}
\end{table*}

\section{Hardware Foundation}

In this section, we provide an overview of the multibeam HTS hardware components including an earth station in the ground segment, on-board payload in the space segment, and ground user equipment in the user segment. In particular, we carefully introduce and compare various antennas and transponders of the on-board payload.

\subsection{The Ground Segment}
The ground segment comprises gateway stations and telemetry, tracking, and control (TT\&C) stations, both of which are called earth stations and administered by operators \cite{nguyen2020overview}. The gateway station is primarily responsible for receiving or transmitting multimedia data, converting frequency, processing baseband signals, and accessing the network. Furthermore, the gateway can establish feeder links with visible satellites. The TT\&C station is responsible for controlling antenna deployment, monitoring satellite location and orbit, and ensuring orbital drift and satellite perturbation are within acceptable limits \cite{maral2020satellite,10256078}.
\begin{figure*}[t]
\begin{center}
\includegraphics[width=17cm]{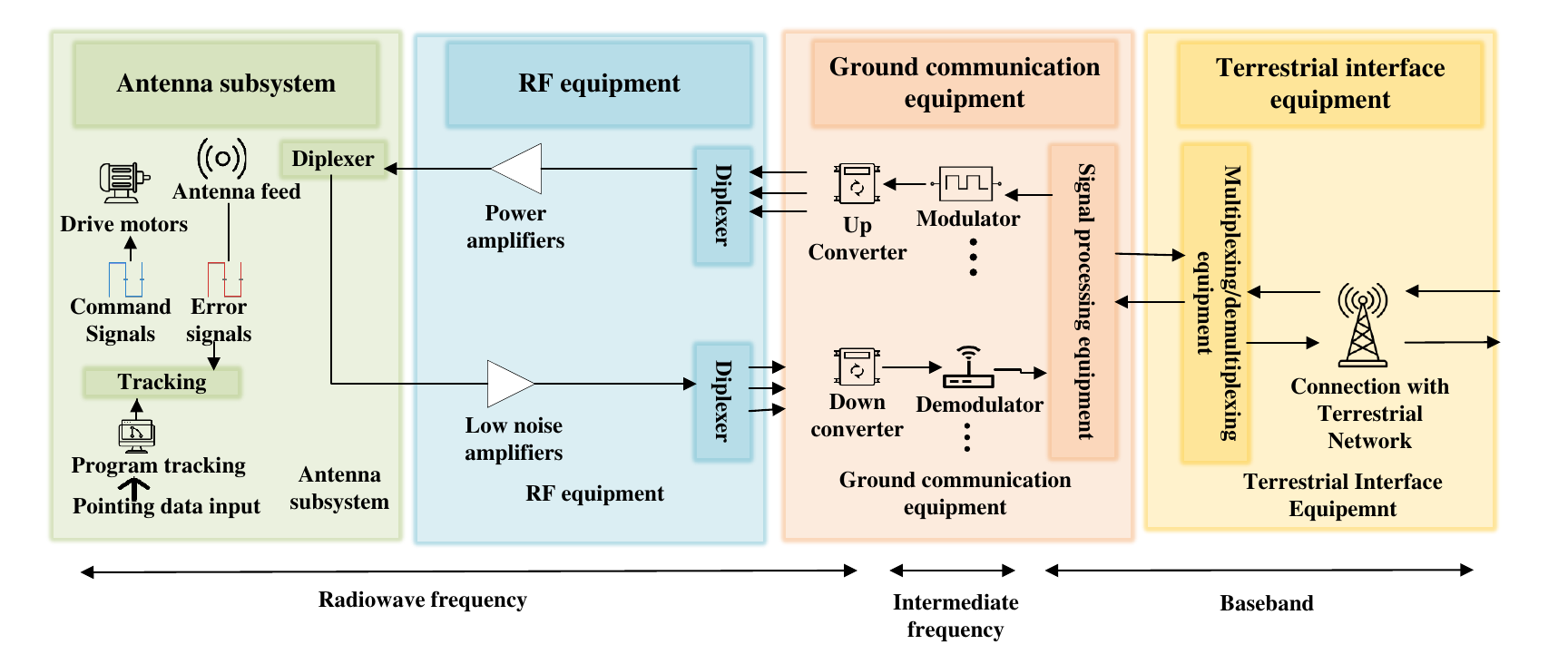}
\end{center}
\caption{  {The model of earth station. It consists of an antenna
subsystem, with an associated tracking system, a transmitting section and a receiving section.
It also includes equipment to interface with the terrestrial network together with various
monitoring and electricity supply installations. This organisation is not, in principle, fundamentally different from that of other telecommunication stations such as those for terrestrial microwave
links. Among the special features, there is the tracking system although it may be rather simple in
certain cases.}}
\label{earthstation}
\end{figure*}

%
The key hardware components of the earth station comprise the RF, IF \cite{maral2020satellite}, and baseband sections, as depicted in Fig. \ref{earthstation}. The RF part includes transceiver antennas, amplifiers, and duplexers. The antennas typically used include axisymmetric parabolic antennas, offset parabolic antennas and Cassegrain antennas. Currently, Cassegrain parabolic antennas are widely used due to their advantages of shorter feed length, less ground clutter absorption, and low noise temperature. Cassegrain parabolic antennas are mainly composed of a main reflector, secondary reflector, and feed source, with the main feed source located at the real focus of the secondary hyperbolic reflector. The amplifiers are designed to amplify signals with the characteristics of low noise and high sensitivity, thereby improving signal quality \cite{elbert2014satellite}. Duplexers are used to differentiate between the received and the transmitted signal, enabling full-duplex communication. The IF part of the earth station primarily consists of frequency converters used to mix the RF or baseband signal with IF signals. The baseband part is composed of baseband processing circuits, modem, and multiplexing/demultiplexing equipment, which provide signal aggregation or distribution as well as support the connection with the ground interface \cite{wang2022capacity}. The focal point for implementing precoding algorithms for multibeam HTS is presently the baseband segment of the gateway station. This choice is driven by the substantial hardware resources within this section, capable of handling the computational intricacies associated with high-dimensional precoding matrices. Additionally, support facilities such as power supplies, monitoring equipment, control equipment, and environmental conditioning devices are also implemented at each earth station to ensure the smooth operation and monitoring of the earth station \cite{elbert2014satellite, maral2020satellite}.

\subsection{The Space Segment}
The space segment refers to the satellite-based components in multibeam HTS systems. It primarily includes communication satellites and their orbital configurations, collectively known as satellite constellations. These constellations are responsible for enabling continuous coverage, maintaining link stability, and ensuring system-wide service performance. Each satellite typically consists of satellite platform and on-board payload. The satellite platform provides basic operational support, such as power supply, thermal regulation, and attitude control. The on-board payload performs communication functions, including signal reception, processing, and transmission.

This subsection is organized into two parts to facilitate a structured understanding. The first part introduces satellite constellations, including both GEO and non-GEO HTS systems, focusing on their orbital configurations, coverage capabilities, and deployment trends. The second part explains the on-board payload, which is further divided into two subsystems: satellite antennas and satellite transponders, detailing their types, functional characteristics, and performance trade-offs.

\emph{1) Satellite Constellations:}
In recent years, the world has witnessed a surge in interest toward satellite constellations, with both traditional and emerging satellite operators aiming to build global HTS constellations to deliver broadband services worldwide. Since the launch of the first HTS in 2004, the industry has evolved beyond the stages of technical validation and market cultivation, moving steadily toward global expansion and industrial-scale deployment \cite{baccelli:hal-04626677}. 

With the rapid development of global mobile internet, non-GEO HTS constellations have emerged in increasing numbers, drawing significant market attention and becoming a key solution for global connectivity.  Unlike large-scale LEO constellations, several GEO HTS systems, such as Viasat-3 and Inmarsat GX, form small-scale constellations with global or regional coverage. These GEO HTS systems rely on multibeam architectures and frequency reuse to deliver high capacity without the need for dynamic orbital coordination. Therefore, GEO HTS constellations are often referred to as GEO HTS systems. The following content presents an overview of GEO HTS systems, non-GEO HTS constellations, and hybrid GEO+LEO/MEO architectures.

\emph{ i) GEO HTS systems:} 
GEO HTS systems operate at an altitude of approximately 36,000 kilometers above the earth's surface.   Owing to their fixed position relative to a point on earth, GEO satellites are particularly well-suited for providing continuous and wide-area coverage. They are extensively used for conventional satellite communication services such as television broadcasting and broadband Internet. The orbital stability of GEO systems enables consistent service without frequent handovers. However, the large propagation distance leads to increased communication latency, which can be a limitation for latency-sensitive applications \cite{10648786}.
GEO HTS systems have the following advantages  \cite{WXYG201706017}:  
\begin{itemize}
    \item  Lower per-satellite deployment costs—typically around \$400–500 million (including launch and insurance). Satellites can be added incrementally, reducing early-stage project risk.
    \item Use of narrow spot beams enables high throughput for localized internet access.
    \item Ground terminals are relatively simple, requiring no tracking antennas, allowing for miniaturized, mass-producible devices.
    \item  GEO satellites operate in a stable space environment and generally have service lives exceeding 15 years, compared to LEO and MEO systems.
\end{itemize}

\begin{figure}
    \centering
    \subfloat[Walker Star]{
    \includegraphics [width=4.0 cm]{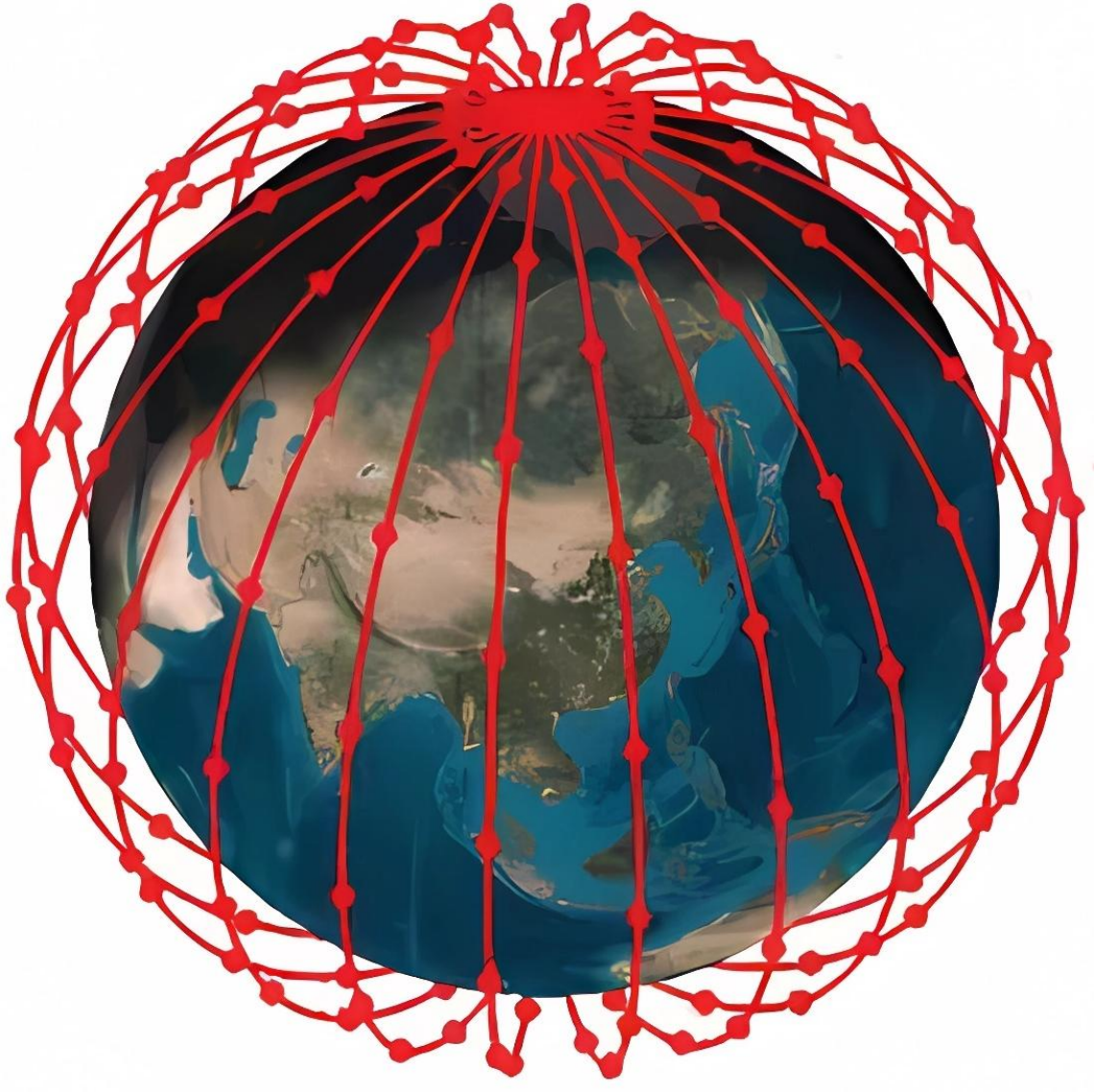}}
    \subfloat[Walker Delta]{
    \includegraphics [width=4.0 cm]{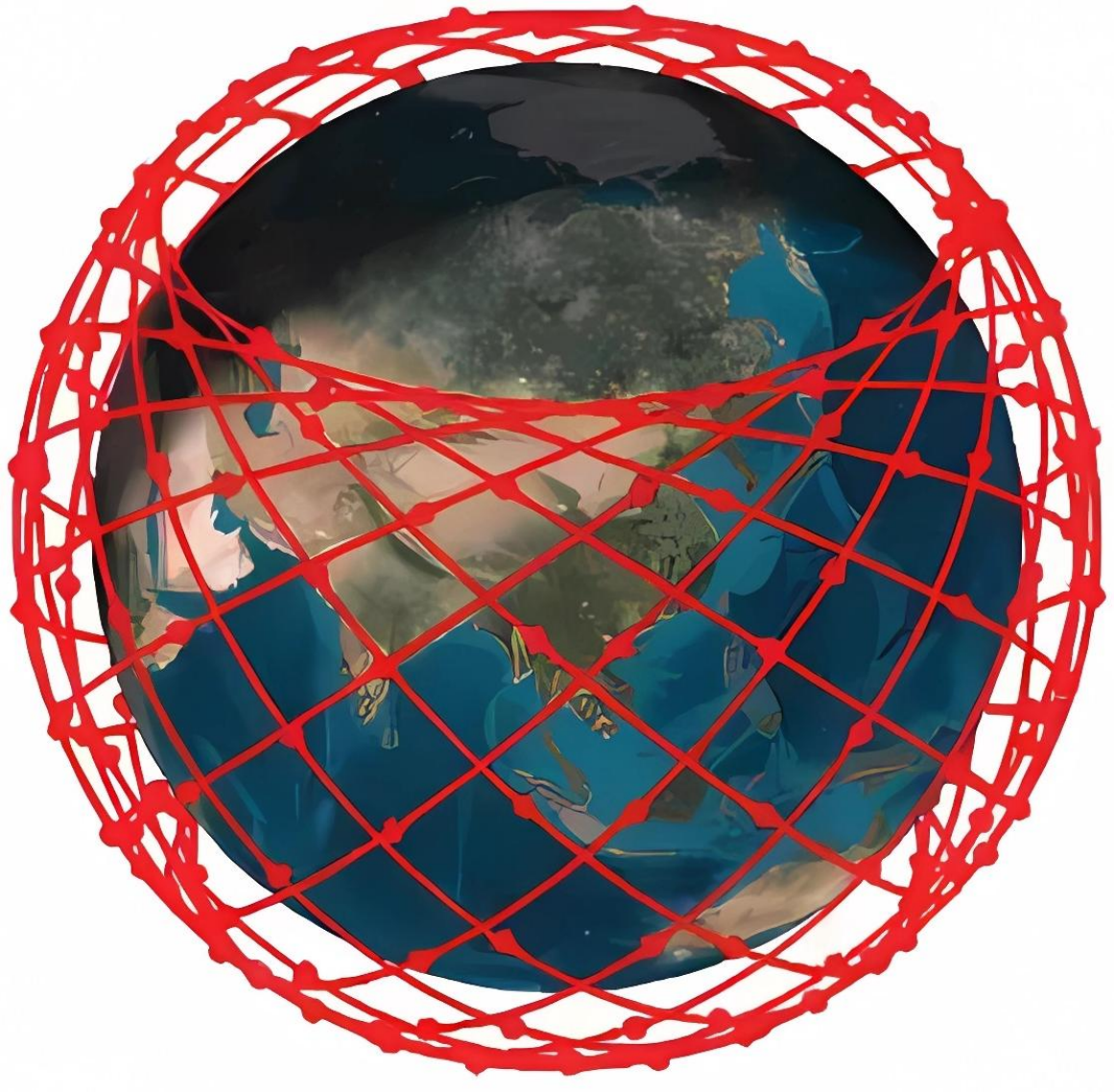}}
    \caption{Walker constellation diagram for HTS systems \cite{wang2022capacity}.}
    \label{Walkercon}
\end{figure}


\emph{ii) Non-GEO HTS Constellations:}  
Non-GEO HTS constellations, primarily operating in LEO and MEO orbits, leverage structured orbital configurations to ensure continuous coverage and low latency. Two representative orbital arrangements widely adopted in such systems are the Walker Delta and Walker Star constellations, as illustrated in Fig. \ref{Walkercon} \cite{wang2022capacity}. The Walker Delta constellation offers advantages in adjusting the beam overlap area and satellite orbit inclination based on specific areas of interest. However, constant changes in relative satellite positions can lead to unstable inter-satellite links, impacting communication quality \cite{wang1993structural}. The Walker Star constellation facilitates easier pointing control of user ground stations but faces interference challenges in the polar regions due to multiple overlapping coverage.

MEO HTS  constellations typically operate between 2,000 and 36,000 km above earth. While their increased altitude results in higher latency compared to LEO, it enables longer visibility windows, simplifying continuity of service. A notable example is SES’s O3b mPOWER system, deployed at 8,062 km in an equatorial Walker Delta configuration with 12 orbital planes, each containing one satellite \cite{10648786}. This architecture delivers low-latency, high-throughput services to equatorial regions. Despite its uniqueness, most MEO orbital shells ($78\%$) remain pending approval, highlighting MEO’s untapped potential \cite{qu2017leo}. 

LEO HTS constellations operate at altitudes from 160 to 2,000 km, offering low-latency services ideal for real-time applications. However, their rapid orbital movement necessitates frequent handovers and complex coordination. Several commercial players adopt Walker-based architectures with varying designs:
\begin{itemize}
\item Telesat adopts a hybrid design, combining a 98.98$^\circ$ high-inclination Walker Star shell at 1,015 km with two additional Walker Delta shells \cite{pachler2021updated}.
\item OneWeb utilizes three uniform Walker Delta shells at 1,200 km \cite{su2019broadband}.
\item SpaceX  employs a 550 km, 53$^\circ$ Walker Delta shell with over 5,500 satellites, alongside planned polar (97.6$^\circ$) and inclined (70$^\circ$) shells \cite{evans1999satellite}.
\item Amazon  plans three Walker Delta shells from 590–630 km, with a primary 610 km shell at 42$^\circ$ inclination, supporting up to 3,236 satellites.
\end{itemize}

Table~\ref{mega-constellations} provides a quantitative comparison of these six major HTS constellations, detailing orbital altitude, inclination, configuration type, and deployment status. Constellation shells marked with A is approved by the Federal Communications Commission (FCC), while P indicates pending applications.  These figures underscore the rapid evolution and vast potential of non-GEO HTS deployments across all orbital layers. Beyond architectural variety, non-GEO HTS constellations offer several distinct advantages over traditional GEO systems \cite{WXYG201706017}:
\begin{itemize}
    \item Lower latency for interactive services due to lower orbits: GEO latency is around 250 ms, MEO about 150 ms, and LEO 30–50 ms, close to terrestrial fiber (10–20 ms) and wireless (10–50 ms).
    \item Higher total system throughput and lower per-user costs from large satellite populations: monthly costs approximate \$500 (GEO), \$150 (MEO), and \$25 (LEO), approaching fiber (\$5/month) and wireless (\$0.5/month).
    \item Global coverage, with systems using inter-satellite links supporting single-hop global communication.
\end{itemize}

\emph{iii) Hybrid GEO+LEO/MEO HTS System:}  
In recent years, hybrid GEO+LEO/MEO HTS systems have attracted growing interest. A prominent example is the 2015 collaboration between Intelsat’s GEO-based Epic system and the LEO-based OneWeb constellation, which marked the early stage of multi-orbit network development \cite{WXYG201611012}. These architectures seek to combine the extensive coverage of GEO satellites with the low latency of non-GEO systems. However, they also introduce new technical challenges, especially for user terminals that must support frequent beam and orbit switching. For instance, O3b mPOWER terminals employ dual tracking antennas to enable seamless satellite handovers. While this approach is effective, it substantially increases terminal cost and complexity, making it more suitable for MEO systems with a relatively limited number of satellites \cite{WXYG201706017}.
\begin{table}[t]
\centering
\renewcommand{\arraystretch}{1.2} 
\caption{Summary of the Orbit Characteristics of the Non-GEO HTS Constellations\cite{pachler2021updated}.}
\resizebox{\linewidth}{!}{
\begin{tabular}{c|c|c|c|c|c|c|c}
\hline\hline
                   \textbf{System} & \textbf{\begin{tabular}[c]{@{}c@{}}Altitude\end{tabular}} & \textbf{Inclination($^{\circ}$)} &\textbf {	Planes} &\textbf{\begin{tabular}[c]{@{}c@{}}Satellites \\per plane \end{tabular}}& \textbf {State}&\textbf {\begin{tabular}[c]{@{}c@{}}Number of \\satellites\end{tabular}}&\textbf {Walker} \\ \hline
\multirow{3}{*}{Telesat}  & 1,015 &98.98&27&13&P& \multirow{3}{*}{1,671}& \multirow{3}{*}{Walker Star} \\ 
                         & 1,325 &50.88&40&33&P& \\ 
                          & 1,200 &87.9&36&49&P& \\   
                         \hline
\multirow{3}{*}{OneWeb} &1,200 &87.9&36&49&P&\multirow{3}{*}{6,375} &\multirow{3}{*}{Walker Delta} \\
                         &1,200 &55&32&72&P& \\
                          &1,200 &40&32&72&P& \\ \hline
\multirow{5}{*}{SpaceX}  &540 &53.2&72&22&P&\multirow{5}{*}{$>$5500} &\multirow{5}{*}{Walker Delta}\\
                          &550 &53&72&22&A& \\
                          &560 &97.6&6&58&P& \\
                          &560 &97.6&4&43&P& \\
                          &570 &70&36&20&P& \\ \hline
\multirow{3}{*}{Amazon} &590 &33&28&28&A&\multirow{3}{*}{3,236} &\multirow{3}{*}{Walker Delta}\\
                          &610 &42&36&36&A& \\
                          &630 &51.9&34&34&A& \\ \hline
\multirow{1}{*}{SES} &8,062 &0&12&1&A&\multirow{1}{*}{20} &\multirow{1}{*}{Walker Delta}\\
                  \hline \hline                
\end{tabular}
}
\label{mega-constellations}
\end{table}

\emph{2) On-Board Payload:}
The on-board payload is the functional core of a communication satellite. It is responsible for receiving, processing, and transmitting signals, thereby determining the satellite’s overall communication capabilities. While the satellite platform provides essential support functions such as power supply, thermal regulation, and attitude control, it is the payload that executes the communication tasks \cite{pratt2019satellite}.
The primary functions of the payload are as follows: 
\begin{itemize}
  \item Capture carrier signals transmitted by ground stations within the coverage area using the given polarisation and frequency band.
  \item Capture as few interfering signals as possible (such as carrier signals from other regions or not having the specified values of frequency or polarisation).
  \item Amplify the received carrier signals while minimizing noise and distortion (the received power is in the tens of picowatts range).
  \item Change the received uplink carrier signals to the downlink carrier frequency.
  \item Provide the power required by the transmitting antenna (ranging from tens to hundreds of watts).
  \item Radiate the carrier signal at the specified frequency and polarization to the given region (service area) on the ground.
\end{itemize}

Regardless of how the payload is organized, these functions must be accomplished. For multibeam satellites, an additional function is to route the carrier from any given uplink beam to any downlink beam. Regenerative repeaters must also provide carrier demodulation and remodulation. The frequency band assigned to the repeater is typically large, ranging from several hundred megahertz to several gigahertz. To facilitate power amplification, this frequency band is usually divided into multiple sub-bands (channels or transponders), with independent amplification chains associated with these sub-bands. The bandwidth of these channels is typically on the order of tens of megahertz.


A typical on-board payload consists of two main subsystems: the satellite antenna and satellite transponders. The satellite antenna focuses on signal transmission and reception, including beam shaping and coverage control. Thesatellite transponders performs signal processing tasks such as amplification, frequency conversion, routing, and, in more advanced designs, demodulation and regeneration. These subsystems vary widely in terms of complexity, performance, and adaptability.

\begin{figure}[t]
\begin{center}
\includegraphics[width=8.0cm]{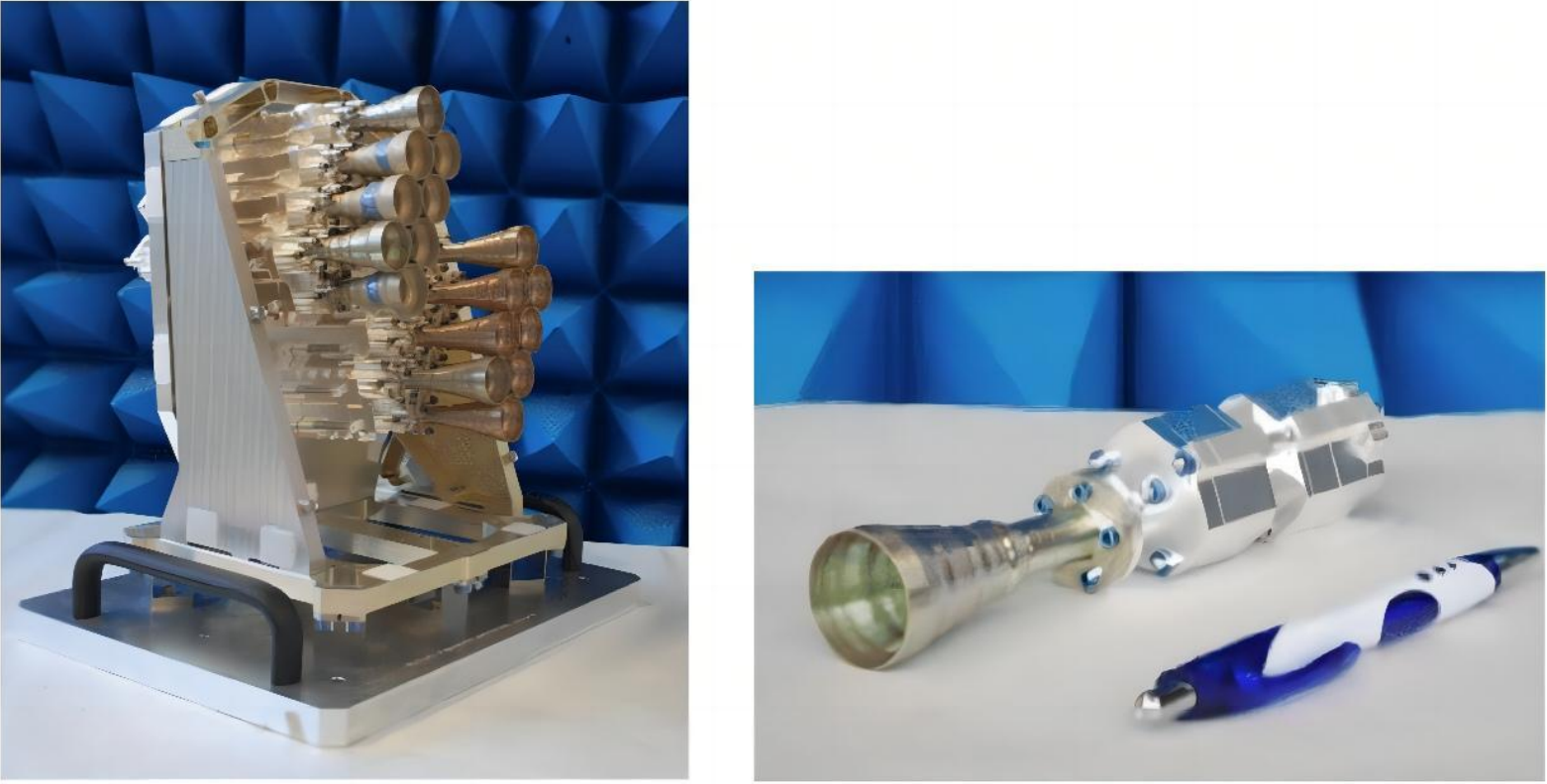}
\end{center}
\caption{Feed cluster (left) and feed with tracking functionality (right)  for SFB antennas developed by Thales Alenia Space – FR \cite{palacin2017multibeam}.}%
\label{SFB}
\end{figure}

\begin{figure}[t]
\begin{center}
\includegraphics[width=6.0cm]{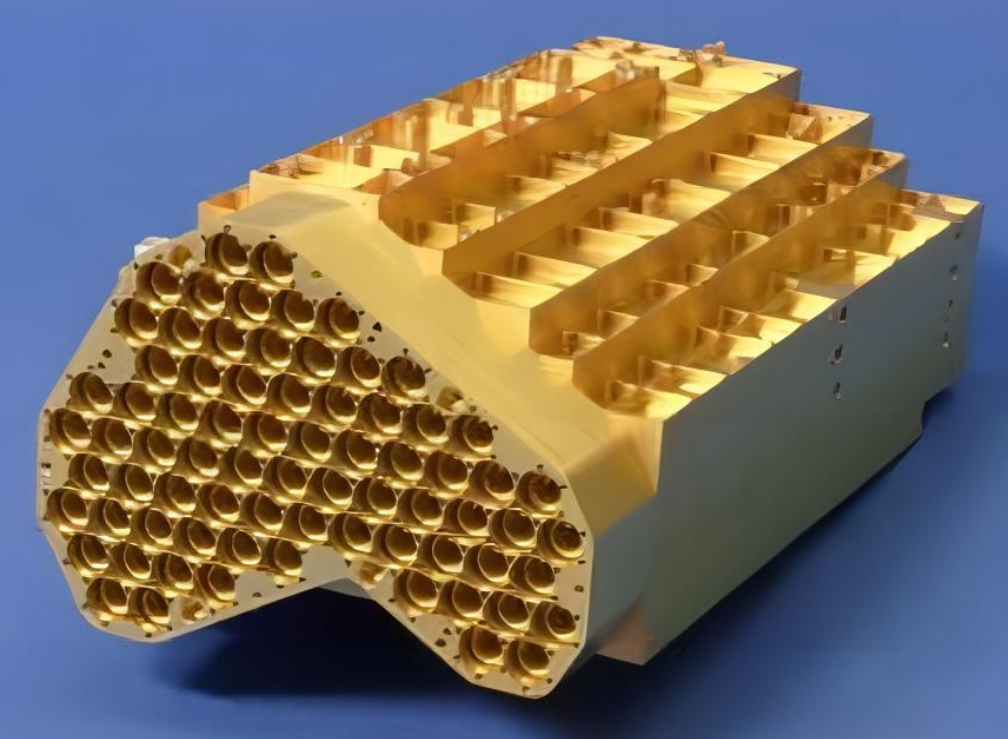}
\end{center}
\caption{Engineering model of MFB antenna developed at Astrium GmbH \cite{ratkorn2008medusa}.}%
\label{MFB}
\end{figure}

\emph{i) Satellite Antennas:} As communication satellite services shift from broadcasting to broadband, there is an increased demand for higher system throughput, which has led to the proliferation of multibeam antenna technology, generating numerous high-gain beams to meet system requirements \cite{rao2015advanced,palacin2017multibeam}.   In multibeam HTS systems, both passive array antennas and active array antennas are typically the focal points of attention \cite{moon2019phased,fenech2016role}.

\begin{figure}[t]
\begin{center}
\includegraphics[width=8.8cm]{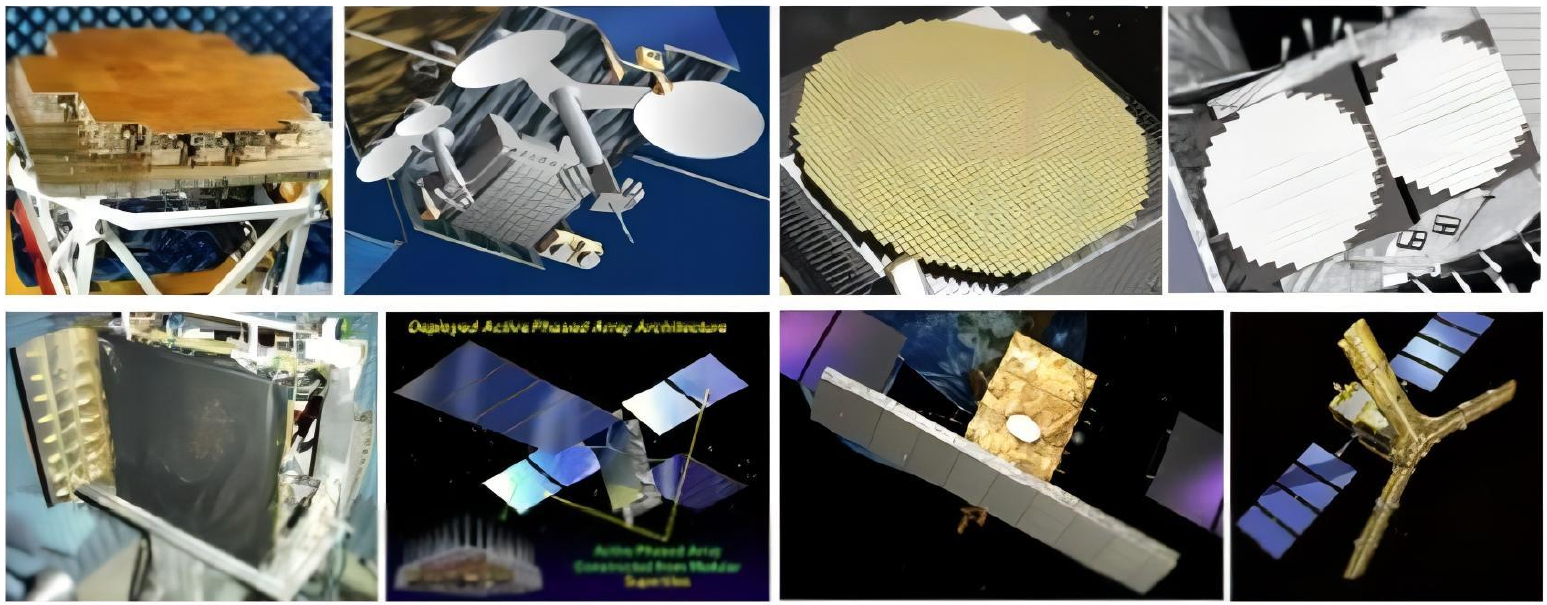}
\end{center}
\caption{Multibeam active array antennas for satellite applications \cite{angeletti2015digital}.}%
\label{phasearray}
\end{figure}
%


\begin{table*}[t]
\centering
\caption{Comparison of Satellite Antennas}
\renewcommand{\arraystretch}{1.2}  
\resizebox{0.8\textwidth}{!}{      
\begin{tabular}{
    >{\centering\arraybackslash}m{2cm} |  
    >{\centering\arraybackslash}m{2cm} |  
    >{\centering\arraybackslash}m{2cm} |    
    >{\centering\arraybackslash}m{2cm} |  
    >{\centering\arraybackslash}m{8cm}   
}
\hline\hline
\multirow{3}{*}{\textbf{Antenna type}} & 
\multicolumn{2}{c|}{\textbf{Passive antenna}} & 
\multirow{3}{*}{\textbf{\makecell{Active\\ antenna}}} & 
\multirow{3}{*}{\textbf{Quantification criteria}} \\ 
\cline{2-3}  
& \textbf{Single beam}& 
  \textbf{Multiple beams} & 
&  \\ 
\hline  
\textbf{Frequency bands} & 
\cellcolor{myyellow}{M} & 
\cellcolor{myblue}{H} & 
\cellcolor{myblue}{H} & 
\makecell[l]{
  L: $\leq$2 bands, $<$1\,GHz bandwidth;\\ 
  M: 3–4 bands, 1–5\,GHz bandwidth;\\ 
  H: $\geq$5 bands or $>$5\,GHz bandwidth \cite{fenech2016role}.} \\ 
\hline
\textbf{Scanning range} & 
\cellcolor{myred}{L} & 
\cellcolor{myblue}{H} & 
\cellcolor{myblue}{H} & 
\makecell[l]{
  L: $<\pm30^\circ$, limited coverage;\\ 
  M: $\pm30^\circ$–$\pm60^\circ$, mixed mechanical-electronic scanning;\\ 
  H: $>\pm60^\circ$, full electronic scanning.} \\
\hline
\textbf{Scanning speed} & 
\cellcolor{myred}{L} & 
\cellcolor{myyellow}{M} & 
\cellcolor{myblue}{H} & 
\makecell[l]{
  L: $>50$\,ms/$^\circ$, mechanical-driven;\\ 
  M: $10$–$50$\,ms/$^\circ$, hybrid mechanical-electronic;\\ 
  H: $<10$\,ms/$^\circ$, digital beamforming.
} \\ 
\hline
\textbf{Pointing accuracy} & 
\cellcolor{myyellow}{M} & 
\cellcolor{myblue}{H} & 
\cellcolor{myblue}{H} & 
\makecell[l]{
  L: $>1^\circ$, significant beam misalignment;\\ 
  M: $0.5^\circ$–$1^\circ$, acceptable for static terminals \cite{7795317};\\ 
  H: $<0.5^\circ$, precision tracking with closed-loop feedback.} \\ 
\hline
\textbf{Directivity} & 
\cellcolor{myyellow}{M} & 
\cellcolor{myyellow}{M} & 
\cellcolor{myblue}{H} & 
\makecell[l]{
  L: Beam width $>10^\circ$, gain $<30$\,dBi;\\ 
  M: Beam width $5^\circ$–$10^\circ$, gain $30$–$40$\,dBi \cite{romier2016overlapping};\\ 
  H: Beam width $<5^\circ$, gain $>40$\,dBi.
} \\ 
\hline
\textbf{Complexity} & 
\cellcolor{myred}{L} & 
\cellcolor{myyellow}{M} & 
\cellcolor{myblue}{H} & 
\makecell[l]{
  L: Simple design, minimal digital processing;\\ 
  M: Standard design, analog signal processing;\\ 
  H: Advanced design.
} \\ 
\hline
\textbf{Cost} & 
\cellcolor{myred}{L} & 
\cellcolor{myyellow}{M} & 
\cellcolor{myblue}{H} & 
\makecell[l]{
  L: $<$100 USD;\\ 
  M: $100$–$300$ USD;\\ 
  H: $>$300 USD.
} \\ 
\hline\hline
\end{tabular}
}
\label{SatAnt}
\end{table*}




 \emph{a) Passive Array Antennas}: Passive array antennas which forge reflector antennas towards multiple feeds per beam (MFB) and single feed per beam (SFB) have gained predominance in cutting-edge technology owing to their extensively refined and efficient models developed over the years \cite{Sharma2021System}. 
 \begin{itemize}
     \item The SFB antenna systems utilize three or four reflectors illuminated by multiple feeds, where each beam is generated by a feed \cite{palacin2017multibeam}. SFB antennas offer slightly better RF performance, low spillover loss, and a simple design, which are suitable for large spacecraft and very large coverage scenarios. However, the SFB antenna requires accommodating more reflectors for seamless coverage, which results in increased total satellite weight, reduced system flexibility and leaving little to no space for accommodating additional tasks on the satellite \cite{montero2015c, angeletti2013recent}.  Furthermore, the scan loss, which refers to the signal loss caused by the change of antenna pointing when the antenna is scanning or pointing, limits the satellite's application scenarios \cite{schneider2011antennas}.  So,  the scan loss is significant.  SFB antennas, developed by Thales Alenia Space, are presented in Fig. \ref{SFB}. Detailed descriptions and analyses of the reflector antenna configuration in SFB can be found in \cite{rao1999design, rao2003parametric}.

     \item The MFB antenna systems utilize one or two reflectors \cite{bosshard2016thales, schneider2013design}, where each beam is generated by a set of feeds with optimized illumination laws \cite{palacin2017multibeam}. Hence, the MFB antenna system achieves continuous coverage with a reduced number of reflectors compared to the SFB antenna system, demonstrating its efficiency in optimizing satellite communication performance. However, this configuration achieves overlap between feed clusters via a relatively complex BFN, producing acceptable crossover levels and enabling shared feeds between adjacent clusters \cite{romier2016overlapping}. Generally, BFN networks provide the required overlap of neighboring feeder clusters in three main ways: duplex sharing, hybrid sharing, and polarization sharing. The polarization sharing scheme allows feed elements to be independently accessible to each polarization port via OMT, without restrictions on amplitude and phase distribution, which is the best choice of the three BFN designs. Due to the complexity of BFN, the selection of MFB may not consistently be the preferred choice in passive antenna architectures \cite{fenech2016role}, emphasizing the nuanced considerations involved in the design and optimization of such systems within the realm of satellite communication. The engineering model of the MFB antenna is presented in the "Medusa" project, as shown in Fig. \ref{MFB} \cite{romier2016overlapping}. Specific configurations and comparisons between different MFBs are described in detail in \cite{gehring2007trade}. Antennas with reflectors typically require a mechanical structure, such as a two-axis steering system, to achieve beam scanning \cite{han2015novel}. However, this mechanical structure adds weight to the satellite payload, occupies significant system space, and results in slow mechanical rotation, leading to increased scanning losses \cite{rostan1995design, egashira1996stacked, nishiyama2003three}. 
 \end{itemize}

\emph{b) Active Array Antennas:}
Active array antennas seamlessly integrate amplifiers with radiating elements, allowing for distributed amplification of the radiating signal. This key distinction sets active array antennas apart from their passive counterparts. Furthermore, the spatial distribution of RF power and the reduction in peak RF power levels contribute to enhanced reliability in active arrays. Moreover, active array antennas can be combined with digital beamforming technology to achieve multiple beam gains and precise beam pointing by adaptive controlling the beam direction map, adjusting the beam side lobes, and mitigating inter-beam interference \cite{ahn2010digital, toso2013recent}, which is also able to increase flexibility for satellite payloads \cite{sikri2019multi}. In addition, active array antennas enable fast beam scanning and minimizes transmission loss from the active components to the radiating devices through the electronic adjustment, resulting in lower beam scanning losses compared to mechanical rotation approaches \cite{chen2022phased,fenn2000development}. While the use of digital devices in active antenna arrays provides superior beam gains and reduces the weight of satellite payloads, it imposes higher demands on the power system and computational capabilities of the satellite payloads. Active array antennas can be deployed in either direct radiation array or array-fed reflector architectures with the choice contingent upon specific system requirements. Currently, mega-constellation satellites commonly employ these two architectures to generate dozens of beams. However, it is seldom discussed that active array antennas can generate hundreds or even thousands of beams in GEO-HTS systems. Fig. \ref{phasearray} illustrates some of the multibeam active array antennas used for satellite applications.
The comparison of the satellite antennas is shown in Table \ref{SatAnt}.

\emph{ii) Satellite Transponders:} There are three main types of transponders commonly used in satellite communication systems: analog bent-pipe transponders, digital bent-pipe transponders, and regenerative transponders. Both analog and digital bent-pipe transponders are classified as transparent transponders. The analog type filters and amplifies the received signals, while the digital type digitizes them. Furthermore, the digital bent-pipe transponders can be divided into conventional types and those enhanced with digital channelizers and beamforming capabilities. Unlike transparent transponders, regenerative transponders demodulate and decode the received signals into digital baseband, which is then re-encoded and re-modulated for retransmission \cite{nguyen2020overview}.

\begin{figure*}[!t]
\centering

\begin{minipage}[t]{0.45\textwidth}
\centering
\includegraphics[width=\textwidth]{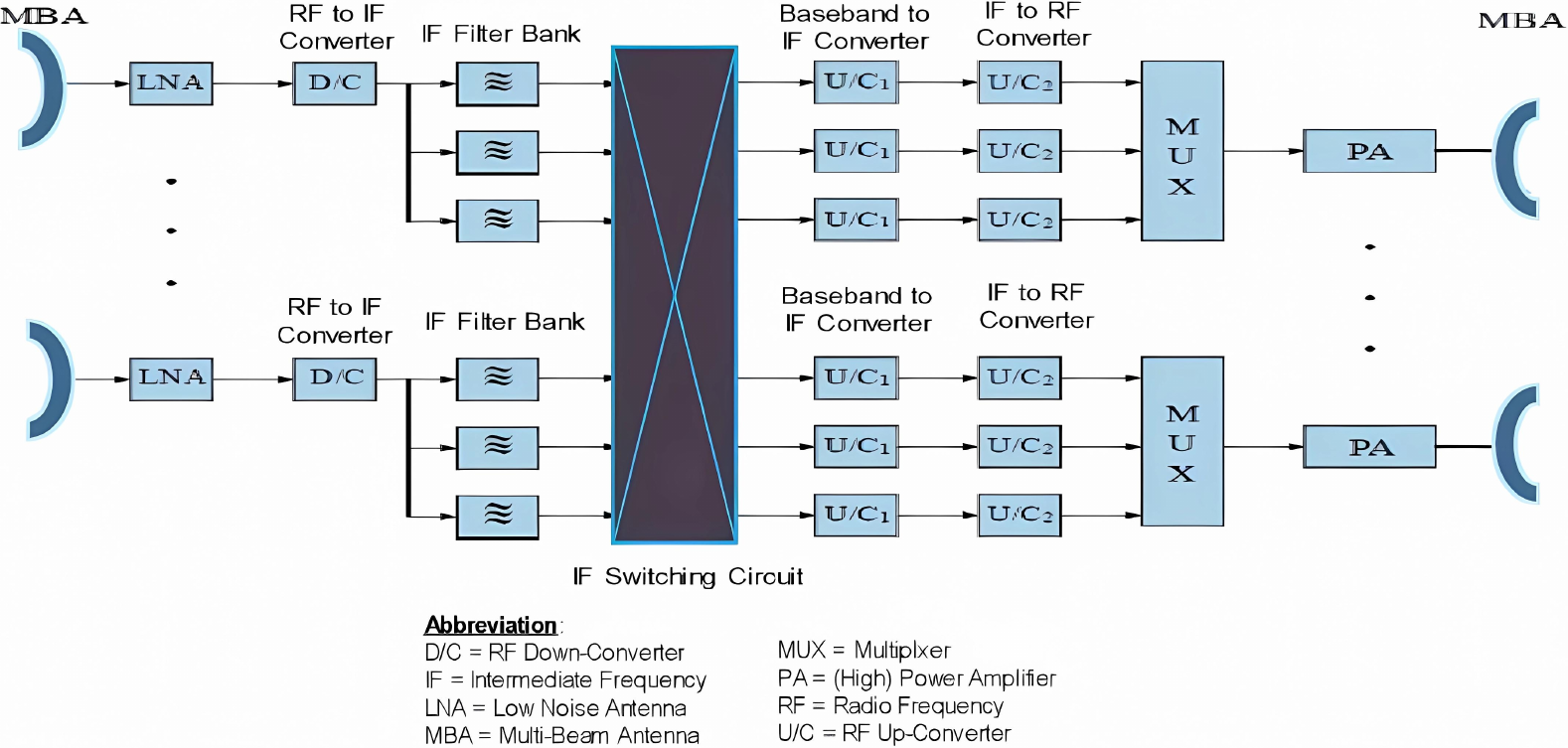}
(a) Analog bent-pipe transponder architecture.
\end{minipage}
\hfill
\begin{minipage}[t]{0.45\textwidth}
\centering
\includegraphics[width=\textwidth]{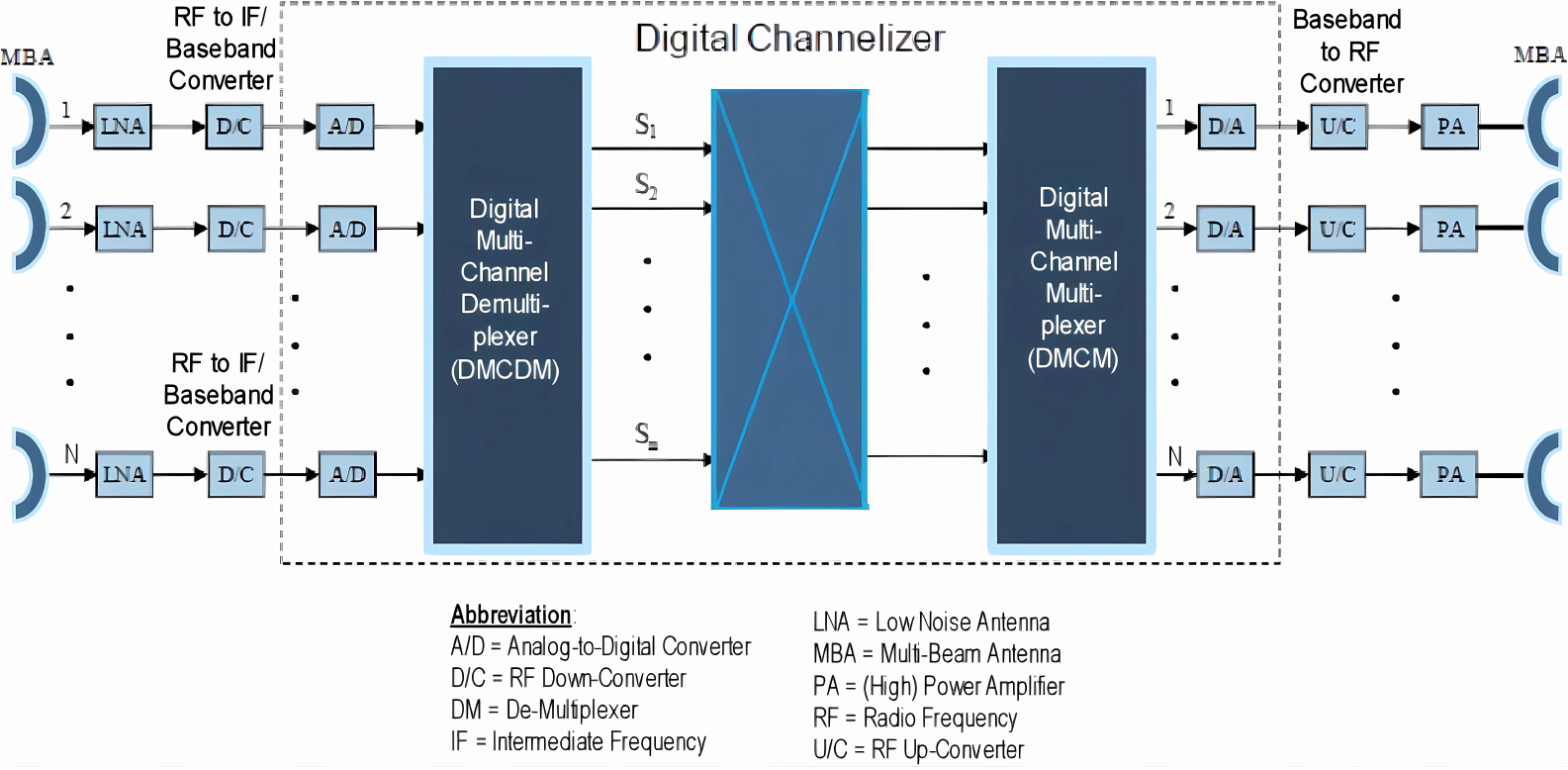}
(b1) Conventional digital bent-pipe transponder architecture.
\end{minipage}

\vspace{1em}

\begin{minipage}[t]{0.45\textwidth}
\centering
\includegraphics[width=\textwidth]{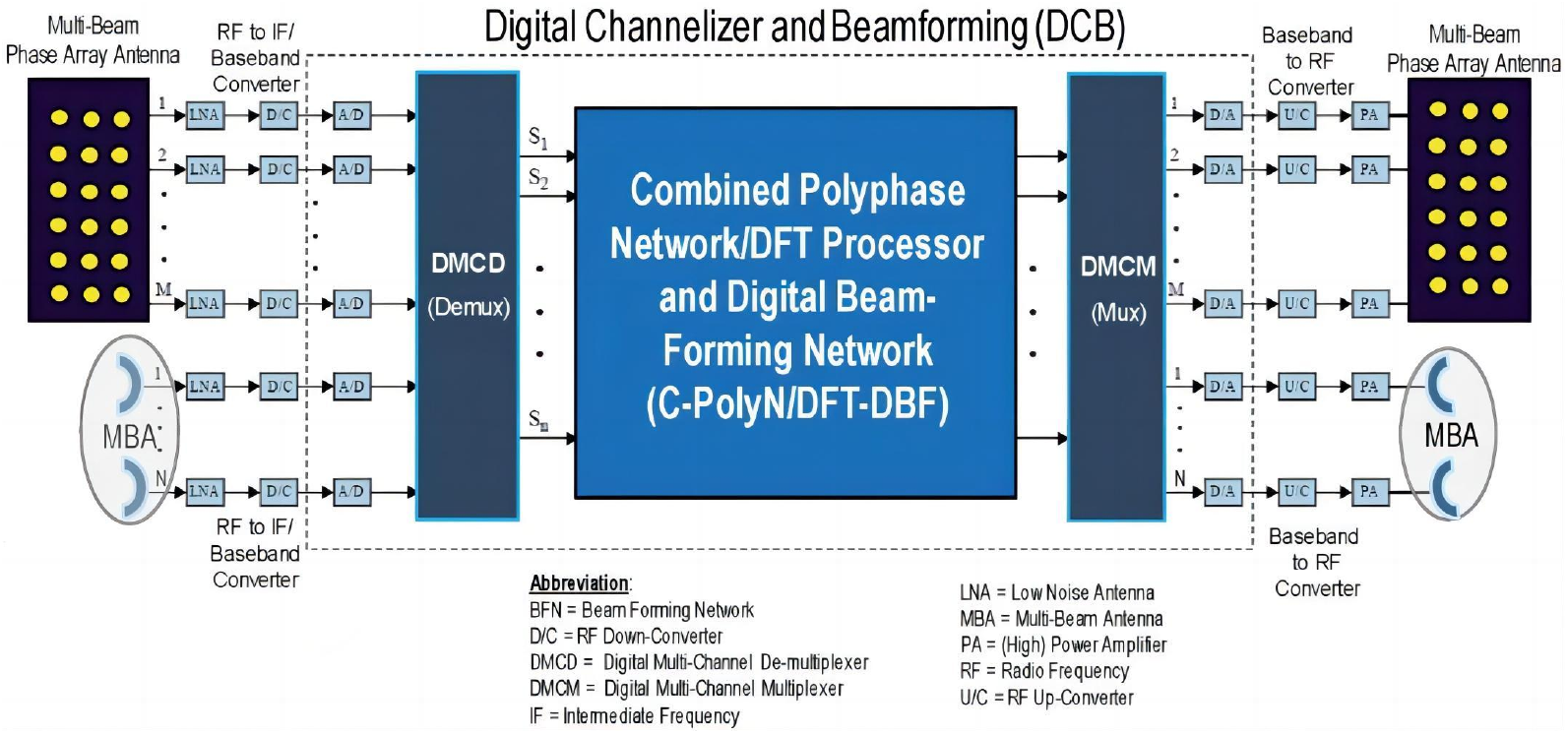}
(b2) Advanced digital bent-pipe transponder with digital channelizer and beamformer.
\end{minipage}
\hfill
\begin{minipage}[t]{0.45\textwidth}
\centering
\includegraphics[width=\textwidth]{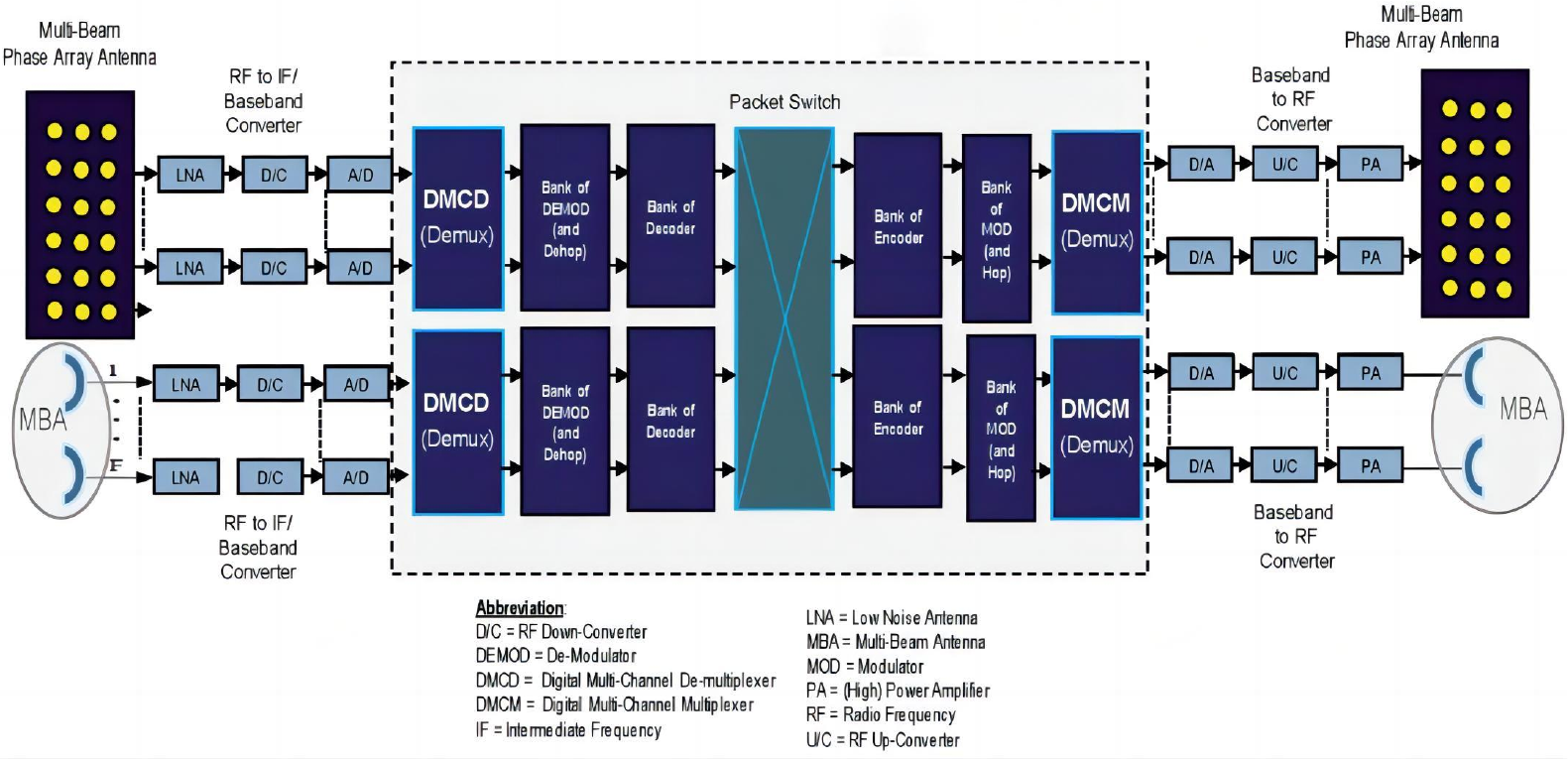}
(c) Regenerative on-board processing transponder architecture.
\end{minipage}

\caption{  {Satellite transponder architectures \cite{nguyen2020overview}.  
(a) Analog  bent-pipe transponders relay signals via analog processing.  
(b) Digital bent-pipe transponders with digital channelization.  
(c) Regenerative bent-pipe transponders support on-board demodulation and decoding for advanced switching.}}
\label{payload architectures}
\end{figure*}


  \emph{a) Analog Bent-Pipe Transponders}: The analog bent-pipe transponder typically consists of LNA, converter, filter bank, and switching circuit, as shown in Fig.\ref{payload architectures} (a). The LNA is responsible for improving the  signal to noise ratio (SNR) of the received RF signal. The converter is employed to obtain IF signals through the process of mixing baseband signals or RF signals with local oscillator signals. Also, the IF signal can be obtained through either primary or secondary frequency conversion. While the primary frequency conversion is advantageous in terms of small size and low power consumption, they may suffer from amplitude and phase imbalances. In contrast, the secondary frequency conversion exhibits excellent selectivity and sensitivity to suppress adjacent interference but come at the cost of increased complexity and power consumption. The filter bank is used to eliminate adjacent band interference and in-band noise. The analog switching circuit is utilized to assign the IF signal to the corresponding channel. In the analog bent-pipe satellite payload architecture, the signal is simply inverted and amplified without further processing. Furthermore, each antenna is composed of multiple analog radio frequency chains, which also limits the number of beams generated by the payload because the total weight of the payload is limited.

\emph{b) Digital Bent-Pipe Transponders:}
Digital bent-pipe transponders are a class of transparent satellite payloads that process signals in the digital domain without performing full signal regeneration. Compared to their analog counterparts, digital transponders offer enhanced flexibility, spectral efficiency, and signal quality through the use of digital signal processing (DSP) techniques. A key feature of these transponders is their ability to digitize received signals and route subchannel data flexibly using digital channelization techniques.
Digital bent-pipe transponders can be broadly classified into two categories, depending on their architectural complexity and functional capabilities.
The first type is conventional digital bent-pipe transponders, which employ digital channelization to enable flexible subband routing and bandwidth allocation.
The second type is advanced digital bent-pipe transponders, which build upon the conventional design by incorporating digital beamforming and multibeam phased array antennas. These enhancements allow for dynamic beam control and high-frequency reuse, significantly improving spatial efficiency and system capacity. The following subsections elaborate on each category in detail.
\begin{itemize}
    \item Conventional Digital Bent-Pipe Transponders:
As depicted in Fig. \ref{payload architectures} (b1), conventional digital bent-pipe transponders digitize incoming signals through digital double conversion and quadrature sampling. Key components include RF bandpass filters, automatic gain controllers, and digital channelization modules. These elements ensure amplitude-phase balance and consistent group delay.
The digital channelizer, typically comprising a polyphase filter network and a DFT processor, divides the wideband signal into uniformly spaced subchannels and later reconstructs the full-band signal. This structure supports flexible bandwidth allocation across users and frequency bands.

Over the decades, advancements such as polyphase decomposition \cite{vaidyanathan1987theory}, FFT-based DFT filter banks \cite{fliege1992orthogonal}, and sampling-rate-reduction techniques \cite{karp1999modified} have improved performance and efficiency. These techniques laid the foundation for non-uniform digital channelization, allowing adaptive bandwidth allocation. Practical implementations of this technology can be found in software-defined radios (SDR) and satellites such as WGS and MUOS \cite{pulliam2008tsat}.

\item  Advanced Digital Bent-Pipe Transponders with Channelizer and Beamformer:
Shown in Fig. \ref{payload architectures} (b2), the advanced type integrates multibeam phased array antennas and adaptive digital beamforming networks. Unlike conventional digital systems, this architecture enables dynamic multibeam formation, precise beam steering, and high-frequency reuse.
The key innovation is the combined use of digital channelization and beamforming. The transponder dynamically adjusts the bandwidth, shape, and direction of each beam while compensating for time, thermal, and amplitude variations \cite{sichi2011beamforming}. Furthermore, it can autonomously regulate gain to counteract environmental changes and analog imperfections \cite{freedman2014advantages}.

\end{itemize}

%

 \emph{c) Regenerative On-Board Processing Transponders}: Fig. \ref{payload architectures} 
 (c) illustrates the architecture of the regenerative on-board processing transponders. The signal received by the multibeam antennas is digitally frequency-decomposed and multiplexed to generate per-channel single-carrier signals. Each of these single-carrier signals is then demodulated and decoded in the respective service channel to recover the originally transmitted information. The information is efficiently packetized and transferred to the downlink using a fast packet switch at the core of the regenerative payload, achieving a statistical multiplexing gain. Additionally, the signals are re-modulated and re-encoded to effectively combat propagation fading and inter-signal interference because the precoding matrix is able to be calculated based on the latest downlink CSI. The regenerative transponder not only allocates bandwidth but also reprocesses the signals to achieve improved system performance. 
The performance comparison of various satellite transponders is listed in Table \ref{transponders}.
%

\begin{table*}[t]
\centering
\caption{Comparison of Satellite Transponders}
\renewcommand{\arraystretch}{1.2} 
\resizebox{\textwidth}{!}{ 
\begin{tabular}{@{}c|c|c|c|l@{}}

\hline \hline
\textbf{\makecell{Transponder\\ type}} & 
\textbf{\makecell{Analog\\ bent-pipe}} & 
\textbf{\makecell{Digital\\ bent-pipe}} & 
\textbf{Regenerative} & 
\textbf{\makecell{Quantification criteria}} \\ 
\hline
\textbf{\makecell{Signal processing\\ capability}} & 
\cellcolor{myred}{L} & 
\cellcolor{myyellow}{M} & 
\cellcolor{myblue}{H} & 
\makecell[l]{
  L: Only analog amplification/frequency conversion;\\ 
  M: Basic digital filtering/modulation adaptation;\\ 
  H: Full-link reconstruction \cite{nguyen2020overview}.
} \\ 
\hline
\textbf{Cost} & 
\cellcolor{myred}{L} & 
\cellcolor{myyellow}{M} & 
\cellcolor{myblue}{H} & 
\makecell[l]{
  L: Single function, reducing manufacturing costs;\\ 
  M: Meet conventional communication needs and control cost investment;\\ 
  H: Achieves complex processing, high-performance transmission, high investment cost.
} \\ 
\hline
\textbf{\makecell{Scalability\\$\&$ \\Flexibility}} & 
\cellcolor{myred}{L} & 
\cellcolor{myyellow}{M} & 
\cellcolor{myblue}{H} & 
\makecell[l]{
  L: Static resource configuration, lacks adaptability;\\ 
  M: Semi-automated parameter adjustment;\\ 
  H:  Fully dynamic multi-dimensional adaptation \cite{butash2010leveraging}.
} \\ 
\hline
\textbf{Reliability} & 
\cellcolor{myyellow}{M} & 
\cellcolor{myblue}{H} & 
\cellcolor{myblue}{H} & 
\makecell[l]{
  L: $< 99.0\%$, no rain fade mitigation;\\ 
  M:  $99.0\%–99.8\%$, partial compensation, availability;\\ 
  H: $\geq 99.9\%$, uses redundancy.
} \\ 
\hline
\textbf{\makecell{Network\\ scale}} & 
\cellcolor{myred}{L} & 
\cellcolor{myyellow}{M} & 
\cellcolor{myblue}{H} & 
\makecell[l]{
  L: Supports $<$100 users, single-channel bandwidth $<$1 Gbps;\\ 
  M: Supports 100–500 users, bandwidth 1–5 Gbps;\\ 
  H: Supports $>$500 users, bandwidth $>$5 Gbps.
} \\ 
\hline
\textbf{\makecell{Transmission\\ rate}} & 
\cellcolor{myred}{L} & 
\cellcolor{myyellow}{M} & 
\cellcolor{myblue}{H} & 
\makecell[l]{
  L: $<$100 Mbps, relies on basic modulation and narrow bandwidth;\\ 
  M: 100–500 Mbps, uses standard modulation;\\ 
  H: $>$500 Mbps, uses high-order modulation and wide bandwidth.
} \\ 
\hline
 \hline
\end{tabular}
}
\label{transponders}
\end{table*}

\subsection{The User Segment}
The user segment comprises high-quality UTs transceivers that transmit and receive satellite signals in the Ka and Ku bands. The user communication terminals can take various forms, including fixed terminals and mobile terminals. 
The fixed terminal typically consists of a very small aperture terminal station (VSAT), which is a two-way satellite earth station. The VSAT station comprises butterfly antennas, LNAs, frequency converters, orthogonal mode-to-conversion converters and an indoor units. The frequency converter mixes the uplink baseband signal or downlink RF signal with local oscillator signals, while the LNA amplifies the received satellite signals in the downlink. The orthogonal mode-to-conversion converter isolates the orthogonal polarization of the signal, and sends the transmitted or received signals to different ports. The indoor unit serves as a satellite modem that interfaces between the outdoor unit and customer-supplied equipment, and controls the transmission or reception of satellite signals.   Willbold et al. \cite{willbold2024vsaster} conducted a systematic security assessment of the VSAT network by analyzing its background, threat taxonomy, experimental analysis, and insecure VAST design practices. It revealed the serious security problems existing in the VSAT network, including denial of service, attacker-in-the-middle, and eavesdropping. Siddiqui et al. \cite{10482090} introduced a  power system simulation for engineers-based VSAT, which realized the evaluation of voltage stability through automated modeling, perturbation analysis, and a python toolkit.
Mobile terminals include handheld terminals, vehicle terminals, naval terminals, and aircraft terminals. Compared to fixed terminals, mobile terminals require smaller antennas to ensure portability, as well as screen display systems to show the user interface and communication status. Mobile terminals also need sufficient memory to store user data and communication records. Moreover, the battery life and power consumption of mobile terminals have higher requirements as they need to maintain continuous and uninterrupted communication for extended periods without frequent charging.
\subsection{Summary \& Lessons Learnt}
 {
The hardware foundation of multibeam HTS systems is the cornerstone for efficient resource allocation and effective precoding. Meanwhile, the hardware foundation provides technical support for resource allocation and precoding. So, we have introduced in detail the hardware foundation of the multibeam HTS system, including the ground segment, space segment and user segment. The structure diagram can be found in Fig.
\ref{Satcom system}.  The ground segment, featuring advanced gateway and TT$\&$C stations, connects terrestrial networks with satellite systems. The space segment includes on-board payload and satellite constellations. These components play a critical role in determining system capacity, operational flexibility, and interference management.  The user segment comprises fixed and mobile terminals, with growing demands for portability, performance, and security.}

 { The multibeam architecture inherently introduces considerable inter-beam interference, and the dynamic and heterogeneous nature of user demand can substantially escalate the complexity and cost of system hardware. Although advanced precoding techniques offer promising theoretical gains, their practical implementation is often hindered by limitations such as on-board memory constraints and restricted processing capabilities.
This insight motivates the integrated view adopted in this paper: hardware foundation, resource allocation, and mulitibeam precoding technique must be co-designed and jointly optimized. In the following sections, we delve deeper into the the two functional domains of resource allocation in Section \RNum{4} and precoding in Section \RNum{5}, with continuous reference to their underlying hardware requirements and limitations. Their interdependencies are summarized in Fig. \ref{SectionContact}, which serves as a structural guide for understanding the coordinated evolution of multibeam HTS systems.}

\section{Hardware Requirements and Limitations}
The design and performance of multibeam HTS systems are heavily influenced by the hardware available and its inherent limitations. As the complexity of these systems increases, the hardware requirements evolve, demanding higher processing power, better efficiency, and more flexible resource management. However, these advancements in hardware also come with significant constraints that must be carefully managed to ensure optimal system performance \cite{elbert2008introduction,6529010}. This section addresses the hardware requirements and limitations of key operations in HTS systems, with a focus on two critical areas: resource allocation and precoding. 
In the first part, we examine the hardware requirements for resource allocation, which are critical for managing power distribution, bandwidth assignment, and beam control across numerous beams  \cite{chen2021system}. We then analyze the hardware constraints affecting resource allocation, addressing challenges arising from power-sharing limitations, beamforming network architectures, and the nonlinear characteristics of power amplifiers.
The second part of this section shifts focus to precoding, where we detail the hardware prerequisites for enabling high-performance signal processing and mitigating inter-beam interference. Subsequently, we discuss the hardware limitations associated with precoding, emphasizing factors such as computational complexity, nonlinear distortion, and phase instability that may impair system performance \cite{sharma2015cognitive}.

\subsection{Hardware Requirements and Limitations for Resource Allocation}
The hardware requirements and limitations for effective resource allocation, which is critical for optimizing power distribution, beam management, and bandwidth allocation in HTS systems, are as shown in Table \ref{hardware_resource}. These systems operate with multiple beams and handle dynamic user traffic.
Each hardware foundation in this stage serves a specific function, and their design requirements are driven by the need to handle high throughput, maintain system stability, and minimize latency. However, practical implementation often reveals several limitations that impact the overall system efficiency.
\begin{table*}[t]
\centering
\caption{Hardware Requirements and Limitations for Resource Allocation.}
\renewcommand\arraystretch{1.2}
\resizebox{0.95\textwidth}{!}{
\begin{tabular}{l|l|l}
\hline \hline
\textbf{Hardware   foundation} & \textbf{Required capability (design)}                              & \textbf{Performance limitation (reality)}                                  \\ \hline
Digital   baseband unit        & Flexible signal format support   and high-speed modulation control & High computational complexity   leads to latency and power consumption     \\ \hline
Data   memory                  & Real-time buffer for   traffic-aware bandwidth allocation          & Limited read/write speed and bandwidth   under concurrent access           \\ \hline
Frequency   shifter            & Precise frequency tuning for   dynamic carrier mapping             & Susceptible to frequency   mismatch and added DSP load                     \\ \hline
BFN    & Rapid phase and amplitude   steering across beams                  & Sensitive to phase calibration   and thermal/aging drift                   \\ \hline
TWTA                           & Per-beam power adjustment based   on user demand                   & Limited dynamic range and power   back-off reduces efficiency              \\ \hline
Beam   scheduling controller   & Adaptive scheduling of beams and   bandwidth                       & Signaling latency and   algorithmic complexity under fast-changing demands \\ \hline \hline
\end{tabular}
}
\label{hardware_resource}
\end{table*}

\emph{1) Digital Baseband Unit:}
The digital baseband unit is required to support flexible signal formats and high-speed modulation control. This capability ensures that the system can accommodate diverse signal types and modulations across multiple beams. 
However, the high computational complexity for real-time signal processing can introduce latency and increase power consumption, adversely affecting system responsiveness. Additionally, the digital baseband unit is responsible for managing the connectivity matrix between beams and amplifiers to facilitate efficient power distribution. However, power-sharing constraints among beams complicate resource allocation, requiring sophisticated power management strategies \cite{vidal2021methodology}. Furthermore, nonlinear characteristics of power amplifiers, such as distortion and leakage, degrade signal quality and complicate interference management, which impacts the effective reuse of beams \cite{kisseleff2020radio}.

\emph{2) Data Memory: }
Real-time data buffering is essential for efficiently allocating bandwidth in response to traffic demands. The memory module needs to provide a buffer that can quickly store and retrieve data while managing bandwidth dynamically. The main limitation here is the limited read/write speed and bandwidth, which becomes apparent under conditions of concurrent access, such as when multiple beams are active simultaneously.

Satellite on-board storage capacity is limited, and different satellites may be equipped with buffers of varying sizes. Therefore, constraints must be applied to ensure that stored data does not exceed capacity limits. Sheng et al.  \cite{sheng2017toward} proposed strategies to manage on-board storage efficiently, ensuring that the system remains operational even under high data load conditions. 

\emph{3) Frequency Shifter:}
The frequency shifter allows for precise frequency tuning, enabling dynamic carrier mapping across multiple beams. This flexibility is essential for ensuring that the system can adapt to changing traffic conditions. However, frequency mismatches can occur due to hardware inaccuracies, which, combined with the additional computational load introduced by signal processing, can compromise system efficiency\cite{mazzali2015board}.



\emph{4) BFN:}
The BFN is responsible for rapidly steering the phase and amplitude of beams to achieve optimal coverage and power distribution. Precise phase and amplitude adjustments are critical; however, BFNs are susceptible to phase calibration errors and thermal or aging-related drift, which can degrade beamforming accuracy and efficiency over time \cite{fenech2016role}.

For beamforming networks supporting over 1000 reconfigurable beams, balancing the trade-offs between analog and digital BFNs is essential. Fully analog BFNs may suffer from quality degradation and higher power consumption, while fully digital BFNs increase the number of ports, leading to higher complexity, power use, and cost. To address these challenges, a hybrid BFN architecture combining analog and digital components has been proposed \cite{de2020future}. This approach leverages the strengths of both technologies to provide a modular, scalable, and flexible solution for beamforming in HTS systems.

\emph{5) TWTA:}  
TWTAs are used for per-beam power adjustment based on user demand, ensuring that each beam receives the appropriate amount of power. They are essential for handling high power levels; however, their limited dynamic range and the requirement for power back-off reduce efficiency, especially in highly dynamic systems where power control must continuously adapt to varying user needs \cite{Sharma2021System}.

In the SFB antenna architecture, an additional constraint arises: each beam is typically connected to a single TWTA, limiting the flexibility of power routing. Cocco et al. \cite{cocco2017radio} investigated radio resource management for flexible satellite payloads, highlighting how the SFB architecture imposes constraints on power distribution and suggesting optimization strategies to address these limitations.

\emph{6) Beam Scheduling Controller:}  
The beam scheduling controller dynamically allocates bandwidth and beam resources to accommodate varying traffic demands. It requires an adaptive scheduling algorithm capable of rapidly reconfiguring beam patterns and power allocation in real-time. However, signaling latency and algorithmic complexity can limit responsiveness under rapidly changing conditions, leading to inefficiencies in resource management.

To enhance scheduling flexibility at the physical layer, ferrite switches are sometimes introduced at the post-amplification stage to enable dynamic slot allocation. While effective, this approach introduces switch losses and increases system complexity. Sheng et al. \cite{sheng2017toward} discussed the inherent trade-offs between flexibility and system complexity in the context of reconfigurable broadband satellite networks.

\subsection{Hardware Requirements and Limitations for Precoding}
The hardware requirements and limitations associated with the implementation of precoding in multibeam HTS systems are summarized in Table \ref{hardware_precoding}. Precoding, as a key enabler of full frequency reuse and interference mitigation, imposes stringent demands on both on-board and ground hardware. Each hardware component plays a critical role in enabling real-time, high-dimensional signal processing, and their performance directly impacts the effectiveness of the overall precoding strategy.

\begin{table*}[t]
\centering
\caption{Hardware Requirements and Limitations for Precoding.}
\renewcommand\arraystretch{1.2}
\resizebox{0.95\textwidth}{!}{
\begin{tabular}{l|l|l}
\hline \hline
\multicolumn{1}{c|}{\textbf{Hardware   foundation}} & \multicolumn{1}{c|}{\textbf{Required capability (design)}} & \multicolumn{1}{c}{\textbf{Performance limitation (reality)}}             \\ \hline
Digital baseband unit                               & Real-time matrix precoding and symbol mapping.              & High computational load under   per-symbol processing constraints.         \\ \hline
Data memory                                         & Fast access to CSI, precoding weights, and symbols.         & Limited access speed degrades   throughput under parallel operation.       \\ \hline
BFN                                                 & Fine-grained per-beam phase  control.                       & Phase error accumulation impacts   precoding accuracy.                     \\ \hline
TWTA                                                & Power-efficient amplification of precoded signals.          & Nonlinear distortion and group   delay affect signal fidelity. \\ \hline
Digital   channelizer                               & Subband separation for frequency-domain precoding.          & Filter bank switching latency   and subband leakage degrade orthogonality. \\ \hline
Time slot control switch                            & Frame-level symbol path  reconfiguration.                   & Switching delay and   synchronization jitter compromise timing precision.  \\ \hline
CSI feedback interface                              & Accurate and timely CSI delivery to the precoding engine.   & Bandwidth and feedback delay   reduce CSI freshness and accuracy.          \\ \hline \hline
\end{tabular}
}
\label{hardware_precoding}
\end{table*}

\emph{1) Digital Baseband Unit:}
This component is responsible for real-time execution of complex precoding matrices and symbol-level processing. The design requires high computational throughput to support multiuser precoding. However, the per-symbol computational complexity significantly increases latency and power consumption, making real-time operation challenging in large-scale beam scenarios.

\emph{2) Data Memory:}
To support fast computation, the system must rapidly access CSI, precoding coefficients, and transmit symbols. In practice, the memory subsystem is limited by read/write speed and bus bandwidth, which constrains throughput, especially during high-load parallel operations.

\emph{3) BFN:}
The BFN provides per-beam phase control for spatial signal shaping and interference suppression. Its precision is essential for maintaining the spatial characteristics of the precoded signal \cite{fenech2016role}. However, hardware imperfections, such as phase mismatches and thermal drift, can accumulate over time, degrading the accuracy and consistency of precoding.

\emph{4) TWTA:}
TWTAs amplify the precoded signals before transmission \cite{cocco2017radio}. Although designed for power efficiency, their inherent nonlinearities and group delay distortions can alter the waveform, reducing the effectiveness of interference mitigation and degrading signal fidelity.

\emph{5) Digital Channelizer:}
In frequency-domain precoding, the digital channelizer is responsible for isolating subbands. However, limitations such as switching latency between filter banks and leakage between subbands reduce the orthogonality of the system and introduce unwanted interference.

\emph{6) Time Slot Control Switch:}
This component dynamically reconfigures signal paths on a per-frame basis to support scheduling and beam hopping. Practical constraints such as switching delay and timing jitter can lead to synchronization issues, which impact symbol alignment and overall system timing precision.

\emph{7) CSI Feedback Interface:}
Accurate and timely CSI is essential for effective precoding. The design assumes low-latency, high-bandwidth feedback links from the user terminal to the precoding engine \cite{luis2020artificial}. However, in reality, feedback channels suffer from limited bandwidth and non-negligible delay, leading to outdated or inaccurate CSI, which impairs the performance of the precoding algorithm \cite{hu2022dynamic}.

Table \ref{hardware_resource} and Table \ref{hardware_precoding} distinguish the functional demands and operational bottlenecks of hardware foundations during the resource allocation and precoding phases. While both phases share components such as the BFN, TWTA, and baseband unit, their roles differ significantly. In resource allocation, emphasis lies in beam-level and bandwidth-level control, whereas precoding requires symbol-level accuracy and low-latency matrix computation. Recognizing these differences is essential for decoupled or joint optimization of real-time payload control.

\section{Resource Allocation}
As radio resources on multibeam HTS are limited, efficient allocation of these resources is essential to enhance system resource utilization, maximize transmission rates, and ensure user QoS. Specifically, effective radio resource allocation for satellites involves the proper allocation of bandwidth and frequency multiplexing to optimize communication capacity, appropriate allocation of multiple point beam power to mitigate interference, and optimal allocation of time slot resources to manage communication duration. In this section, we will provide a review of bandwidth, power, time slot and joint resource allocation for multibeam HTS.

\subsection{The Specificity of Resource Allocation in Multibeam HTS Systems}
 {Compared to traditional FSS satellites, resource allocation pays special attention to the dynamic variations in multibeam HTS, where the interplay between multibeam patterns and time-varying user demands introduces unprecedented optimization challenges \cite{tani2017flexibility}. Firstly, as users move within the system's coverage area, their associated beams and resource demands change continuously. Secondly, in multibeam HTS systems, the increased number of beams necessitates more sophisticated management of inter-beam interference. Traditional FSS satellites, with fewer beams, face relatively simpler interference issues, whereas multibeam HTS systems require complex interference mitigation strategies. Lastly, multibeam HTS supports diverse service types, each with distinct QoS demands \cite{chen2021system}. Meeting these heterogeneous requirements significantly increases the complexity of resource allocation. In contrast, traditional FSS satellites, often handling more uniform service types, face simpler resource allocation scenarios.}

\subsection{Bandwidth Allocation}

In multibeam HTS systems, efficient bandwidth allocation must dynamically adapt to varying traffic demands across beams, aiming to maximize overall throughput while ensuring equitable service delivery. Existing approaches include deterministic algorithms \cite{hong2008optimal}, flexible allocation schemes, and data-driven methods based on neural networks and deep learning (DL). The following discussions elaborate on these key trade-offs, and trace the evolution of bandwidth allocation techniques \cite{hong2008optimal}.

\emph{1) Key Trade-offs:}
Bandwidth allocation in multibeam HTS systems involves managing trade-offs across capacity, fairness, complexity and adaptability.
Firstly, this includes balancing system capacity and user fairness, where maximizing throughput must not exclude users with poor channel conditions \cite{chen2021system}. Secondly, another key trade-off is between algorithm complexity and adaptability. While simple algorithms work well in static settings, dynamic traffic environments require more adaptive solutions such as deep reinforcement learning (DRL). Effective strategies should account for user diversity, fairness, and responsiveness to changing conditions.

\emph{i) Balancing System Capacity and User Fairness:}
Adopting a maximum sum-rate objective can significantly increase system throughput by prioritizing users with good channel conditions, but this often leaves users in poor conditions underserved \cite{hong2008optimal}. In contrast, a rate-balancing objective aims to improve service to the worst-off users, enhancing fairness at the cost of reduced total throughput. A more nuanced approach, rate-matching, seeks a compromise by adjusting resource allocation to balance performance and fairness depending on system conditions \cite{XTYD2025052200W}. These trade-offs reflect the need to consider user heterogeneity and service equity in resource management.


\emph{ii) Balancing Algorithm Complexity and Adaptability:}
Traditional heuristic or optimization-based methods perform effectively in static environments due to their structural clarity and low computational overhead. However, they struggle to handle dynamic traffic patterns and heterogeneous service requirements. On the other hand, DL-based algorithms such as DRL demonstrate strong adaptability in non-stationary environments \cite{hu2018deep}, enabling real-time responses to spatial-temporal traffic fluctuations. Nevertheless, such models typically demand high training costs and computational resources. Therefore, selecting a suitable algorithm necessitates a careful balance between operational efficiency and environmental responsiveness.

\emph{2) The Evolution of Bandwidth Allocation Techniques:}
To address the complex trade-offs involved in bandwidth allocation for multibeam HTS systems, a wide range of algorithmic approaches have been developed, mainly in the following three directions.  This progression reflects the growing need for flexible and real-time solutions in increasingly dynamic satellite communication environments.

\emph{i) Optimization-Theoretic Methods Based on Deterministic Modeling:}
Early work by Takeshi Mizuikjz et al. introduced a deterministic allocation method aimed at minimizing co-channel interference, identifying the problem as NP-complete, hence computationally intractable in its general form \cite{hong2008optimal}. This recognition catalyzed the pursuit of more tractable solutions. Later methods introduced flexible allocation strategies—for example, assigning more bandwidth to beams with higher traffic demand to improve fairness, albeit at the cost of total capacity \cite{park2012dynamic}. To refine allocation granularity, dichotomous search and subgradient algorithms were introduced, enabling more precise and efficient resource distribution \cite{park2012flexible, wang2013optimal}.

\emph{ii) Heuristic and System-Aware Methods for Practical Allocation:}
Model-based approaches such as model predictive control and sparse optimization have been validated using realistic traffic patterns, improving practical deployment feasibility \cite{abe2018frequency}. In environments with overlapping bandwidth constraints, a non-dominated beam coding algorithm was developed to enhance solution quality while reducing computational complexity \cite{gao2022research}. These efforts demonstrate a shift from purely theoretical models toward algorithms that better accommodate operational requirements in HTS systems.

\emph{iii) Neural Network-Based Methods for Adaptive Bandwidth Allocation:}
Kyrgiazos et al. applied neural networks to estimate beam-level traffic distribution and jointly optimize bandwidth and beam size \cite{kyrgiazos2013irregular}. Building on this, Hu et al. proposed a DRL-based framework that dynamically allocates bandwidth while accounting for both spatial and temporal traffic characteristics \cite{hu2018deep}. In parallel, attention has turned to feeder link bandwidth management, with methods developed to jointly optimize user and feeder link utilization, thereby avoiding congestion and improving localized throughput \cite{tani2017flexibility}. These innovations represent a paradigm shift toward self-learning, real-time optimization frameworks.



\subsection{Power Allocation}
 {
In multibeam HTS systems, power allocation faces multiple complex challenges primarily due to the system’s multibeam structure and its high-capacity demands. Firstly, traffic load varies significantly across beams because of differences in time, geography, and user activity \cite{durand2017power}. This dynamic and uneven demand requires power allocation strategies that can respond in real-time to ensure efficiency without wasting resources. However, achieving real-time responsiveness is computationally difficult, as it involves solving high-dimensional nonlinear optimization problems under the constraint of limited on-board computing capacity \cite{hong2008optimal,aravanis2015power}.
Secondly, co-channel interference among adjacent beams that share the same frequency band further complicates power allocation. Therefore, power optimization alone is inadequate and must be coordinated with frequency and time allocation to effectively mitigate interference \cite{aravanis2015power}.
Thirdly, nonlinear distortions introduced by satellite power amplifiers, such as traveling wave tube amplifiers, further intensify inter-beam interference and add complexity to power management.
Finally, power allocation itself is an NP-complete problem that requires a careful balance between maximizing overall system throughput and maintaining fairness among beams \cite{wang2014optimization}.}

 {To address these challenges, existing research can be broadly categorized into three groups: 1)  power optimization in static network environments, 2) power optimization with interference awareness, and 3) power optimization in dynamic network environments \cite{qi2011optimum}. The following discussion provides a detailed review of representative studies in each category.}

\emph{1) Power Optimization in Static Network Environments:}
These studies generally assume a static network environment and formulate convex optimization problems to align power supply with heterogeneous beam demands. In \cite{choi2005optimum, hong2008optimal, qi2011optimum, wang2014optimization, aravanis2015power}, power allocation is modeled as an NP-hard problem with the objective of minimizing the mismatch between capacity and demand. For example, Choi et al. \cite{choi2005optimum} proposed optimal beam and power allocation schemes, although their approach did not resolve the Lagrange multiplier. To address this, Hong et al. \cite{hong2008optimal} introduced a heuristic dichotomy method to determine the optimal Lagrange multiplier. Similarly, Qi et al. \cite{qi2011optimum} applied a subgradient method to identify optimal solutions while accounting for system capacity distribution under varying channel conditions.
In the work by Wang et al. \cite{wang2014optimization}, the focus shifted to intra-beam (user-level) optimization, where reduced system capacity was observed for users with high traffic demand. Aravanis et al. \cite{aravanis2015power} proposed a two-stage optimization approach: the first stage employed a meta-heuristic method to maximize throughput, while the second stage used a dedicated algorithm to minimize total power consumption.

\emph{2) Power Optimization with Interference Awareness:}
Inter-beam interference and additional physical-layer constraints, such as rain fading and beam placement, are explicitly incorporated into the power optimization process. To mitigate rain-induced signal degradation and inter-beam interference, Durand et al. \cite{durand2017power} proposed a particle swarm optimization (PSO)-based algorithm that reallocates power from clear-sky beams to those affected by rain, thereby ensuring compliance with minimum SINR requirements.

Subsequent studies \cite{takahashi2019adaptive, takahashi2020adaptive, takahashi2021dbf} further enhanced the optimization framework by accounting for beam positioning and interference from both co-frequency and cross-frequency beams. Although these works improved spatial modeling in power control, they did not provide explicit algorithms for determining the optimal power allocation values.

\emph{3) Power Optimization in Dynamic Network Environments:}
To overcome the limitations of static resource allocation, recent studies have adopted AI-based approaches to manage time-varying and spatially non-uniform traffic demands. In \cite{destounis2011dynamic}, power was dynamically reallocated based on rain fading predictions using physical models and DRL policies to minimize unmet demand. In \cite{dai2021deep}, DRL was combined with real-time CSI and cache awareness to enhance robustness under channel variation. A systematic comparison of AI algorithms, including PSO, SA, genetic algorithms (GA), and DRL, was conducted in \cite{luis2020artificial}, focusing on scalability and convergence characteristics. To address the issue of local optima in DRL, Hu et al. \cite{hu2022dynamic} proposed a hybrid DRL and SA framework that adapts to environmental changes and improves optimization performance through annealing-based exploration.


\subsection{Time Slot Allocation}
Time slot resource allocation in multibeam HTS systems involves scheduling on-board radio resources such as power and bandwidth for each beam within limited communication time intervals. This allocation aims to maximize resource utilization while avoiding signal collisions. This process is especially important in systems employing the beam hopping technique, where only a subset of beams is activated at any given moment to dynamically respond to varying traffic demands \cite{lei2010frequency}. In beam hopping, power and bandwidth resources are exclusively assigned to the illuminated beams. Moreover, the bandwidth allocated to each beam can be either full or partial, requiring a careful balance between efficient frequency reuse and minimizing interference.

The beam hopping technique was initially proposed in 2006 as a solution for accommodating irregular and time-varying traffic distributions, with the corresponding payload architecture described in \cite{angeletti2006beam}. Subsequent studies \cite{anzalchi2010beam} empirically demonstrated the superior performance of multibeam satellite systems employing beam hopping compared to conventional non-hopping systems, highlighting the importance of judicious time slot allocation in enhancing communication capacity.

For clarity, the extant literature on time slot resource allocation is categorized into three principal domains: 1) foundational algorithms developed in early research, 2) heuristic and optimization-based methodologies, and 3) learning-based approaches approaches for dynamic environments. These categories are elaborated below.

\emph{1) Foundational Algorithms  Developed in Early Research:}
Early contributions to time slot allocation in multibeam systems are exemplified by the work of Gopal et al. \cite{gopal1982optimal}, who formulated an optimization algorithm aimed at minimizing total transmission time. This algorithm employed random matrices combined with substitution matrices to reduce the complexity of switching configurations, accommodating arbitrary numbers of uplink and downlink beams, bandwidth allocations, and traffic demands. Although originating several decades ago, this work laid a foundational basis for subsequent investigations into time slot allocation for multibeam HTS.

\emph{2) Heuristic and Optimization-Based Algorithms:} 
Building on foundational work, considerable research has developed heuristic and iterative optimization methods to address the complexity of time slot allocation in multibeam HTS systems. These approaches strive to balance computational efficiency with allocation effectiveness, tackling challenges such as interference reduction, capacity matching, and fairness.

Lei et al. \cite{lei2010frequency} proposed an iterative algorithm that allocates time slots based on beams’ unassigned SINR values, efficiently leveraging channel conditions to improve system performance. Alegre et al. \cite{alegre2011heuristic} refined this by assigning beams to specific time slots within a fixed window, aligning capacity with demand while reducing co-channel interference.
Han et al. \cite{han2015qos, han2015qos1} incorporated QoS and fairness, proposing a framework that balances throughput with inter-beam fairness and addresses delay constraints using stochastic gradient and time-sharing methods.
Zhong et al. \cite{zhong2017joint} applied subgradient optimization to maximize weighted throughput under fairness constraints. Extending this, Shi et al. \cite{shi2017joint} and Wang et al. \cite{wang2019resource} introduced clustering techniques that partition coverage areas to optimize power allocation and time slot scheduling within clusters. Wang et al. further improved clustering by optimizing cluster formation based on capacity disparities.
More recently, Zhang et al. \cite{zhang2021dynamic} developed a genetic algorithm for beam-splitting clusters that optimizes time slot allocation considering rain attenuation, though focusing solely on time slots.

Overall, these methodologies demonstrate a clear evolution from basic SINR-based heuristics to sophisticated frameworks integrating QoS, fairness, clustering, and environmental factors. Nevertheless, challenges persist in jointly optimizing multiple resource dimensions under real-time constraints.

\emph{3) Learning-Based Approaches for Dynamic Environments:} 
To address the challenges posed by rapidly varying traffic patterns and channel conditions, recent research has explored DL approaches. Zhang et al. \cite{zhang2019dynamic} and Lei et al. \cite{lei2020deep} demonstrated that DL-based frameworks can significantly accelerate time slot allocation procedures, effectively adapting to dynamic traffic distributions and channel state variations. These approaches present promising avenues for real-time, adaptive resource allocation by obviating the need for computationally intensive optimization during operation.

\subsection{Joint Resource Allocation}
In multibeam HTS systems, limited on-board resources and diverse user demands make joint resource allocation crucial for maximizing system capacity and efficiency. Compared to optimizing individual resources separately, joint  power, bandwidth, and time slot allocation more effectively addresses interference management, dynamic channel conditions, and variable traffic demands. Existing research on joint resource allocation can be categorized into three main areas: 1) joint power and bandwidth allocation, 2) joint power and time slot allocation, and 3) joint  power, bandwidth, and time slots. The following sections review key contributions within each category.

\emph{1) Joint Power and Bandwidth Allocation:}
This category focuses on simultaneously optimizing power and bandwidth to maximize system throughput under payload resource constraints. Early works such as \cite{lei2010joint, lei2011multibeam} formulated joint allocation by minimizing the ratio of maximum demanded capacity to offered SINR. Heuristic algorithms including simulated annealing \cite{cocco2017radio}, GA \cite{paris2019genetic}, and hybrid methods combining PSO with GA \cite{pachler2020allocating} have improved convergence speed and solution quality. More recent approaches leverage successive convex approximation \cite{abdu2021flexible} and DRL\cite{he2022multi} to enable flexible, on-demand resource allocation that adapts to dynamic traffic and channel variations. Notably, adaptive allocation schemes inspired by the water-filling algorithm \cite{li2014research, 6519406} enable real-time adjustment of resources based on instantaneous channel state information, effectively balancing capacity demands and fading effects.

\emph{2) Joint Power and Time Slot Allocation:}
This group of methods addresses the joint scheduling of time slots and power allocation, particularly within beam hopping frameworks to enhance spectral efficiency and meet QoS requirements. For example, Li et al. \cite{202205005} designed an adaptive power allocation algorithm using PSO within dynamic beam hopping to maximize system QoS. While much research focuses on GEO satellites, recent studies explore LEO systems where transient and delay-sensitive services require specialized strategies. Liang  et al. \cite{DZYX202302006} proposed an energy-efficient joint beam hopping scheduling and power allocation approach tailored for LEO payload constraints. Chen et al. \cite{DZYX202302003} developed a DRL-based beam hopping resource allocation scheme for LEO satellites, emphasizing delay minimization. Liu  et al. \cite{Ziyi} extended resource allocation to long-term scenarios using convex optimization for multibeam hopping across orbits, addressing satellite mobility and varying coverage.

\emph{3) Joint Power, Bandwidth, and Time Slot Allocation:}
Comprehensive joint optimization of power, bandwidth, and time slots aims to further enhance resource utilization by coordinating all critical dimensions simultaneously. Shi and Li et al. \cite{shi2017joint} focused on joint power and time slot allocation for beam hopping downlinks in intelligent gateway systems. Wang Lin et al. \cite{wang2019resource} proposed a hierarchical optimization framework, first optimizing cluster sizes for beam hopping and then jointly allocating power and time slots to maximize bandwidth efficiency. Reviews by Tang Jingyu and Wang Lin et al. \cite{YDTX201905008, NJYD201903005} summarize such multi-resource allocation algorithms, highlighting their shared goal of improving system throughput, spectrum efficiency, and latency. Moreover, Liu and Hu et al. \cite{liu2018deep, hu2019deep} applied deep reinforcement learning to beam hopping resource allocation, achieving significant improvements in capacity and delay through coordinated optimization of beams, bandwidth, power, and time slots.

\subsection{Summary \& Lessons Learnt}
\begin{figure*}[t]
\begin{center}
\includegraphics[width=1\textwidth,height=0.5\textwidth]{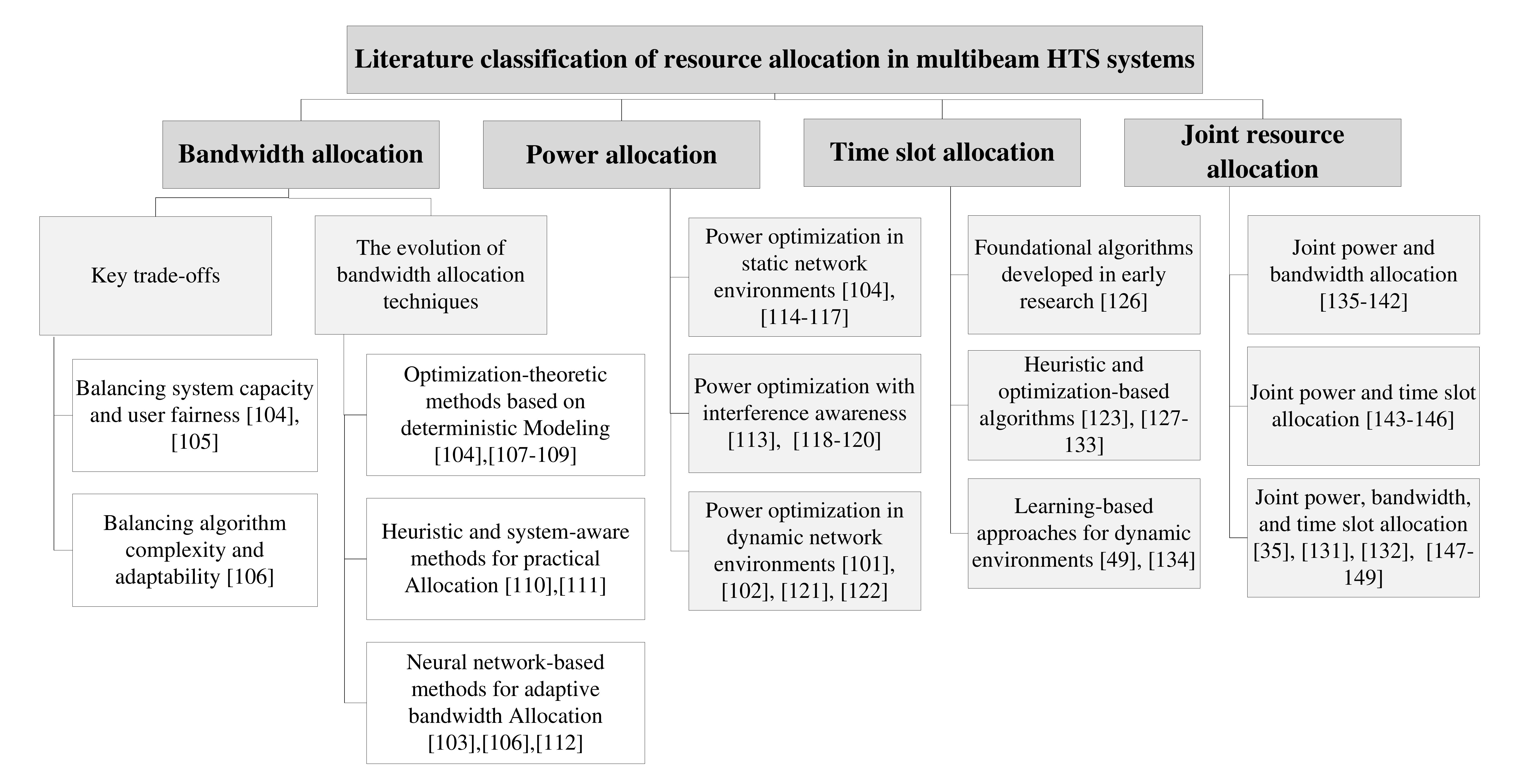}
\end{center}
\caption{Literature classification of resource allocation in multibeam HTS.}%
\label{resource}
\end{figure*}

Resource allocation plays a pivotal role in multibeam HTS systems, ensuring efficient utilization of limited resources, maximizing system throughput, and maintaining the desired QoS for users. To provide a structured overview, existing research is categorized into four primary areas: bandwidth allocation, power allocation, time-slot allocation, and joint resource allocation, as illustrated in Fig. \ref{resource}.

Single-dimensional resource allocation methods such as bandwidth, power, and time slot allocation \cite{angeletti2006beam} offer simplicity and are well-suited for systems with static traffic patterns and limited reconfigurability. For instance, fixed bandwidth or power schemes offer low complexity but often struggle to adapt to dynamic traffic conditions, potentially resulting in resource underutilization. Time-slot allocation introduces a degree of flexibility but can suffer from signaling latency and coarse resolution, limiting its effectiveness in rapidly changing environments.

In contrast, joint resource allocation schemes aim to optimize multiple dimensions simultaneously, yielding greater throughput and improved fairness among beams. These methods, however, often involve non-convex or NP-hard problems that require heuristic algorithms, problem decomposition, or AI-based solutions to reach near-optimal results. While more computationally demanding \cite{shi2017joint}, such approaches are well-suited for systems featuring flexible payloads and advanced onboard processing capabilities \cite{YDTX201905008, NJYD201903005}. Therefore, selecting an appropriate allocation strategy requires a careful trade-off between algorithmic complexity, hardware constraints, and the system’s operational dynamics.

\section{Multibeam Precoding Technique}

In multibeam HTS systems, precoding plays a pivotal role in managing inter-beam interference and enhancing overall spectral efficiency, thereby significantly improving system throughput and link reliability \cite{li2021downlink}. Multibeam precoding in HTS faces unique challenges due to the satellite’s limited on-board processing power and stringent payload constraints \cite{chen2021system}. The complexity of precoding algorithms, especially those involving large-scale matrix operations, directly impacts on-board resource consumption and real-time processing capability. Consequently, the development and optimization of low-complexity precoding techniques tailored for multibeam HTS are critical for enabling high-performance satellite communications. This section provides a comprehensive review of precoding algorithms classified according to different system architectures, including single gateway, multiple gateways, on-board precoding, and hybrid on-board/on-ground precoding models.

\subsection{The Specificity of Precoding in Multibeam HTS Systems}
Traditional FSS systems typically employ only 1 to 20 wide beams, which limits their ability to manage interference and results in relatively low data throughput. In contrast, multibeam HTS systems adopt spot beam technology and support substantially higher beam counts, typically ranging from 100 to over 3000. This represents a scale increase by a factor of 10 to 150 compared to traditional satellites \cite{ahmad2021zero,wang2019multicast}. The increased beam density enables higher antenna gain and efficient spatial frequency reuse, significantly enhancing system capacity, spectral efficiency, and user coverage.

To mitigate co-channel interference in multibeam satellite communications, several key techniques are commonly employed. Firstly, frequency reuse schemes partition the available spectrum into sub-bands that are reused by beams sufficiently separated in space to minimize interference; however, this approach compromises spectral efficiency \cite{li2008performance, sharma2015cognitive}. Secondly, multi-user detection, primarily applied in satellite CDMA or multibeam MIMO systems, performs interference cancellation at the receiver but requires real-time channel estimation and entails significant computational complexity \cite{li2006satellite, wang2015research}. Thirdly, precoding jointly processes user signals at the ground station or on-board payload, shifting computational complexity away from user terminals. When accurate CSI is available, it can significantly suppress inter-beam interference, thereby improving the effective SINR at receivers \cite{zhao2021research}. However, its performance is constrained by CSI aging, hardware limitations, and nonlinear distortions in practical systems \cite{joroughi2018board,chatzinotas2011joint,ahmad2021zero,wang2019multicast}.


\subsection{Single Gateway Precoding}

\begin{figure}[t]
\begin{center}
\includegraphics[width=8.5cm]{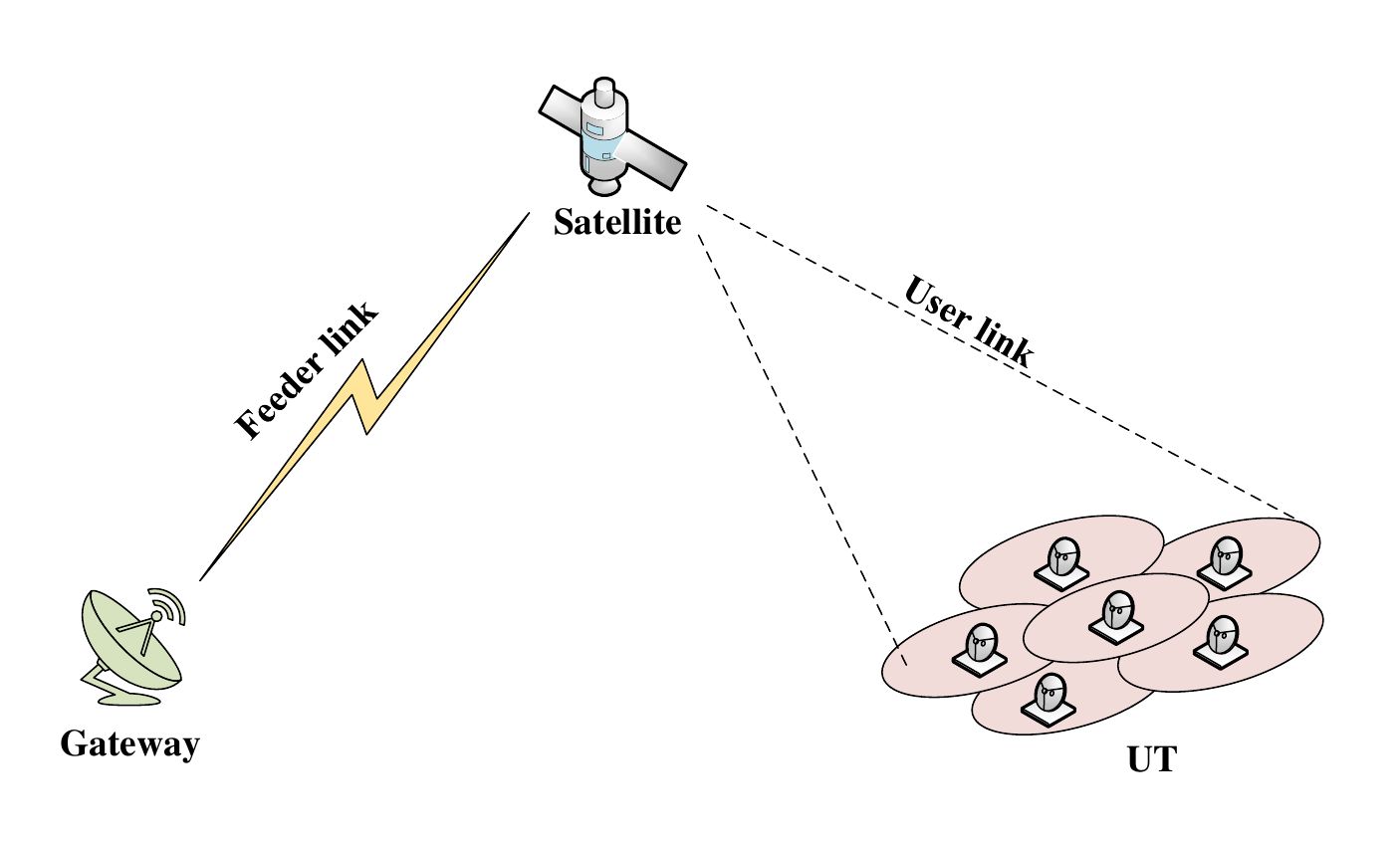}
\end{center}
\caption{  {Single  gateway system model consists of one gateway station, one satellite, and multiple user terminals.  The gateway estimates the CSI, generates the precoding matrix, and processes the transmit signal accordingly.  The satellite relays the precoded signal via multibeam antennas to UTs, which act as passive receivers. Both unicast and multicast precoding algorithms (e.g., MMSE, ZF) are employed to mitigate inter-beam interference and improve resource utilization.}}%
\label{single-gateway}
\end{figure}

The single gateway system model is depicted in Fig. \ref{single-gateway}, illustrating the framework for the two main precoding approaches discussed here: 1) unicast precoding, where each beam serves one user per time slot, and 2) multicast precoding, which enables simultaneous transmission to multiple users within the same beam.

\begin{figure}[t]
\begin{center}
\includegraphics[width=8.0cm]{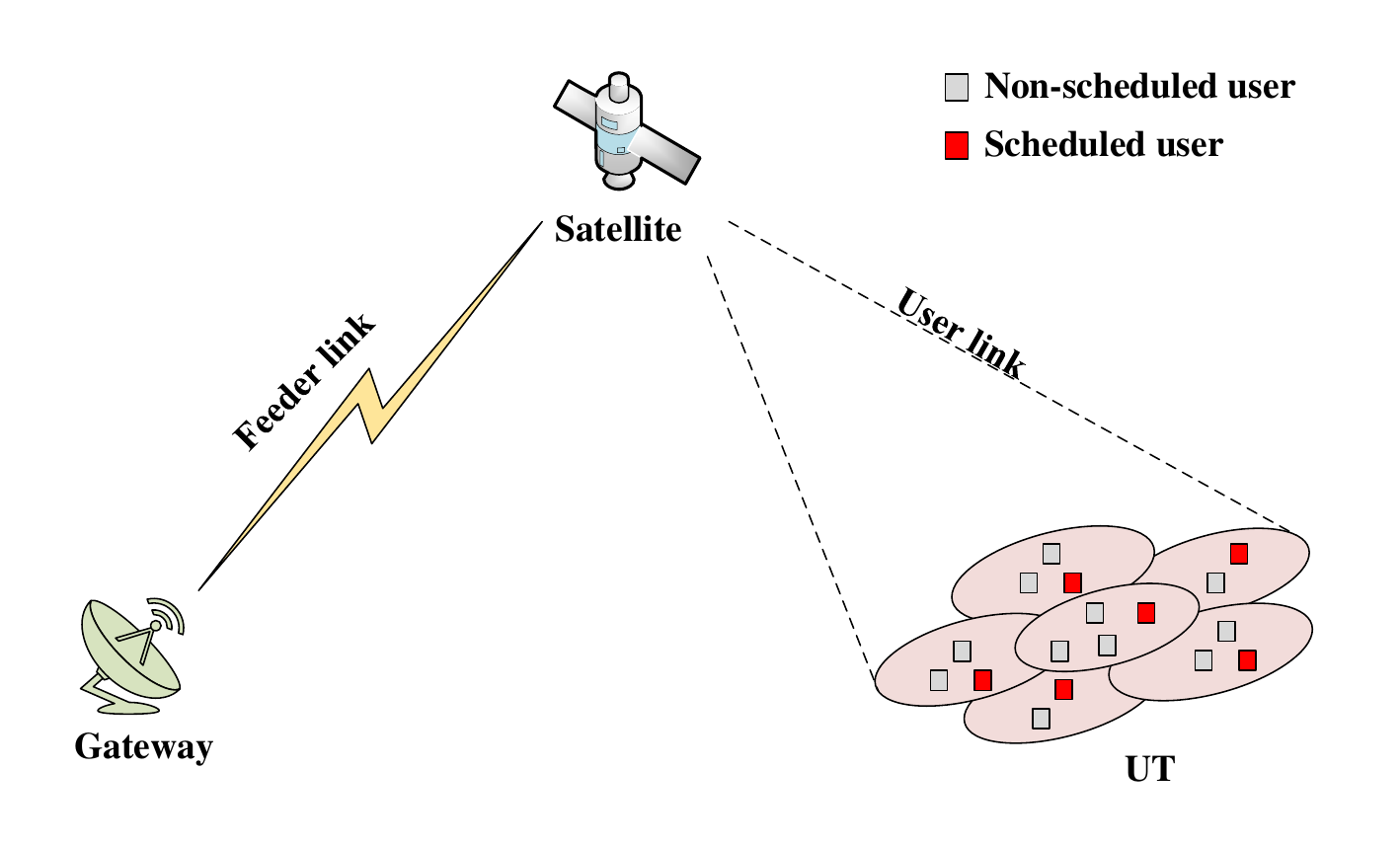}
\end{center}
\caption{Unicast scheduling for a multibeam satellite \cite{10256078}. Each beam serves only one user per time slot, offering high precision and strong interference suppression.}%
\label{unicast}
\end{figure}

\emph{1) Unicast Precoding:}
Unicast precoding in multibeam HTS systems typically operates under a full frequency reuse assumption, with each time slot serving one UT per beam. This scheduling strategy ensures that the number of UTs served in each time slot matches the number of beams, effectively utilizing time division multiplexing, as illustrated in Fig. \ref{unicast}.
 Early studies by Cottatellucci et al. \cite{cottatellucci2006interference} and Gallinaro et al. \cite{gallinaro2005perspectives} demonstrated that applying precoding in this context can enhance system throughput by up to 55$\%$ compared to traditional three-color reuse schemes. These pioneering works, particularly the MMSE-based precoding approach, laid the groundwork for exploring trade-offs between total throughput and user fairness.

\begin{table}[t]
\centering
\captionsetup{justification=centering, labelfont=bf}
\caption{Comparison of Multibeam Precoding Algorithms \cite{zorba2008improved}.}
\scalebox{0.85}{
\begin{tabular}{ccccc}
\hline\hline
\textbf{Algorithm} & \textbf{\begin{tabular}[c]{@{}c@{}}Beam \\ number\end{tabular}} & \textbf{\begin{tabular}[c]{@{}c@{}}Rate\\ (bps/Hz)\end{tabular}} & \textbf{Availability} & \textbf{Variance} \\ \hline
                   & Beam 1                 & 0.86                 & 43\%                  & 0.11             \\
MOB                & Beam 4                 & 0.86                 & 42.5\%                & 0.11             \\
                   & Aggregate              & 6.04                 & 42.7\%                & 0.11              \\ \hline
                   & Beam 1                 & 3.16                 & 84.9\%                & 4.24            \\
MMSE               & Beam 4                 & 1.89                 & 74.8\%                & 1.63             \\
                   & Aggregate              & 20.9                 & 83.7\%                & 3.89            \\ \hline
                   & Beam 1                 & 8.09                 & 100\%                 & 3.74               \\
IMOB               & Beam 7                 & 2.19                 & 87.6\%                & 0.74               \\
                   & Aggregate              & 24.4                 & 95.5\%                & 1.12               \\ \hline\hline
\end{tabular}}
\label{Mulpre}
\end{table}

Subsequent research focused on improving performance under various assumptions of user feedback and system constraints.
Table \ref{Mulpre} collects all the performance for all the
transmission techniques in a scenario with EIRP=72dB and
for the city of Rome. When the MOB precoder is employed,it is observed that even the precoding matrix is generated such
that its columns are orthogonal to each other, the orthogonality
is lost when the signals are transmitted through the channel \cite{zorba2008improved}.
The availability of a large number of users in the system does
not compensate for the random precoding vectors generation
drawback. 

The table shows the outstanding behaviour of the IMOB
proposal in comparison with MMSE beamforming and the
reference scenario \cite{zorba2008improved}. A remarkable conclusion is that the IMOB technique outperforms the MMSE technique in terms of
sum rate and availability while its main target was to improve
the rate variance, which is also enhanced. Therefore, IMOB
outperforms the MMSE beamformer in all the system metrics
under study, emerging as an attractive candidate to exploit the
MIMO benefits in satellite systems. However, its performance degrades with fewer UTs and it incurs higher feedback overhead. More recently, Ahmad et al. \cite{ahmad2021zero} introduced a ZF precoding algorithm with partial CSI, offering reduced delay and overhead, but its effectiveness is limited to high-SINR scenarios.


While the early studies primarily focused on validating the feasibility of precoding under idealized assumptions \cite{ahmad2021zero} . More recent research emphasizes designing precoding schemes under practical system constraints, such as linear and nonlinear power constraints \cite{chatzinotas2011joint}. Optimization objectives in such systems typically fall into two major groups based on design priorities:
i) rate-centric optimization, which aims to satisfy user rate demands; and
ii) energy efficiency optimization, which focuses on maximizing throughput per unit of power consumption.
The following discussion is organized around these two optimization objectives, with attention to how different types of power constraints affect the design in each case.

\emph{ i) Rate-Centric Optimization}: The assumption is made that the traffic demand of the k-th UT is denoted as $F_k$, and its available Shannon rate is denoted as $r_k$. Furthermore, it is essential to ensure that the objective function is continuous. The function primarily includes the following points:

\begin{itemize}
  \item Maximum Sum Rate : $\max\sum\nolimits_{k = 1}^K {r_k}$, which is widely used in terrestrial communication systems for maximizing the total system throughput but users with poor channel quality cannot be served resulting in a lack of inter-user fairness. Ahmad et al. \cite{ahmad2021zero} employed iterative vector normalization to optimize a ZF precoding matrix targeting the maximum sum-rate under linear power constraints.
  \item Rate Balancing: $\max \ \min_k \frac{r_k}{F_k}$, of which the objective is to maximize the rate of the worst user and ensures fairness among users but does not maximize the total system rate. Chatzinotas et al. \cite{chatzinotas2011joint} proposed a precoding method achieving rate balancing under linear power constraints, offering fairness and superior spectral efficiency compared to traditional frequency reuse methods.
  \item Rate Matching: $\min\sum\nolimits_{k = 1}^K {\left|F_k - r_k\right|} ^n$, where $n$ is the predefined order that can be adjusted to match different system requirements. In addition, the objective of the rate matching is equivalent to that of the maximum total rate on the condition of $n=1$. Zheng et al. \cite{zheng2012generic} developed an alternating optimization algorithm to address non-convexity under both linear and nonlinear power constraints, ensuring convergence and applicability across various polarization configurations. 
\end{itemize}



  \emph{ii) Energy Efficiency Optimization}:  Given the limited power supply of satellites, which typically depend on solar panels or on-board batteries, energy efficiency has emerged as a critical design objective. Energy efficiency is commonly defined as the ratio of total user throughput (or the worst-case user rate) to total power consumption. In contrast to rate-centric approaches, energy-efficient precoding in multibeam HTS systems aims to minimize overall satellite power consumption while ensuring that all served users meet their specified QoS requirements.
In \cite{chatzinotas2011energy}, the authors proposed an energy-efficient MMSE precoding design using the steepest descent method, achieving local optimality without severely degrading performance in low-SNR regimes. Later, Qi et al. \cite{qi2018precoding} employed a sequential convex approximation (SCA) method to design ZF and SCA-based precoders under total power constraints. Their results show that the SCA-based design outperforms ZF due to the latter's noise amplification effects. Additionally, their work highlights the value of using real satellite channel measurements to improve both energy efficiency and system robustness.

\begin{figure}[t]
\begin{center}
\includegraphics[width=8.0cm]{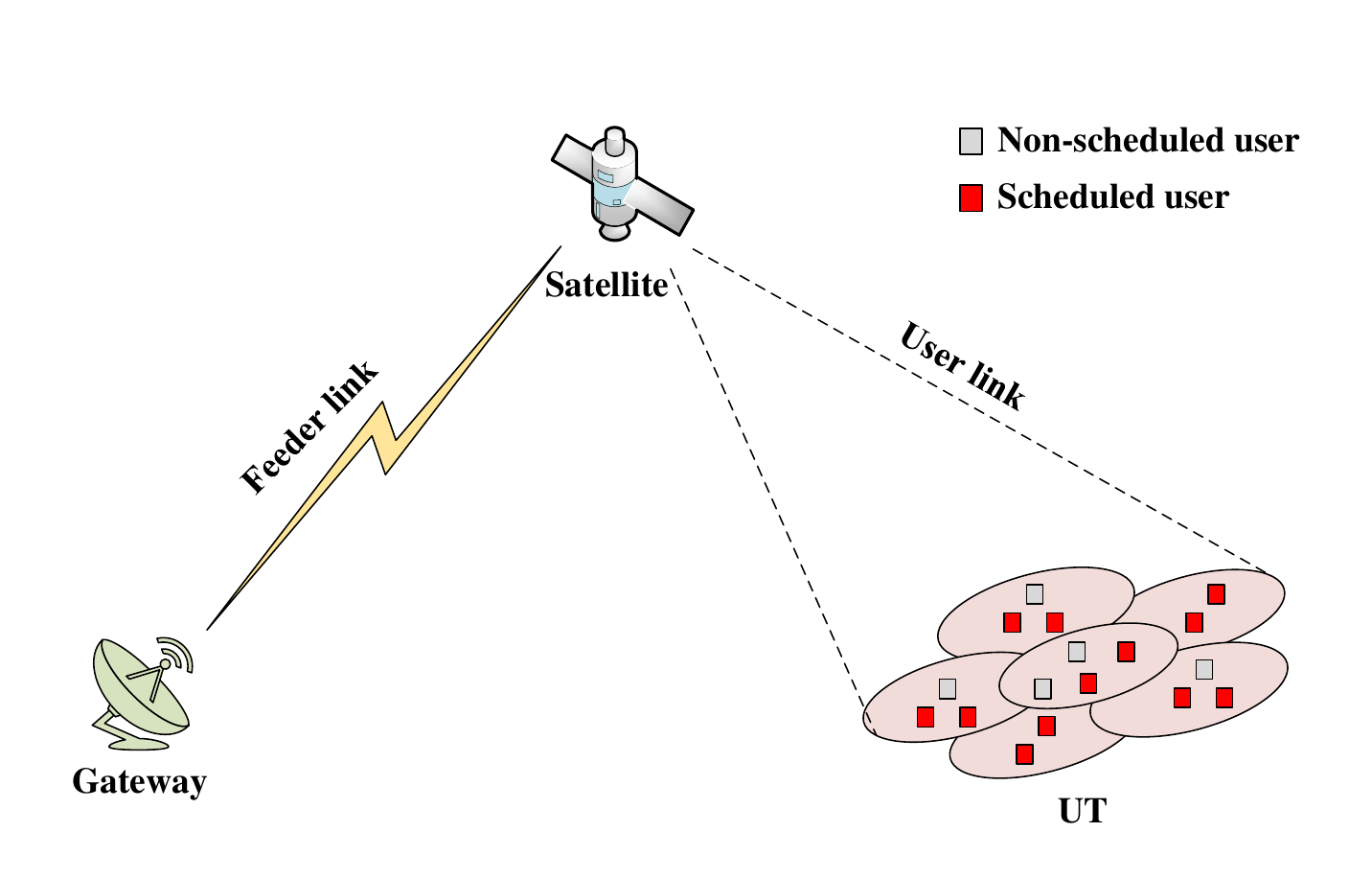}
\end{center}
\caption{Multicast scheduling for a multibeam satellite \cite{10256078}. Each beam serves multiple users per time slot (sharing a common frame), which better aligns with standards such as DVB-S2X and offers higher spectral efficiency.}%
\label{multicast}
\end{figure}

\emph{2) Multicast Precoding:}
Multicast precoding is a key technique in multibeam HTS systems, allowing a common signal to be transmitted simultaneously to multiple users grouped within the same beam, as shown in Fig. \ref{multicast}.
Compared to unicast, multicast offers significant spectral efficiency gains and aligns better with practical framing mechanisms in DVB-S2X standards. However, the design of efficient multicast precoding requires addressing several key algorithmic strategies, including:  
i) frame-aware design of multicast precoding algorithms,  
ii) user scheduling and channel similarity grouping,  
iii) multicast precoding optimization under system constraints, and  
iv) feeder link limitations and system-level precoding strategies. 
The following discussion is structured around these four core challenges to clarify their technical motivations, representative methods, and performance implications.

  \emph{i) Frame-Aware Design of Multicast Precoding Algorithms}: 
  Serving only one user per beam is not practical in the DVB-S2 satellite standard \cite{DVBS2}, which maked coding multiple users in a frame more efficient than unicast. Therefore, multicast is more practical and efficient way where each beam is able to serve multiple users composing a data frame. However, designing precoding based on the DVB-S2 standard \cite{arapoglou2016dvb} presented a new challenge, which involves the variable length frame structure that causes the estimated pilots to no longer align. 
  
  As a result, the precoding matrix needs to be recalculated in the case of a change in the forward error correction code (FEC) block, thereby increasing complexity \cite{vazquez2016precoding}. To address this issue, the DVB-S2 extension standard \cite{DVBS2X}, which extended DVB-S2, proposed a superframe structure with a constant frame length and reserves precoding byte segments \cite{arapoglou2016dvb}. Furthermore, the multicast precoding rate is equal to the framing rate, which helps reduce the complexity of precoding.

\emph{ii) User Scheduling and Channel Similarity Grouping}: 
 In multicast transmission, multiple users are interleaved in the same physical layer frame, sharing the same radio resources and decoding identical frames. Consequently, all users must adopt the same modulation and coding scheme (MODCOD), which is determined by the worst channel quality user. Since precoding performance is highly sensitive to CSI , grouping users with similar channels into the same frame is crucial to maximizing precoding gain.   User grouping strategies can be broadly categorized based on geographic location or CSI-based metrics.  These strategies fall into four general categories:  geographical user clustering algorithm, multicast-aware user scheduling algorithm, heuristic scheduling algorithm  and geographical scheduling algorithm. The comparison of user scheduling algorithms is shown in Table \ref{usersch}.
\begin{itemize}
\item \emph{Geographical User Clustering Algorithm}: This strategy involves scheduling users into frames based on their geographical proximity and channel similarity, considering the channel matrix without phase effects. Users that are geographically close and have similar channels are grouped into the same frame using the k-means algorithm \cite{taricco2014linear}. Though this approach is simple to implement, it may result in the poor similarity of user channels within the same frame due to neglecting the phase effect, which can enlarge the differences between user channel vectors.

\item  \emph{Multicast-Aware User Scheduling Algorithm}: This strategy involves scheduling users into frames based on their measurement of channel similarity with known CSI at the transmitter. Firstly, each group is assigned a user so that no interference users are assigned between different groups, using the semi-orthogonal criterion \cite{yoo2006optimality}. Secondly, for the order of each group, the user that is most parallel to the previously selected user is scheduled in the same frame to maximize channel similarity within the same frame \cite{christopoulos2015multicast}.

\item \emph{Heuristic Scheduling Algorithm}: This algorithm constructs similar channel groups using the cosine similarity measure. The optimal spectral cluster is obtained by using a heuristic algorithm \cite{pothen1990partitioning} to complete the user scheduling, aiming to make the user channels in the adjacent synchronized frame as orthogonal as possible \cite{lagunas2018cross}. Additionally, a weighted round-robin approach is used by the system scheduler to fulfill different QoS requirements within the same frame.

\item  \emph{Geographical Scheduling Algorithm}: This strategy involves scheduling users into frames from the perspective of the system scheduler. The method divides the same number of sectors in each beam and schedules users in the same sectors of each beam to the same time frame, avoiding degradation of precoding performance as users in adjacent beams and closely located are scheduled in the same time frame \cite{guidotti2018geographical}. 
\end{itemize}

\begin{table*}[t]
\centering
\caption{Comparison of User Scheduling Algorithms.}
\resizebox{0.9\textwidth}{!}{
\begin{tabular}{c|l|l|l|l}
\hline\hline

    \textbf{Algorithm}         & \makecell[c]{\textbf{Core Mechanism}}             & \makecell[c]{\textbf{Advantages}}                                                                   &  \makecell[c]{\textbf{Disadvantages} }                  & \makecell[c]{\textbf{Reference} }    
        \\ \hline
\makecell[c]{Geographical \\user clustering}     &\makecell[l]{Groups users by physical proximity \\using location data.} &        Simple to implement                                                                  & Neglect phase effects                    & \cite{taricco2014linear}        \\ \hline
\makecell[c]{Multicast-aware \\user scheduling}    & \makecell[l]{Maximizes channel similarity within \\transmission frames.}&      \begin{tabular}[c]{@{}l@{}}Maximize intra-frame\\ channel similarity\end{tabular}    & Require accurate CSI                     & \cite{yoo2006optimality}         \\ \hline
\makecell[c]{Heuristic \\scheduling}     & \makecell[l]{Maximizes channel orthogonality across\\ frames using rule-based optimization.}               & \begin{tabular}[c]{@{}l@{}}Maximize inter-frame\\ channel orthogonality\end{tabular} & Equalize weight parameters               &\cite{pothen1990partitioning}     \\ \hline
\makecell[c]{Geographical \\scheduling }  & \makecell[l]{Assigns transmission slots based on \\spatial distribution.}              &\begin{tabular}[c]{@{}l@{}}Streamline user\\ scheduling operation\end{tabular}        & \begin{tabular}[c]{@{}l@{}}Imbalanced user allocation\\ across frames\end{tabular} & \cite{guidotti2018geographical}  \\ \hline\hline
\end{tabular}
}
\label{usersch}
\end{table*}

\emph{iii) Multicast Precoding Optimization Under System Constraints:} With user groups established, designing a suitable precoding matrix involves solving optimization problems under various system-level constraints, such as per-antenna power limits, rate fairness, and robustness to CSI errors. Existing work can be categorized into three main algorithmic approaches:
\begin{itemize}
    \item \emph{Linear Precoding Under Simplified Constraints:} Taricco et al. \cite{taricco2014linear} introduced the use of geographic user clustering and the MMSE algorithm to calculate the precoding matrix for the total system rate problem under the per antenna power constraint, which was the first work addressing the importance of user scheduling and the feasibility of precoding in the multigroup multicast scenario. However, the authors only briefly analyzed the impact factors on CSI and simplified the power constraint and optimization problem.
    
    \item \emph{Nonlinear Optimization with Fairness and Robustness:}  To overcome these limitations, Christopoulos et al. \cite{christopoulos2014frame} proposed weighted fair multigroup multicast precoding that combines the principle of semidefinite relaxation and Gaussian randomization with bisection to solve the optimal rate balance problem under the actual power constraint of each antenna. Although this method has higher complexity compared to MMSE, it outperforms MMSE in terms of system rate and takes into account the actual power constraint. In \cite{christopoulos2014multicast}, Gaussian randomization-based precoding was proposed to solve the rate matching problem with per-antenna power constraint, which has been shown to provide a higher accuracy solution and effectively improve the minimum SINR with relatively low complexity. Christopoulos et al. \cite{christopoulos2014weighted} further improved upon \cite{christopoulos2014multicast} by adding a weighting factor to discriminate different user QoS weights, making the algorithm adaptable to heterogeneous services. The algorithm is also extended to be robust to non-perfect CSI, effectively mitigating the impact of non-perfect CSI on precoding gain.
    
    \item \emph{Low-Complexity and One-Time Precoding Approaches:}
    In the case of per-antenna power constraint, the optimal multicast multibeam satellite system sum rate problem, as mentioned in \cite{christopoulos2015multicast,joroughi2016generalized}, Jouroghi et al. \cite{joroughi2016generalized} proposed a one-time precoding design method that suppresses interference and maximizes intra-frame rate, while keeping the complexity low compared to iterative algorithms. Furthermore, the algorithm achieves higher spectral efficiency compared to frame-based R-ZF and average MMSE, and maintains acceptable spectral efficiency even with 5 users per frame. In \cite{christopoulos2015multicast}, the authors proposed a subgradient precoding optimization algorithm that allows for redistribution of residual power among groups to maximize the total system throughput. Moreover, the CSI-based multicast-aware user scheduling proposed in \cite{christopoulos2015multicast} not only improved intra-frame user channel similarity but also reduced algorithm complexity, making it a well-performing algorithm.
\end{itemize}

 \emph{iv) Feeder Link Limitations and System-Level Precoding Strategies}: 
Beyond algorithmic design, practical multicast precoding in HTS systems is constrained by feeder link capacity—especially in Ka-band systems (limited to ~2 GHz) \cite{de2017network}. To address these limitations, two main strategies have been explored in recent research:

\begin{itemize}
    \item \emph{Band Migration to Q/V Band:} Migrating feeder links to the Q/V band (40/50 GHz) offers up to 5 GHz bandwidth, supporting aggressive frequency reuse and over 200 beams \cite{kyrgiazos2012gateway,arapoglou2013gateway}. However, Q/V bands suffer severe rain fading, with propagation losses up to 15–20 dB, risking link outages \cite{vidal2012next,de2018adaptive}.
    \item  \emph{Alternative System-Level Precoding Architectures:} To mitigate these challenges, the following precoding techniques have been explored: multiple gateway precoding, on-board precoding and hybrid on-board/on-ground precoding. The following subsections will provide a detailed introduction to these precoding approaches.

\end{itemize}

\emph{3) Summary: }
In the single gateway precoding scenario, we explored various precoding techniques to enhance the throughput of multibeam HTS systems. The MMSE precoding algorithm was identified as a baseline, which maximizes the system throughput. However, it was noted that MMSE may not always ensure user fairness. The IMOB algorithm was highlighted for its sequential user selection process, which demonstrated superior performance over MMSE in certain aspects, despite its higher overhead and potential performance degradation with a small number of users. ZF precoding, with its ability to eliminate interference, was also discussed, particularly in high SINR scenarios. Overall, the choice of the optimal precoding algorithm in a single gateway scenario depended on the specific requirements for system throughput, user fairness, and computational complexity. IMOB stood out when high performance was required, despite its increased complexity.
\subsection{Multiple Gateways Precoding} 
\begin{figure}[t]
\setlength{\belowcaptionskip}{-0.25cm}
\begin{center}
\includegraphics[width=8.0cm]{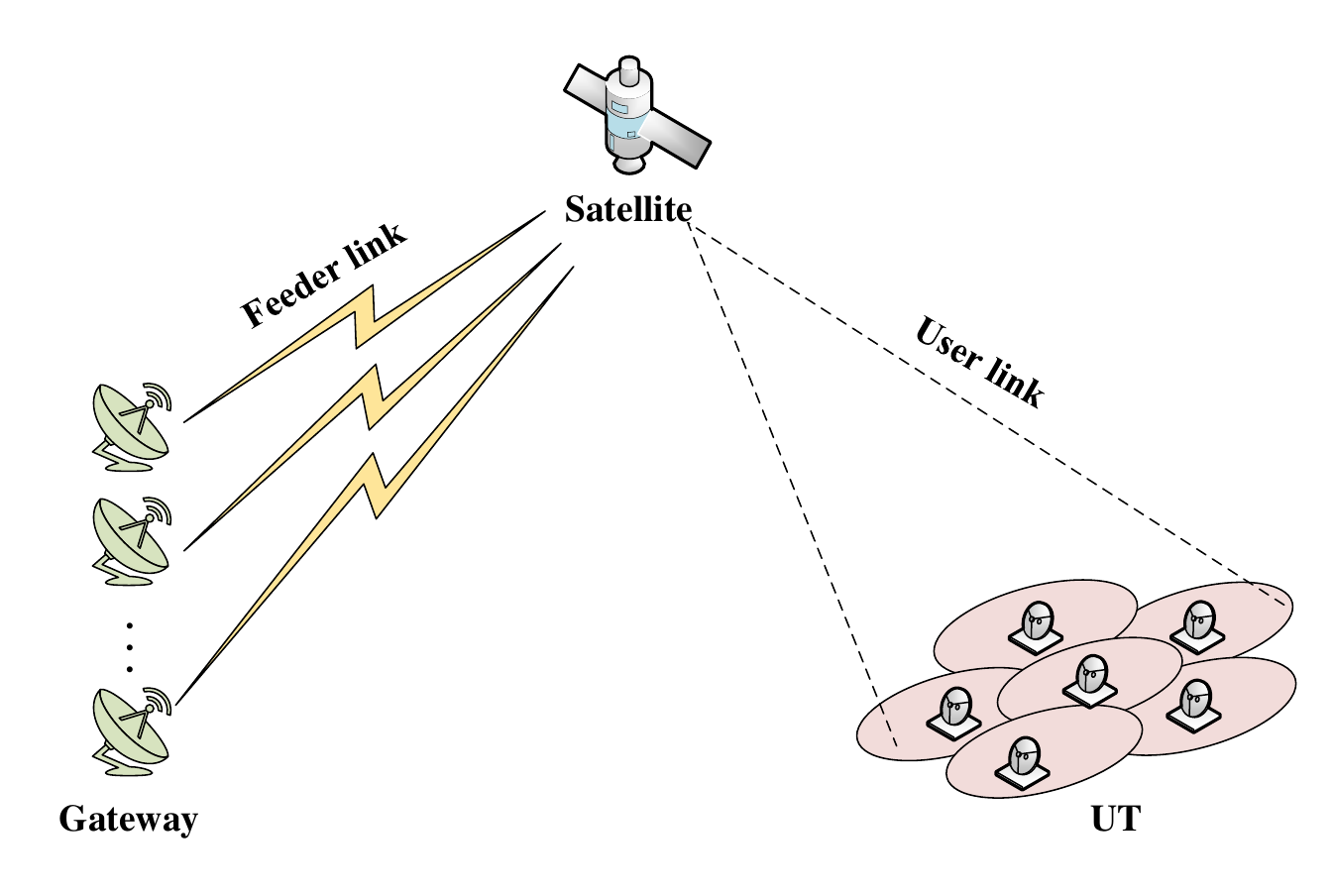}
\end{center}
\caption{Multiple gateways system model. 
Each ground gateway  independently computes the precoding matrix for the beams it serves, which are typically divided by region or coverage area. This reduces the load and feeder link bandwidth required for each gateway. However, due to the lack of CSI sharing between gateways, the effectiveness of interference mitigation is limited.}%
\label{multiple-cluster}
\end{figure}

A feasible and practical approach was to deploy multiple gateways in the Ka band due to lower multibeam signal processing complexity, and shorter network loops\cite{zheng2012multi}. Specifically, a smaller number of beams handled by each gateway reduced signal processing complexity and minimized feeder link bandwidth usage\cite{joroughi2016impact}. Furthermore, traffic could be rerouted through other gateways in case of gateway failure, avoiding service interruption\cite{zheng2012multi}. Satellite service providers, such as internet providers and data centers, strategically laid out gateways closer to them based on valuable geographic distribution to reduce backhaul cost\cite{joroughi2016impact}. The basic structure of the multiple gateways model is shown in Fig. \ref{multiple-cluster}.
However, the multiple gateway architecture introduces new interference challenges that can significantly degrade the performance of precoding strategies. Specifically, two main issues arise: 1) inter-cluster interference caused by overlapping coverage of adjacent gateways, and 2) feeder link interference due to misaligned or uncoordinated feeder transmissions. This discussion is organized based on the proposed classification, providing a comprehensive review of state-of-the-art interference mitigation strategies within each category.

\emph{1) Inter-Cluster Interference Mitigation:}
Inter-cluster interference arises when multiple gateways serve adjacent beam clusters without effective coordination. Inspired by cooperative processing in terrestrial cellular networks \cite{ng2008distributed,bjornson2010cooperative,huang2010distributed}, inter-gateway cooperation has been proposed as a key technique to mitigate such interference. Zheng et al. \cite{zheng2012multi} introduced the concept of hyper clusters, where neighboring clusters share partial CSI or user data to enable cooperative precoding. Using an R-ZF algorithm, gateways independently optimize their intra-cluster precoding under power constraints, improving signal-to-leakage-and-noise ratio (SLNR) and overall throughput.

Later works such as \cite{christopoulos2016multicast} advanced this idea by introducing centralized gateway coordination based on global CSI, allowing for more effective SINR balancing and enhanced fairness. While this improves performance, full cooperation among gateways increases hardware cost and demands high-speed fiber connectivity to exchange data in real-time \cite{yang2014free, ansari2015ergodic}. To address the practical limits of full cooperation, Devillers et al. \cite{devillers2011advanced} proposed a ZF-based solution using block decomposition of the channel matrix, which avoids the need for data exchange and still mitigates inter-cluster and intra-cluster interference.

In parallel, user grouping algorithms have been proposed to enhance multicast precoding by selecting users with similar channel characteristics. For instance, \cite{8761875} proposed a sum-rate maximization algorithm, while \cite{10433460} presented a channel-correlation-based grouping scheme that separates users with similar channels to reduce nonlinear interference and improve overall system performance.
 
\emph{2) Feeder Link Interference Mitigation:}
Even though feeder link antennas are highly directional, misalignment or imperfect calibration can still lead to inter-feeder link interference \cite{joroughi2016precoding, joroughi2016impact}. Although this interference is often assumed negligible \cite{1146175}, practical systems must still account for it, especially under rain fade or high-load conditions \cite{6843130}. Therefore, precoding design in multiple gateway scenarios must jointly address both feeder and user link interference to ensure robust system performance \cite{wang2019multicast}.

Several algorithms have been developed to mitigate feeder link interference. For example, the M-ZF and M-MMSE algorithms in \cite{joroughi2016impact} extend traditional ZF/MMSE techniques to account for inter-feeder and inter-cluster interference. Notably, M-MMSE achieves performance close to the ideal case with no feeder link interference, outperforming ZF-based methods in practical conditions.

Building on these methods, Wang et al. \cite{wang2019multicast} developed centralized and decentralized precoding algorithms for multicast multigroup transmissions under feeder link interference. Both approaches aim to maximize sum rate while satisfying user SINR and power constraints. The decentralized algorithm, in particular, reduces coordination overhead by relying only on local CSI. Furthermore, Wang et al. \cite{wang2019sum} extended the joint power control and beamforming (JPCB) algorithm, which was originally proposed for terrestrial and single-gateway systems, by introducing a geometric programming-based solution. This approach avoids step-size tuning and reduces computational complexity, offering improved throughput and scalability. However, due to shared precoding across users in a frame, system performance still suffers when users with highly dissimilar channels are grouped together.

\emph{3) Summary:} The multiple gateways precoding section delved into the challenges of inter-cluster interference and the mitigation of feeder link interference. It was concluded that inter-gateway cooperation, either through sharing CSI or limited collaboration schemes, can significantly reduce inter-cluster interference. The M-ZF and M-MMSE algorithms were identified as effective in diminishing interference across feeder links. Despite the benefits of multiple gateways in reducing signal processing complexity and feeder link bandwidth usage, the increased system costs and the need for effective cooperation strategies present challenges. The optimal precoding strategy in this scenario involves a balance between interference mitigation and the practicality of inter-gateway cooperation, with M-MMSE showing promise due to its effectiveness in interference reduction.

\subsection{On-Board Precoding}
\begin{figure}[t]
\begin{center}
\includegraphics[width=8.0cm]{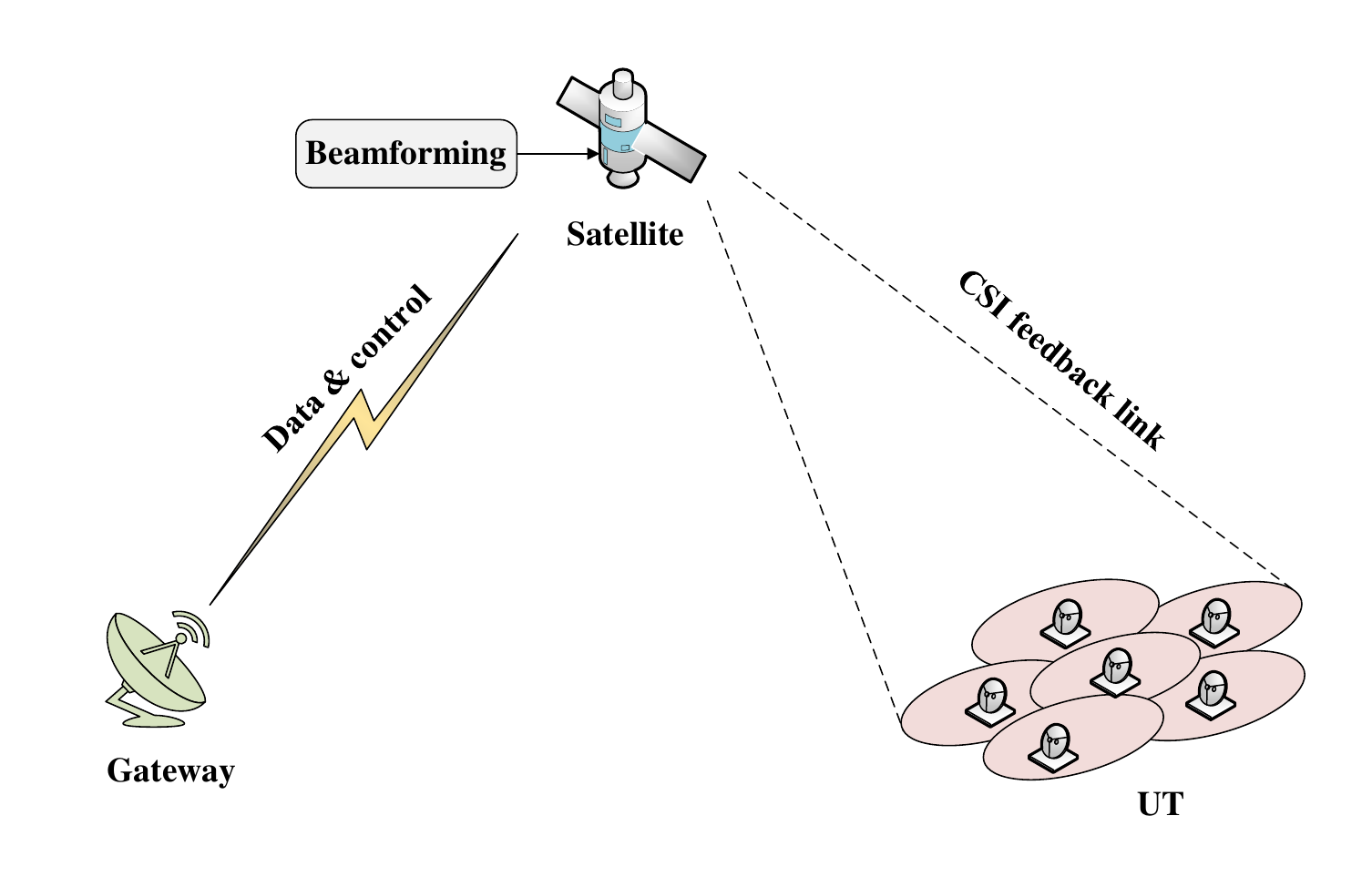}
\end{center}
\caption{On-board precoding for a multibeam satellite. The precoding matrix is computed and applied directly on the satellite rather than on the ground, thereby reducing the computational burden on ground stations. The gateway only transmits user data and simple control instructions. The satellite obtains CSI (via user feedback), calculates the precoding matrix, and performs multibeam downlink transmission. UTs receive the precoded signals.}%
\label{onboard}
\end{figure}

As multibeam satellite systems become increasingly complex, the limitations of ground-based precoding, including CSI feedback delays, gateway coordination overhead, and feeder link congestion, have become more pronounced. In response, OBP has emerged as a promising alternative, as shown in Fig. \ref{onboard}.
 By relocating precoding computations from ground gateways to the satellite payload, OBP can reduce feeder link bandwidth requirements, avoid CSI exchange between gateways, and enhance flexibility in interference mitigation. However, due to limited on-board processing resources, OBP must rely on low-complexity algorithms that maintain acceptable performance under system constraints.

Existing research on OBP can be broadly categorized into three main approaches:  
1) sparse beamforming-based OBP, which simplifies on-board computations using sparse matrices;  
2) joint resource management-based OBP, which integrates precoding with feed selection and switching mechanisms;  
3)  robust fixed-matrix OBP, which uses precomputed matrices designed to tolerate channel variations.  
The following discussions present representative works under each category.

\emph{1) Sparse Beamforming-Based OBP:}
To address the computational burden of on-board matrix calculations, Bandi et al. \cite{bandi2018sparsity} proposed a sparse beamforming approach. In this method, the precoding matrix is designed to be sparse while still satisfying users' SINR constraints and total transmit power limits. This sparsity significantly reduces computational complexity and implementation cost, while maintaining performance similar to traditional power minimization techniques. However, the approach does not account for additional overhead that may arise when CSI changes, as recalculating the sparse matrix can still be complex under dynamic conditions.

\emph{2) Joint Resource Management-Based OBP:}
Joroughi et al. \cite{joroughi2019deploying} introduced an integrated on-board design that combines precoding with feed selection and signal switching. This joint approach aims to make efficient use of limited feeder link resources while maintaining acceptable on-board complexity. The proposed algorithm minimizes the sum mean square error (SMSE) under time-varying and perturbed CSI assumptions. Although this method improves link utilization and processing efficiency, it overlooks interference between feeder links, which may affect system performance in high-density deployments.

\emph{3) Robust Fixed-Matrix OBP:}
To further reduce on-board processing requirements, Joroughi et al. \cite{joroughi2018board} also proposed a fixed on-board precoding matrix based on MMSE design. This matrix is robust to channel variations and capable of suppressing both user link and feeder link interference without requiring frequent updates. While this approach drastically lowers the satellite's computational load, it sacrifices adaptability. Since the matrix cannot be adjusted in real-time, its performance may degrade under rapidly changing channel conditions.

\emph{4) Summary:}
On-board precoding offers an effective means to bypass the limitations of ground-based precoding, particularly in systems with multiple gateways. This section categorized existing OBP approaches into three types based on their design strategies and complexity-performance trade-offs. Sparse beamforming prioritizes simplicity while retaining reasonable performance; joint resource management enhances feeder link efficiency through integrated control; and robust fixed-matrix designs ensure stability under strict on-board constraints. A practical OBP solution should strike a balance between adaptability to CSI variations and on-board computational feasibility, with sparse matrix-based designs standing out for their practical value.


\begin{figure}[t]
\begin{center}
\includegraphics[width=8.5cm]{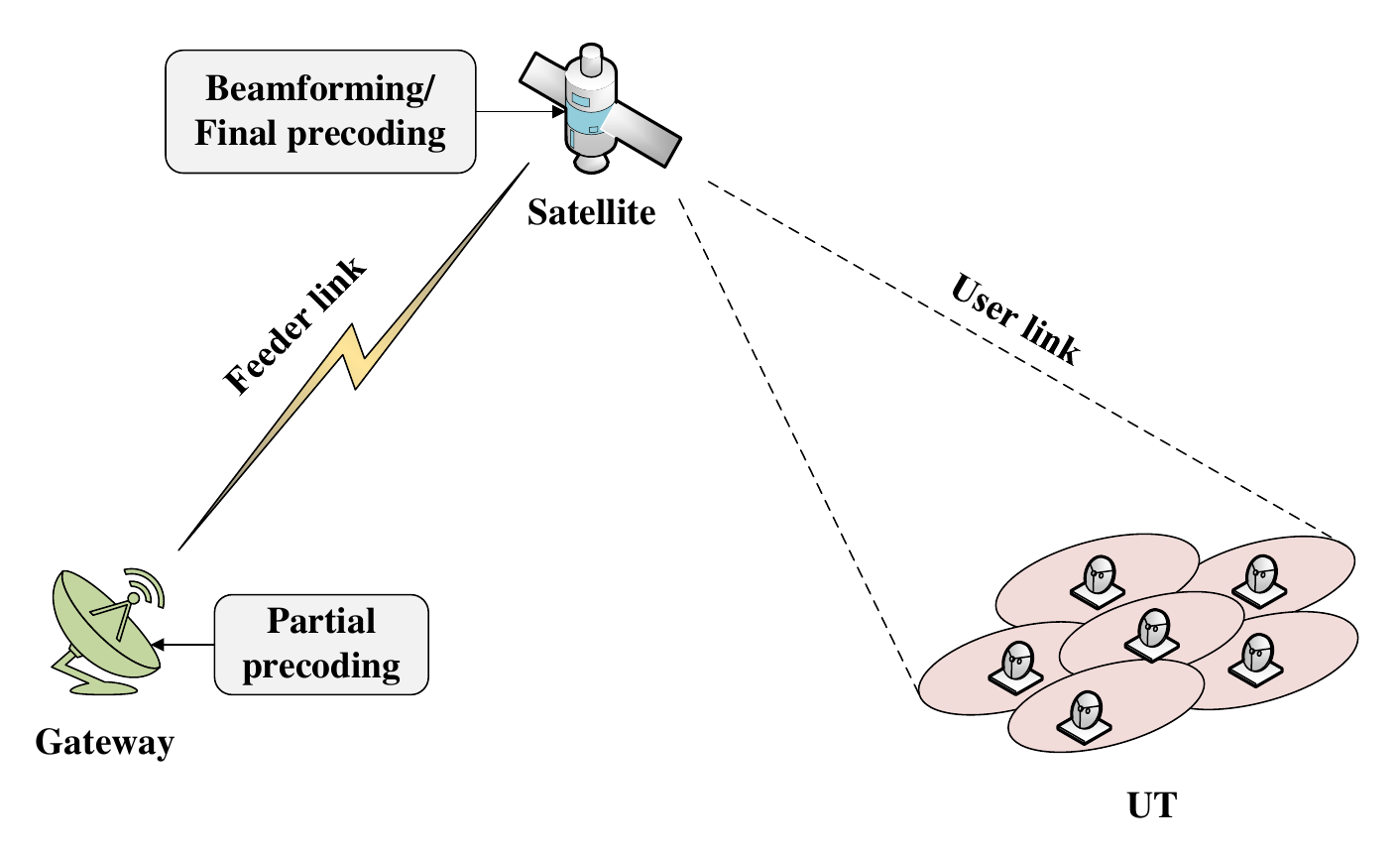}
\end{center}
\caption{Hybrid  on-board/on-ground  precoding system model.  The precoding task is divided between the ground and the on-board: the ground station performs coarse-grained processing (e.g., clustering, weight assignment), while the on-board handles fine-grained beamforming and dynamic adjustment. This approach reduces the on-board computational burden but increases system complexity, requiring careful coordination between the ground and on-board components.}%
\label{Hybird-processing}
\end{figure}

\begin{table*}[t]
\centering
\caption{Comparison of On-Board Beamforming Matrix Design Algorithms.}
\resizebox{0.8\textwidth}{!}{
\begin{tabular}{c|c|c|c|c|c|l}
\hline\hline
\textbf{Algorithm} & \textbf{SVD } &\textbf{PCA}  & \textbf{DPSS}& \textbf{FAPI } & \textbf{ Bartlett} & \makecell[c]{\textbf{Quantification criteria}} \\ \hline

Reference & \cite{joroughi2017onboard}& \cite{thibault2014joint}&\cite{song2017efficient}&
  \cite{yue2017space}&\cite{zheng2005adaptive}\\ \hline
  
Scalability       & \cellcolor{myyellow}{M}     & \cellcolor{myblue}{H} & \cellcolor{myyellow}{M}          & \cellcolor{myblue}{H}           & \cellcolor{myblue}{H}    &   \makecell[l]{L: Limited scalability, suitable only for small systems;\\
       M: Handles medium-scale problems;\\
        H: Efficient for large-scale systems \cite{joroughi2017onboard}. } \\ \hline
Robustness       & \cellcolor{myblue}{H}          & \cellcolor{myred}{L}           & \cellcolor{myblue}{H}           & \cellcolor{myblue}{H}            & \cellcolor{myblue}{H}    & \makecell[l]{L: Performance degrades significantly with imperfections;\\
        M: Some tolerance to imperfections;\\
        H: Maintains performance under various imperfections.}    \\ \hline
Accuracy       & \cellcolor{myblue}{H}       & \cellcolor{myyellow}{M}              & \cellcolor{myblue}{H}               & \cellcolor{myblue}{H}         & \cellcolor{myblue}{H}   & \makecell[l]{ L: Significant deviation from optimal;\\
        M: Acceptable deviation from optimal;\\
        H: Very close to optimal solution.}        \\ \hline
\makecell[c]{Convergence\\ speed }       & \cellcolor{myblue}{H}          & \cellcolor{myblue}{H}              & \cellcolor{myyellow}{M}           & \cellcolor{myblue}{H}            & \cellcolor{myblue}{H} & 
\makecell[l]{L: Slow convergence (many iterations);\\
        M: Moderate convergence speed;\\
        H: Fast convergence (few iterations).}
\\ \hline

\makecell[c]{Processing \\capacity }  & \cellcolor{myyellow}{M}        & \cellcolor{myblue}{H}               & \cellcolor{myyellow}{M}              & \cellcolor{myblue}{H}         & \cellcolor{myblue}{H}     &
\makecell[l]{L: Limited processing capability;\\
        M: Handles moderate processing loads;\\
        H: Capable of heavy processing loads.}
\\ \hline

\makecell[c]{ Computational \\complexity} & \cellcolor{myblue}{H}  &\cellcolor{myred}{L}   &\cellcolor{myblue}{H}  &\cellcolor{myred}{L}  &\cellcolor{myred}{L} &
\makecell[l]{L: Minimal computational resources;\\
         M: Moderate resource requirements;\\
        H: Significant computational resources \cite{thibault2014joint}.}
\\

\hline\hline
\end{tabular}
}
\label{Beamforming Matrix Design}
\end{table*}

\subsection{Hybrid On-Board/On-Ground Precoding}
Hybrid on-board/on-ground precoding has emerged as a practical and effective solution to overcome the limitations of fully terrestrial or fully on-board precoding in multibeam satellite systems. While on-ground precoding with multiple gateways offers high spectral efficiency, it requires strict calibration and extensive CSI exchange, which may not be feasible in large-scale deployments \cite{angeletti2010space}. Conversely, on-board precoding reduces feeder link usage but is constrained by limited satellite payload capabilities. To balance these trade-offs, hybrid on-board/on-ground precoding combines adaptive on-ground processing with simplified on-board beamforming, as shown in Fig. \ref{Hybird-processing}, achieving a balance between performance, flexibility, and complexity \cite{angeletti2009space, 2010Hybrid}.
To clarify the current research landscape, we categorize hybrid on-board/on-ground precoding studies based on their application scenarios into two main groups: 1) single gateway scenarios and 2) multiple gateway scenarios.

\emph{1) Single Gateway Scenarios:}
In single gateway satellite systems, hybrid on-board/on-ground precoding focuses on efficiently balancing on-board processing constraints with the complexity and flexibility of terrestrial precoding \cite{arnau2011hybrid}. The key challenge lies in designing on-board beamforming and ground precoding schemes that reduce feeder link bandwidth requirements while maintaining robust interference mitigation and system performance \cite{devillers2011joint}. To address this, existing research primarily falls into two categories: i) fixed on-board beamforming combined with adaptive ground precoding, which simplifies on-board hardware at the expense of some performance loss; ii) advanced on-board beamforming matrix design techniques that seek to optimize the beamforming process itself to approach the performance of fully terrestrial solutions \cite{joroughi2013design, pham2021control}. The following subsections explore these two aspects in detail, highlighting their respective design principles, algorithmic strategies, and performance trade-offs.

\emph{i) Fixed Beamforming and Precoding Schemes:}  
This category focuses on hybrid architectures where the on-board beamforming matrix is fixed and non-adaptive, primarily performing signal compression to reduce feeder link load. The adaptive terrestrial precoding then handles user interference mitigation, balancing complexity and performance.
In one of the earliest implementations, fixed on-board beamforming combined with adaptive terrestrial precoding was proposed in \cite{arnau2011hybrid, devillers2011joint}. In this approach, the on-board matrix performs non-adaptive signal compression to reduce feeder link load, while the ground precoder mitigates inter-user interference. \cite{arnau2011hybrid} adopted regularized channel inversion, whereas \cite{devillers2011joint} employed UpConst MMSE precoding based on uplink-downlink duality. The latter proved more robust to imperfect CSI and delivered improved throughput \cite{arnau2012performance}. Though hybrid designs may incur minor throughput loss compared to full terrestrial precoding, they effectively reduce feeder link bandwidth requirements, offering a compelling trade-off.

\emph{ii) Advanced On-Board Beamforming Matrix Design Techniques:}  
This category addresses the design and optimization of adaptive on-board beamforming matrices to improve spectral efficiency and robustness. These techniques aim to enhance overall system performance by jointly optimizing on-board and terrestrial processing. In \cite{joroughi2013design}, a fixed on-board matrix reduced SMSE compared to ESA's default method under channel variations. Later, Joroughi et al. \cite{joroughi2017onboard} introduced an optimal beamforming design that accounted for eigenvalue perturbations and, when combined with MMSE terrestrial precoding, approached the performance of fully terrestrial systems.  
Thibault et al. \cite{thibault2014joint} compared SVD, DFT, and PCA-based beamforming methods, demonstrating the superiority of SVD in terms of spectral efficiency and robustness.  
Subsequent work proposed several on-board matrix generation techniques—e.g., DPSS beamspace, Bartlett beamforming, and the greedy sparse recovery (GSR) algorithm \cite{song2017efficient}. Among them, GSR showed higher throughput and reasonable complexity. The FAPI method in \cite{yue2017space} further reduced computational cost while improving signal orthogonality and convergence. These studies emphasize the importance of low-complexity, high-performance matrix design for efficient hybrid systems.  
A comparative analysis of these methods is provided in Table \ref{Beamforming Matrix Design}.

\emph{2) Multiple Gateway Scenarios:}  
Most early works assumed a single gateway, but scaling hybrid  on-board/on-ground  precoding to multiple gateways raises new challenges. In \cite{8377156}, the authors addressed this by combining on-board SVD beamforming with ZF terrestrial precoding. While effective in mitigating intra- and inter-cluster interference, this method required full CSI sharing between gateways.  
\cite{mosquera2018distributed, ramirez2018two} proposed a joint design using on-board SVD beamforming and MMSE gateway precoding to eliminate interference without requiring inter-cluster CSI exchange.  
A control-theoretic method was proposed in \cite{pham2021control}, where a proportional-differential-integral  controller at the gateway minimized interference and on-board beamforming reduced cross-cluster leakage.  
Further, Joroughi et al. \cite{joroughi2017joint} explored robustness to channel disturbances using fixed on-board matrices, while Ram\'{\i}rez
 et al. \cite{ramirez2020two} extended joint designs to frame-level processing with separate adaptation rates for ground and on-board precoders.  
\cite{wang2019multicast} introduced a two-level frame-based scheme where on-board leakage-aware MMSE beamforming mitigated feeder link interference, and ground-side SCA-based precoding addressed user interference. This approach allowed flexible updates to the on-board matrix, improving efficiency without significant on-board complexity.

\emph{3) Summary:}  
Hybrid  on-board/on-ground  precoding leverages the strengths of both on-ground and on-board processing to meet the needs of high-throughput satellite systems. Fixed beamforming schemes reduce bandwidth but may sacrifice robustness. on-board beamforming matrix design plays a pivotal role, with methods like SVD, PCA, and FAPI offering different trade-offs in scalability, complexity, and spectral efficiency. Finally, expanding hybrid architectures to support multiple gateways introduces coordination challenges, but also unlocks system-wide performance gains. An effective hybrid strategy requires balancing complexity, CSI availability, and interference control across both the space and ground segments.

\subsection{Summary \& Lessons Learnt}
In this comprehensive review, we have explored the intricacies of precoding techniques within different deployment scenarios for multibeam HTS systems. Precoding is a critical technology for enhancing system capacity and mitigating interference in HTS systems. The analysis of various precoding algorithms across single gateway, multiple gateways, on-board, and hybrid  on-board/on-ground  precoding models provides insights into their performance and suitability for different operational environments.

The choice of the best precoding algorithm is scenario-dependent and influenced by factors such as system throughput requirements, user fairness, computational complexity, and adaptability to dynamic channel conditions. IMOB is preferred in single gateway scenarios for high throughput, while M-MMSE is suitable for multiple gateways due to its interference mitigation capabilities. For on-board precoding, sparse beamforming is practical for its simplicity, and in hybrid scenarios, SVD and PCA offer robust performance across varying channel conditions. This review provides a clear understanding of the performance of different precoding algorithms, enabling readers to make informed decisions based on their specific satellite communication requirements.

\begin{figure*}[t]
\begin{center}
\includegraphics[width=1\textwidth,height=0.48\textwidth]{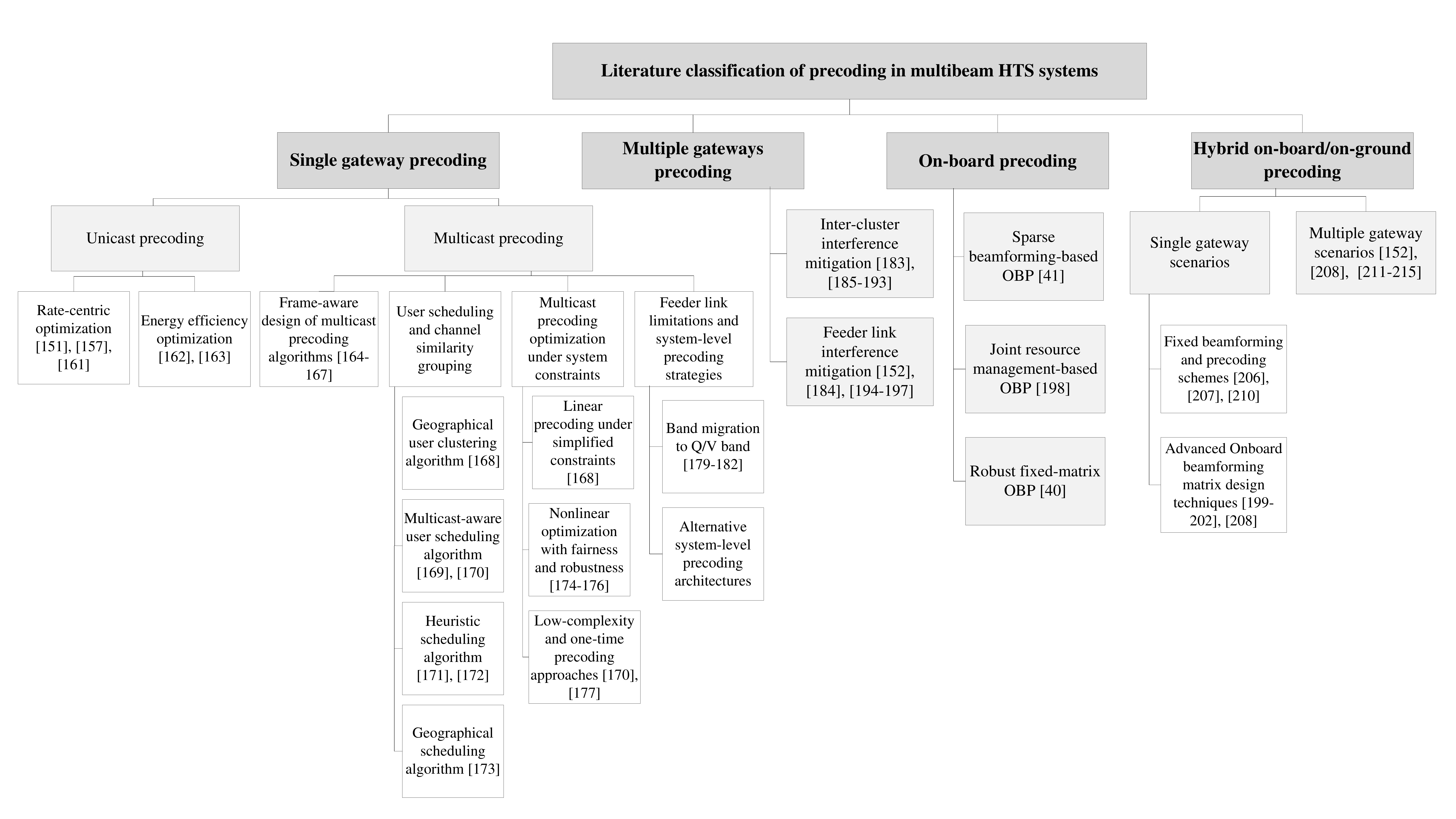}
\end{center}
\caption{Literature classification of precoding in multibeam HTS.}%
\label{precoding}
\end{figure*}

Fig. \ref{precoding} provides a structured overview of the development and interrelations of precoding research in HTS systems. Each reference is categorized, and their contributions and relationships to other works in the field are highlighted.

\section{Coordination Mechanism of Resource Allocation and Precoding}

In multibeam HTS systems, adjacent beams may share the same frequency resources, leading to inter-beam interference. Resource allocation must coordinate frequency reuse and power control to mitigate such interference \cite{hong2008optimal,chatzinotas2011joint}. Proper bandwidth and resource allocation help ensure precoding performance, while effective time slot allocation, when combined with precoding, enhances overall system efficiency.  Precoding can enhance the rationality and flexibility of resource allocation \cite{zheng2012generic, taricco2014linear}. These two components are tightly coupled, with each influencing the other’s design space, operational flexibility, and overall system performance.

To better understand this mutual dependence, we categorize their interactions into six key aspects: 1) interference management, 2) power constraints, 3) spectral resource allocation, 4) user scheduling, 5) on-board processing capability, and 6) feedback link efficiency. Each aspect reveals how design choices or limitations in one domain shape the optimization process of the other, providing guidance for joint system-level design.
The bidirectional relationship between resource allocation and precoding in multibeam HTS systems is illustrated in Table \ref{mutual_precoding_resource}. 

\begin{table*}[t]
\centering
\caption{Mutual Impacts Between Resource Allocation and Precoding.}
\renewcommand\arraystretch{1.2}
\resizebox{0.97\textwidth}{!}{
\begin{tabular}{l|l|l}
\hline \hline
\textbf{Aspect} & \textbf{\begin{tabular}[c]{@{}c@{}}Impact of resource allocation on precoding \end{tabular}} & \textbf{Impact  of precoding on resource allocation} \\ \hline

\makecell[l]{Interference management} &  
\makecell[l]{Defines the interference matrix that \\ precoding must resolve \cite{chen2021system}.} &
\makecell[l]{Effective interference mitigation enables \\ higher frequency reuse levels \cite{hong2008optimal}.} \\ \hline

\makecell[l]{Power constraints} &  
\makecell[l]{Precoding design must optimize beamforming \\ gain within the power budget.} &
\makecell[l]{Precoding performance can guide reverse \\ optimization of power allocation \cite{zheng2012generic}.} \\ \hline

\makecell[l]{Spectral resource allocation} & 
\makecell[l]{Determines the dimensionality and spatial \\ reuse strategy of the precoding matrix.} &
\makecell[l]{Helps evaluate which beams are suitable \\ for dense frequency reuse.} \\ \hline

\makecell[l]{User scheduling} &  
\makecell[l]{Provides the column vectors for constructing \\ the precoding matrix \cite{hong2008optimal}.} &
\makecell[l]{Precoding favors user groups with low \\ channel correlation, influencing scheduling.} \\ \hline

\makecell[l]{on-board processing capability} &  
\makecell[l]{Precoding must adapt to on-board or \\ ground-based hardware constraints.} &
\makecell[l]{Enables more flexible and elastic \\ resource allocation strategies.} \\ \hline

\makecell[l]{Feedback link efficiency} &  
\makecell[l]{Determines whether precoding can be \\ updated in real-time \cite{chatzinotas2011joint}.} &
\makecell[l]{Supports iterative optimization of allocation \\ via intelligent algorithms \cite{taricco2014linear}.} \\ \hline \hline

\end{tabular}
}
\label{mutual_precoding_resource}
\end{table*}


\emph{1) Interference Management:}  Resource allocation defines the interference pattern that precoding must mitigate. Conversely, effective precoding can significantly reduce co-channel interference, thereby enabling more aggressive frequency reuse across beams \cite{chen2021system, hong2008optimal}.

\emph{2)  Power Constraints:} Resource allocation sets the power budget for each beam, within which the precoder must operate. On the other hand, feedback from precoding performance—such as achieved signal quality—can guide power redistribution to maximize system throughput \cite{zheng2012generic}.

\emph{3) Spectral Resource Allocation:}  The allocation of frequency bands and reuse factors affects the structure and rank of the precoding matrix. At the same time, effective precoding can help determine which beams can tolerate tighter spectral reuse, influencing future resource planning.

\emph{4) User Scheduling:}  The user set selected for transmission in each frame directly determines the input vectors for precoding. Precoding performance is highly sensitive to the spatial correlation among users, thereby influencing which users are scheduled simultaneously \cite{hong2008optimal}.

\emph{5) On-Board Processing Capability:}  Hardware constraints (e.g., on the satellite payload) limit the complexity and flexibility of precoding. In return, advanced precoding schemes can unlock more flexible and efficient resource allocation strategies, such as adaptive bandwidth or beam shaping.

\emph{6) Feedback Link Efficiency:} The capacity and latency of CSI feedback channels affect the timeliness and accuracy of precoding updates. Moreover, precoding systems that incorporate real-time or predictive feedback can facilitate more intelligent and dynamic resource allocation decisions \cite{chatzinotas2011joint, taricco2014linear}.

In conclusion, this mutual dependency between resource allocation and precoding underscores the necessity of joint optimization frameworks in the design of future HTS systems. Viewing these elements as interconnected rather than independent modules enables the development of more flexible, intelligent, and performance-optimized satellite communication architectures.




\section{Challenges and Research Perspectives}

\subsection{The Application of Artificial Intelligence}

In Section \RNum{5}, all the referenced works assume perfect CSI. However, in satellite communication systems—especially in HTS networks—the rapid movement of satellites and long round-trip delays render the acquired CSI outdated by the time it is used. This channel aging issue severely degrades the effectiveness of precoding matrices and thus significantly reduces the overall system throughput.

To tackle this problem, several studies have proposed DL-based channel estimation techniques in MIMO systems, aiming to achieve more accurate and timely CSI with reduced estimation latency \cite{ye2017power,he2018deep,jin2019channel,guo2020deep,zhang2021deep,lu2021mimo,10534904}. These DL-based methods have shown strong potential in satellite environments as well, where conventional estimation techniques struggle with non-stationary channels and delayed feedback. Their ability to generalize over time-varying channel conditions makes them well-suited for the unique challenges of satellite communications.

Moreover, in addressing the non-convex optimization problems discussed in Section \RNum{5}, traditional iterative algorithms used for calculating the optimal precoding matrix typically involve high computational complexity and latency. To alleviate this, recent research has explored the application of DL-based algorithms for downlink beamforming matrix design, replacing the need for computationally intensive iterative procedures \cite{huang2018unsupervised,huang2019fast,zhang2020deep,zhang2021model,xia2019deep}. While these methods were initially developed for terrestrial MIMO systems, their low-latency and data-driven characteristics make them highly attractive for satellite scenarios—particularly for on-board processing or near-real-time adaptation in GEO and LEO HTS systems. These approaches have demonstrated promising results in terms of achieving competitive spectral efficiency, reducing computational delay, and mitigating the impact of outdated CSI in dynamic satellite environments.

In distributed mega constellation networks, each satellite operates with independent data processing capabilities, contributing to increased system capacity and reduced transmission power through a distributed approach. However, maintaining synchronization among satellites is crucial, necessitating precise inter-satellite timing and phase alignment. Conventional synchronization methods involve processing receiver feedback either through open-loop or closed-loop strategies.  The superior efficacy of AI algorithms are highlighted in addressing synchronization challenges within end-to-end communication systems compared to traditional methods \cite{wang2013learning,wang2017detecting,lee2019synchronization,zibar2015application,wu2019deep,tong2020enhanced}. Notably, these algorithms primarily cater to ground communication scenarios rather than satellite communication environments. In addition, each satellite communicates information through routing algorithms in distributed mega constellation networks. The continuous increase in the number of satellites within these massive constellations results in a rise in both node count and path count in the network, escalating the complexity of routing algorithms due to the variable path quantity. The superior efficacy and reduced load are achieved by employing AI algorithms to address routing concerns among satellites  \cite{kato2019optimizing,cigliano2020machine,qiu2019deep}.

While the papers on employing AI to tackle issues in satellite communication is expanding \cite{fourati2021artificial}, current surveys indicate that AI algorithms have not been practically integrated into satellite communication systems. Numerous challenges persistently hinder this integration. Although deploying AI algorithms at ground stations presents advantages due to the ample computing resources supporting extensive data training, issues arise from potential obsolescence caused by round-trip delays in decision-making \cite{fontanesi2023artificial}. While satellites capture ample data in space, the power and computational requirements for AI algorithms exceed current satellite capabilities. Despite the evident advantages of AI algorithms in satellite communication, a critical challenge remains how to effectively apply these AI algorithms within operational communication satellite systems.

\subsection{Extremely High Frequency Communication}
Currently, the primary frequency band for satellite communication is in the Ka band. With the increasing number of satellites in orbit, the available Ka band spectrum is becoming scarce. Consequently, the extremely high frequency (EHF)  band is poised to emerge as the predominant frequency range for future communications, offering higher communication rates and larger bandwidth. Moreover, the migration of feeder links towards EHF band not only aligns with the requirements of multibeam HTS systems but also frees the scarce and valuable Ka band, providing increased flexibility for the user segment \cite{stallo2009transponders}. 

Despite these advantages, communicating in the EHF band still faces several challenges. The foremost challenge lies in the rapid attenuation of communication signal energy or even potential disruptions in communication links due to rainfall, stemming from the close resemblance in wavelength between electromagnetic waves in the EHF band and raindrop diameters. Therefore, it is an essential research topic to solve the link outages in the EHF band. Accurate and adaptive prediction of link outage models enable more efficient adjustment of user access to maintain high quality communication services \cite{de2018adaptive}. Accurate estimation of rain attenuation models in the Q/V band is essential, as regression models that approximate the standard international telecommunication union (ITU) model can help reduce system errors\cite{rossi2022experimental}. Moreover, the deployment of intelligent gateways ensures the availability of necessary links and robustness against rain attenuation while maintaining optimal performance because redundant intelligent gateways can effectively route essential information to gateways affected by rainfall \cite{kyrgiazos2012gateway,codispoti2018validation,kyrgiazos2014gateway,delamotte2019smart,H2020-COMPET-2016}. It remains an area of future research to address interruptions in EHF band communication, focusing on the feasibility of implementing adaptive prediction model algorithms and balancing system cost due to the number of gateway deployments against system performance. This aims to facilitate the earlier utilization of the EHF band, ensuring a better user experience.

\subsection{Further Development of Communication Hardware}

Currently, the production of high-power transmitters in the EHF band poses challenges, characterized by high noise coefficients and subpar performance. Consequently, future research endeavors focus on developing high-power EHF band transmitters, highly sensitive receivers, and adaptive antenna arrays. This effort aims to render satellite communication feasible in the EHF band from a hardware feasibility perspective \cite{yu2021generation}. Fully digital payloads offer heightened flexibility to meet the growing demands for dynamic allocations \cite{braun2021processing}. Within these payloads, employing higher-order modulation and coding schemes helps reduce the per-bit communication costs. However, achieving fully digital payloads necessitates higher-power amplifiers and enhanced power generation capabilities on-board satellites to meet computational resource demands. Therefore, future research and breakthroughs focus on developing fully digital payloads with increased power generation capacity and efficiency, aiming to reduce satellite manufacturing costs while maintaining high computational capabilities within satellite payloads.

In addition to the payload advancements, the integration of on-board AI presents both opportunities and challenges. AI algorithms can enhance satellite autonomy \cite{sun2024knowledge,sun2025acomprehesive}, reduce dependence on ground control, and offer improved privacy protection. However, their deployment is constrained by limited on-board computational resources and long convergence times \cite{fontanesi2022transfer}. Although the current embedded AI chip market is dominated by NVIDIA, AMD, Intel, and Qualcomm, commercialization is gradually advancing \cite{pang2022ai}. Consequently, the potential emergence of spaceborne AI embedded chips with acceptable computational complexity may pave the way for satellite communications to enter new frontiers \cite{10443615}. Ensuring uninterrupted satellite communication services for users requires their devices to track satellite movements and execute seamless transitions between satellites, which imposes higher demands on the complexity of these user devices \cite{jakoby2020microwave}. Therefore, the future research focus for satellite communication terminal manufacturers lies in designing next-generation antennas and terminal equipment with increased flexibility and lower costs \cite{guidotti2019lte}. In conclusion, only through further advancements in communication hardware can users truly experience more convenient and comfortable satellite communication data services.

\section{Conclusions}


This aticle has provided a comprehensive review of the research progress in the field of multibeam HTS systems over the past two decades, focusing on hardware foundations, resource allocation, and precoding techniques. With the advancement of satellite communication technologies, the hardware foundation of HTS systems has undergone significant evolution. Technological progress in ground stations, on-board payloads, and UTs has provided the potential for higher communication capacity and efficiency. In particular, the development of phased array antennas and digital transparent processors has laid the groundwork for the flexibility and reconfigurability of satellite communication. Resource allocation techniques are crucial for enhancing spectral efficiency and QoS in HTS systems. Research on dynamic bandwidth allocation, power allocation, and time slot allocation has enabled satellite systems to adapt to varying traffic demands and environmental conditions, thereby optimizing system performance. Precoding technology plays an essential role in HTS systems by effectively mitigating multi-user interference and enhancing system capacity. The evolution from single-gateway precoding to multi-gateway precoding, and further to on-board precoding and hybrid  on-board/on-ground  precoding, offers a variety of options to meet the needs of different application scenarios.

Despite the progress in HTS research, numerous challenges remain, such as the accuracy of CSI, synchronization of inter-satellite links, and the complexity of hardware implementation. The application of AI in channel estimation, resource management, and routing optimization provides new avenues to address these challenges. With the evolution towards 6G communication technologies, HTS systems are expected to play a more significant role in providing global coverage, supporting IoT, and realizing integrated on-board/on-ground networks. Additionally, the development of EHF communication technologies, such as Q/V band applications, will bring new opportunities and challenges for satellite communication.

\bibliographystyle{IEEEtran}
\bibliography{IEEEabrv,main}

\end{document}